\documentclass[10pt]{article}
\pdfoutput=1
\usepackage{mathStyle}
\usepackage{amsmath,amssymb,amsthm,latexsym,bbm,calc,wasysym,mathtools,empheq, yfonts,upgreek}
\usepackage[english]{babel}
\usepackage[usenames,dvipsnames]{xcolor}
\usepackage{relsize}
\usepackage{graphicx}
\usepackage{array}
\usepackage{tabularx,float}
\usepackage[export]{adjustbox}
\usepackage[mathscr]{euscript}
\usepackage[ampersand]{easylist}

\usepackage{tikz-cd}
\usepackage{tikz}
\usetikzlibrary{shapes.geometric}
\usetikzlibrary{arrows,matrix,calc,scopes,decorations.markings,intersections}
\usetikzlibrary{arrows.meta}
\usetikzlibrary{decorations.pathreplacing,angles,quotes,patterns}

\usepackage{pifont,dsfont}
\usepackage{marvosym}
\usepackage{slashed}
\usepackage{subfigure}
\usepackage{sidecap}
\usepackage{stmaryrd}
\usepackage{empheq}
\usepackage{placeins}
\usepackage{enumitem}
\usepackage{verbatim}
\usepackage{upgreek}
\newcolumntype{Y}{>{\centering\arraybackslash}X}

\DeclareMathAlphabet{\mathpzc}{OT1}{pzc}{m}{it}

\usepackage{amsthm}
\newtheoremstyle{mydef}%
	{0.9em} %Space above
	{0.7em}% Space below
	{\itshape\hangindent=2em}% Body font 
	{1.8em}% Indent amount
	{\scshape}% ⟨Theorem head font⟩
	{.}% ⟨Punctuation after theorem head ⟩
	{.5em}% Space after theorem head 
	{}%
	
\theoremstyle{mydef} 
\newtheorem{definition}{Definition}
\numberwithin{definition}{section}

\theoremstyle{mydef}
\newtheorem{lemma}{Lemma}
\numberwithin{lemma}{section}

\theoremstyle{mydef}
\newtheorem{theorem}{Theorem}
\numberwithin{theorem}{section}

\theoremstyle{mydef}
\newtheorem{example}{Example}
\numberwithin{example}{section}

\theoremstyle{mydef}
\newtheorem{property}{Property}
\numberwithin{property}{section}

\theoremstyle{mydef}

\numberwithin{proposition}{section}

\theoremstyle{mydef}

\numberwithin{remark}{section}

\theoremstyle{mydef}

\numberwithin{conjecture}{section}

\renewenvironment{proof}{{\scshape Proof.}}{\qed}

\newcommand{\rU}{{\rm U}}

\newcommand{\la}{\langle}
\newcommand{\ra}{\rangle}
\newcommand{\q}{\quad}
\newcommand{\nn}{\nonumber}
\newcommand{\sss}{\scriptstyle}
\newcommand{\ssss}{\scriptscriptstyle}
\newcommand{\bul}{{\sss \bullet}}

\newcommand{\mc}[1]{\mathcal{#1}}
\newcommand{\ms}[1]{\mathscr{#1}}
\newcommand{\fr}[1]{\mathfrak{#1}}
\newcommand{\msf}[1]{\mathsf{#1}}

\newcommand{\Ob}{{\rm Ob}}
\newcommand{\End}{{\rm End}}
\newcommand{\Hom}{{\rm Hom}}
\newcommand{\HOM}{\mathsf{Hom}}
\newcommand{\Vect}{\mathsf{Vec}}
\newcommand{\TVect}{\mathsf{2Vec}}
\newcommand{\Mod}{\mathsf{Mod}}
\newcommand{\MOD}{\mathsf{MOD}}
\newcommand{\GrAlg}{\mathbb C[\Lambda^2  G]^{\msf t^2(\pi)}}
\newcommand{\VectGr}{\mathsf{Vec}^{{\msf t(\pi)}}_{\Lambda  G}}
\newcommand{\TVectGr}{\mathsf{2Vec}^{\pi}_{G}}
\newcommand{\btimes}{\text{\small{\raisebox{-0.4pt}[0pt]{$ \; \boxtimes \; $}}}}
\newcommand{\btimesFt}{\text{\scriptsize{\raisebox{-0.3pt}[0pt]{$ \,  \boxtimes \, $}}}}
\newcommand{\bigbplus}{
	\mathop{
		\vphantom{\bigoplus} 
		\mathchoice
		{\vcenter{\hbox{\resizebox{\widthof{$\displaystyle\bigoplus$}}{!}{$\boxplus$}}}}
		{\vcenter{\hbox{\resizebox{\widthof{$\bigoplus$}}{!}{$\boxplus$}}}}
		{\vcenter{\hbox{\resizebox{\widthof{$\scriptstyle\oplus$}}{!}{$\boxplus$}}}}
		{\vcenter{\hbox{\resizebox{\widthof{$\scriptscriptstyle\oplus$}}{!}{$\boxplus$}}}}
	}\displaylimits 
}

\tikzset{->1-/.style={decoration={
			markings,
			mark=at position #1 with {\arrow{Triangle[fill=black, width=3.5pt, length=3.5pt]}}},postaction={decorate}}}
\tikzset{->2-/.style={decoration={
					markings,
					mark=at position #1 with {\arrow{Triangle[fill=white, width=5pt, length=5pt]}}},postaction={decorate}}}

\tikzset{->-/.style={decoration={
			markings,
			mark=at position #1 with {\arrow{latex}}},postaction={decorate}}}

\tikzset{-<-/.style={decoration={
			markings,
			mark=at position #1 with {\arrow{latex reversed}}},postaction={decorate}}}

\tikzset{
	every picture/.append style={
		line join=round,
		line width = 0.6pt
	}
}

\tikzstyle{dot}=[dotted, line width=0.8pt]
\tikzstyle{op}=[opacity=0.4]
\tikzstyle{op8}=[opacity=0.8]
\tikzstyle{op7}=[opacity=0.7]
\tikzstyle{op6}=[opacity=0.6]
\tikzstyle{op5}=[opacity=0.5]
\tikzstyle{op4}=[opacity=0.4]
\tikzstyle{op3}=[opacity=0.3]
\tikzstyle{op2}=[opacity=0.2]
\tikzstyle{op1}=[opacity=0.1]
\tikzstyle{sha}=[shorten >= 0.3em]
\tikzstyle{shb}=[shorten <= 0.3em]
\tikzstyle{shab}=[shorten >= 0.3em, shorten <= 0.3em]
\tikzstyle{fat}=[white, double = black, line width = 2pt, double distance = 0.6pt]

\def\centerarc[#1](#2)(#3:#4:#5)
{ \draw[#1] ($(#2)+({#5*cos(#3)},{#5*sin(#3)})$) arc (#3:#4:#5); }

\tikzset{
	partial ellipse/.style args={#1:#2:#3}{
		insert path={+ (#1:#3) arc (#1:#2:#3)}
	}
}

\newcolumntype{M}{>{\centering\arraybackslash}m{1cm}}

\newcommand\tikzmark[2]{%
	\tikz[remember picture,baseline] \node[] (#1){#2};%
}

\newcommand\linkTab[2]{%
	\begin{tikzpicture}[remember picture, overlay, shift={(0,0)}]
	\draw[|->] (#1) to node[pos=0.3, above, yshift=-1pt, sloped] {\textsf{\scriptsize Dim}} (#2);
	\end{tikzpicture}%
}

\newcommand\curvTabL[2]{%
	\begin{tikzpicture}[remember picture, overlay, shift={(0,0)}]
	\draw[->] (#1) to[out=180, in=180, looseness=1.5] node[pos=0.5, right, xshift=1pt] {${\sss \times \mathbb S^1}$} (#2) ;
	\end{tikzpicture}%
}

\newcommand\curvTabR[2]{%
	\begin{tikzpicture}[remember picture, overlay, shift={(0,0)}]
	\draw[->] (#1) to[out=0, in=0, looseness=1.5] node[pos=0.5, left, xshift=-1pt] {${\sss \times \mathbb S^1}$} (#2) ;
	\end{tikzpicture}%
}

\newcommand\mapTab[2]{%
	\begin{tikzpicture}[remember picture, overlay, shift={(0,0)}]
	\draw[|->, shorten <= 15pt, shorten >= 15pt] (#1) --node[pos=0.5, above] {${\sss \mc Z^\pi_G}$} (#2) ;
	\end{tikzpicture}%
}

\newcommand{\braidGroup}[3]{
	\begin{tikzpicture}[scale=#1,baseline={([yshift=0.3ex]current bounding box.center)}]
	\ifthenelse{#2 = 1}{
		\coordinate (0) at (0,0);
		\coordinate (1) at (0,1);
		\draw[] (0) -- (1);
		\node[below] at (0) {${\sss #3}$};
	}{}
	\ifthenelse{#2 = 2}{
		\coordinate (0) at (0,0);
		\coordinate (1) at (0,1);
		\coordinate (2) at (0.5,0);
		\coordinate (3) at (0.5,1);
		\draw[] (0) -- (1);
		\draw[] (2) -- (3);
		\node[below] at (0) {${\sss #3}$};
		\node[below] at (2) {${\sss #3+1}$};
	}{}
	\ifthenelse{#2 = 3}{
		\coordinate (0) at (0,0);
		\coordinate (1) at (0,1);
		\coordinate (2) at (0.5,0);
		\coordinate (3) at (0.5,1);
		\draw[] (1) -- (2);
		\draw[fat] (0) -- (3);
		\node[below] at (0) {${\sss #3}$};
		\node[below] at (2) {${\sss #3+1}$};
	}{}
	\ifthenelse{#2 = 4}{
		\coordinate (0) at (0,0);
		\coordinate (1) at (0,1);
		\coordinate (2) at (0.5,0);
		\coordinate (3) at (0.5,1);
		\draw[] (0) -- (3);
		\draw[fat] (1) -- (2);
		\node[below] at (0) {${\sss #3}$};
		\node[below] at (2) {${\sss #3+1}$};
	}{}
	\end{tikzpicture}
}

\newcommand{\loopTube}[2]{
	\begin{tikzpicture}[scale=#1,baseline={([yshift=0.7ex]current bounding box.center)}]
	\coordinate (4) at (-1.5,5);
	\coordinate (5) at (1.5,5);
	\coordinate (8) at (-1.5,-5);
	\coordinate (9) at (1.5,-5);
	\draw (4) to[out=-90, in=-90, looseness=0.5] (5);
	\draw (4) to[out=90, in=90, looseness=0.5] (5);
	\draw (8) to[out=-90, in=-90, looseness=0.5] (9);
	\draw[op6] (8) to[out=90, in=90, looseness=0.5] (9);
	\draw (4) -- (8);
	\draw (5) -- (9);
	\node[below, yshift=-3.5pt] at (0,-5) {${\sss #2}$};
	\end{tikzpicture}
}

\newcommand{\loopBraid}[2]{
	\begin{tikzpicture}[scale=#1,baseline={([yshift=0.7ex]current bounding box.center)}]
	\coordinate (4) at (-1.5,5);
	\coordinate (5) at (1.5,5);
	\coordinate (8) at (-1.5,-5);
	\coordinate (9) at (1.5,-5);
	\coordinate (6) at (2.5,5);
	\coordinate (7) at (5.5,5);
	\coordinate (10) at (2.5,-5);
	\coordinate (11) at (5.5,-5);
	\ifthenelse{#2 = 1}{
		\coordinate (0) at (0.25,0);
		\coordinate (1) at (3.75,0);
		\coordinate (2) at (0.75,0);
		\coordinate (3) at (3.25,0);
		\draw (0) to[out=-90, in=-90, looseness=0.5] (1);
		\path[name path= P01] (0) to[out=90, in=90, looseness=0.5] (1);
		\draw[op6] (2) to[out=-90, in=-90, looseness=0.5] (3);
		\draw[op3] (2) to[out=90, in=90, looseness=0.5] (3);
		\path[name path = P62] (6) to[out=-95, in=90, looseness=1.5] (2);
		\path[name path = P73] (7) to[out=-95, in=90, looseness=1.5] (3);
		\path[name path = P28] (2) to[out=-90, in=85, looseness=1.5, edge node={coordinate[pos=0.18] (sp3)}, edge node={coordinate[pos=0.33] (sp5)}] (8);
		\path[name path = P39] (3) to[out=-90, in=85, looseness=1.5, edge node={coordinate[pos=0.63] (sp4)}, edge node={coordinate[pos=0.78] (sp6)}] (9);
		\draw[name path = P40] (4) to[out=-85, in=90, looseness=1.5] (0);
		\path[name path = P51] (5) to[out=-85, in=90, looseness=1.5] (1);
		\draw[name path = P010] (0) to[out=-90, in=95, looseness=1.5] (10);
		\draw[name path = P111] (1) to[out=-90, in=95, looseness=1.5] (11);
		\path [name intersections={of=P62 and P51,by={sp1}}];
		\path [name intersections={of=P73 and P51,by={sp2}}];
		\draw[name path=Psp12] (sp1) to[out=140, in=-65, looseness=2] (sp2);
		\draw[op6] (sp3) to[out=-100, in=-120, looseness=0.3] (sp4);
		\draw[op3] (sp3) to[out=80, in=60, looseness=0.3] (sp4);	
		\draw (sp5) to[out=-100, in=-120, looseness=0.3] (sp6);
		\draw (sp5) to[out=80, in=60, looseness=0.3] (sp6);	
		\path [name intersections={of=Psp12 and P62, by={sp7}}];
		\path [name intersections={of=Psp12 and P73, by={sp8}}];
		\path [name intersections={of=P01 and P62, by={sp9}}];
		\path [name intersections={of=P01 and P73, by={sp10}}];		
		\begin{scope}
			\clip (6) -- (sp7) -- (sp8) -- (5.5,1) -- (7) -- (6);
			\draw (6) to[out=-95, in=90, looseness=1.5] (2);
			\draw (7) to[out=-95, in=90, looseness=1.5] (3);
		\end{scope}
		\begin{scope}
			\clip (sp7) -- (sp8) -- (3.5,0) -- (0.6,0) -- (0.75,2) -- (sp7);
			\draw[op6] (6) to[out=-95, in=90, looseness=1.5] (2);
			\draw[op6] (7) to[out=-95, in=90, looseness=1.5] (3);
		\end{scope}
		\begin{scope}
			\clip (0.6,0) -- (sp3) -- (sp4) -- (3.25,-2) -- (3.5,0) -- (0.6,0);
			\draw[op6] (2) to[out=-90, in=85, looseness=1.5] (8);
			\draw[op6] (3) to[out=-90, in=85, looseness=1.5] (9);
		\end{scope}
		\begin{scope}
			\clip (8) -- (-1.5,-3) -- (sp5) -- (sp6) -- (9) -- (8);
			\draw[op6] (2) to[out=-90, in=85, looseness=1.5] (8);
			\draw[op6] (3) to[out=-90, in=85, looseness=1.5] (9);
		\end{scope}
		\begin{scope}
			\clip (0) rectangle (sp9);
			\draw[op6] (0) to[out=90, in=90, looseness=0.5] (1);
		\end{scope}
		\begin{scope}
			\clip (1) rectangle (sp10);
			\draw[op6] (0) to[out=90, in=90, looseness=0.5] (1);
		\end{scope}
		\begin{scope}
			\clip (sp9) -- (sp10) -- (3.5,1) -- (0.5,1) -- (sp9);
			\draw[op2] (0) to[out=90, in=90, looseness=0.5] (1);
		\end{scope}
		\begin{scope}
			\clip (5) rectangle (sp1);
			\draw[] (5) to[out=-85, in=90, looseness=1.5] (1);
		\end{scope}
		\begin{scope}
			\clip (sp1) rectangle (sp2);
			\draw[op3] (5) to[out=-85, in=90, looseness=1.5] (1);
		\end{scope}
		\begin{scope}
			\clip (sp2) rectangle (4,0);
			\draw (5) to[out=-85, in=90, looseness=1.5] (1);
		\end{scope}
	}{}	
	\ifthenelse{#2 = 2}{
		\path[name path=P410] (4) to[out=-85, in=95, looseness=1] (10);
		\path[name path=P511] (5) to[out=-85, in=95, looseness=1] (11);
		\draw[name path=P68] (6) to[out=-95, in=85, looseness=1] (8);
		\draw[name path=P79] (7) to[out=-95, in=85, looseness=1] (9);	
		\path [name intersections={of=P511 and P68, by={sp1}}];	
		\path [name intersections={of=P511 and P79, by={sp2}}];
		\path [name intersections={of=P410 and P68, by={sp3}}];	
		\path [name intersections={of=P410 and P79, by={sp4}}];
		\begin{scope}
			\clip (5) -- (sp1) -- (sp3) -- (-1.5,2) -- (-1.6,5) -- (5);
			\draw (4) to[out=-95, in=85, looseness=1] (10);
			\draw (5) to[out=-95, in=85, looseness=1] (11);	
		\end{scope}	
		\begin{scope}
			\clip (sp1) -- (sp2) -- (sp4) -- (sp3) -- (sp1);
			\draw[op3] (4) to[out=-95, in=85, looseness=1] (10);
			\draw[op3] (5) to[out=-95, in=85, looseness=1] (11);	
		\end{scope}	
		\begin{scope}
			\clip (10) -- (sp4) -- (sp2) -- (5.5,-2) -- (5.6,-5) -- (10);
			\draw (4) to[out=-95, in=85, looseness=1] (10);
			\draw (5) to[out=-95, in=85, looseness=1] (11);		
		\end{scope}		
	}{}
	\draw (4) to[out=-90, in=-90, looseness=0.5] (5);
	\draw (4) to[out=90, in=90, looseness=0.5] (5);
	\draw (8) to[out=-90, in=-90, looseness=0.5] (9);
	\draw[op6] (8) to[out=90, in=90, looseness=0.5] (9);
	\draw (6) to[out=-90, in=-90, looseness=0.5] (7);
	\draw (6) to[out=90, in=90, looseness=0.5] (7);
	\draw (10) to[out=-90, in=-90, looseness=0.5] (11);
	\draw[op6] (10) to[out=90, in=90, looseness=0.5] (11);
	\node[below, yshift=-3.5pt] at (0,-5) {${\sss i}$};
	\node[below, yshift=-3.5pt] at (4,-5) {${\sss i+1}$};
	\end{tikzpicture}
}

\newcommand{\movieLoop}[2]{
	\begin{tikzpicture}[scale=#1,baseline={([yshift=0.7ex]current bounding box.center)}]
	\coordinate (4) at (-1.5,5);
	\coordinate (5) at (1.5,5);
	\coordinate (8) at (-1.5,-5);
	\coordinate (9) at (1.5,-5);
	\coordinate (6) at (2.5,5);
	\coordinate (7) at (5.5,5);
	\coordinate (10) at (2.5,-5);
	\coordinate (11) at (5.5,-5);
	\coordinate (0) at (0.25,0);
	\coordinate (1) at (3.75,0);
	\coordinate (2) at (0.75,0);
	\coordinate (3) at (3.25,0);
	\coordinate (20) at (-0.625,2.5);
	\coordinate (21) at (2.875,2.5);
	\coordinate (22) at (1.625,2.5);
	\coordinate (23) at (4.125,2.5);
	\coordinate (24) at (1.125,-2.5);
	\coordinate (25) at (4.625,-2.5);
	\coordinate (26) at (-0.375,-2.5);
	\coordinate (27) at (2.125,-2.5);
	
	\ifthenelse{#2 = 1}{
		\draw (0) to[out=-90, in=-90, looseness=0.5] (1);
		\draw (0) to[out=90, in=90, looseness=0.5] (1);
		\draw (2) to[out=-90, in=-90, looseness=0.5] (3);
		\draw (2) to[out=90, in=90, looseness=0.5] (3);
		
		\draw (20) to[out=-90, in=-90, looseness=0.5] (21);
		\draw (20) to[out=90, in=90, looseness=0.5] (21);
		\draw[fat] (22) to[out=-90, in=-90, looseness=0.5] (23) to[out=90, in=90, looseness=0.5] (22);

		\draw (26) to[out=-90, in=-90, looseness=0.5] (27) to[out=90, in=90, looseness=0.5] (26);
		\draw[fat] (24) to[out=-90, in=-90, looseness=0.5] (25) to[out=90, in=90, looseness=0.5] (24);
		
		\draw (4) to[out=-90, in=-90, looseness=0.5] (5);
		\draw (4) to[out=90, in=90, looseness=0.5] (5);
		\draw (8) to[out=-90, in=-90, looseness=0.5] (9);
		\draw (8) to[out=90, in=90, looseness=0.5] (9);
		\draw (6) to[out=-90, in=-90, looseness=0.5] (7);
		\draw (6) to[out=90, in=90, looseness=0.5] (7);
		\draw (10) to[out=-90, in=-90, looseness=0.5] (11);
		\draw (10) to[out=90, in=90, looseness=0.5] (11);
		\node[below, yshift=-3.5pt] at (0,-5) {${\sss i}$};
		\node[below, yshift=-3.5pt] at (4,-5) {${\sss i+1}$};
	}{}
	\ifthenelse{#2 = 2}{
		\draw (0) to[out=-90, in=-90, looseness=0.5] (1);
		\draw (0) to[out=90, in=90, looseness=0.5] (1);
		\draw (2) to[out=-90, in=-90, looseness=0.5] (3);
		\draw (2) to[out=90, in=90, looseness=0.5] (3);
		
		\draw (20) to[out=-90, in=-90, looseness=0.5] (21);
		\draw (20) to[out=90, in=90, looseness=0.5] (21);
		\draw[fat] (22) to[out=-90, in=-90, looseness=0.5] (23) to[out=90, in=90, looseness=0.5] (22);

		\draw (24) to[out=-90, in=-90, looseness=0.5] (25) to[out=90, in=90, looseness=0.5] (24);
		\draw[fat] (26) to[out=-90, in=-90, looseness=0.5] (27) to[out=90, in=90, looseness=0.5] (26);
		
		\draw (4) to[out=-90, in=-90, looseness=0.5] (5);
		\draw (4) to[out=90, in=90, looseness=0.5] (5);
		\draw (8) to[out=-90, in=-90, looseness=0.5] (9);
		\draw (8) to[out=90, in=90, looseness=0.5] (9);
		\draw (6) to[out=-90, in=-90, looseness=0.5] (7);
		\draw (6) to[out=90, in=90, looseness=0.5] (7);
		\draw (10) to[out=-90, in=-90, looseness=0.5] (11);
		\draw (10) to[out=90, in=90, looseness=0.5] (11);
		\node[below, yshift=-3.5pt] at (0,-5) {${\sss i}$};
		\node[below, yshift=-3.5pt] at (4,-5) {${\sss i+1}$};
	}{}
	\end{tikzpicture}
}

\newcommand{\necklace}[1]{
	\begin{tikzpicture}[scale=#1,baseline=0pt]
	\coordinate (0) at (0,0);
	\coordinate (1) at (0.5,0);
	\coordinate (5) at (1.5,-0.9);
	\coordinate (8) at (1.5,0.9);
	\coordinate (6) at (3,-0.9);
	\coordinate (9) at (3,0.9);
	\coordinate (7) at (4.5,-0.9);
	\coordinate (10) at (4.5,0.9);
	\coordinate (3) at (5.5,0);
	\coordinate (4) at (6,0);

	\draw[dotted] (0) -- (1)
		  (3) -- (4);
	\draw[] (5) to[out=180, in=-180, looseness=0.8] (8);
	\draw[] (6) to[out=180, in=-180, looseness=0.8] (9);
	\draw[] (7) to[out=180, in=-180, looseness=0.8] (10);
	\draw[fat] (1) -- (3);	
	\begin{scope}
		\clip (1.5,0.92) rectangle (4.5,-0.92);
		\draw[fat] (5) to[out=0, in=0, looseness=0.8] (8);	
	\end{scope}
	\begin{scope}
		\clip (4.5,0.92) rectangle (5.5,-0.92);
		\draw[fat] (7) to[out=0, in=0, looseness=0.8] (10);	
	\end{scope}
	\fill[white, opacity=0.5] (10) to[out=-168, in=-12, looseness=1] (8) to[out=0, in=0, looseness=0.8] (5) to[out=12, in=168, looseness=1] (7) to[out=0, in=0, looseness=0.8] (10);
	\draw[line width=0.4pt, ->1-=0.25, ->1-=0.75] (10) to[out=-168, in=-12, looseness=1] (8);
	\draw[line width=0.4pt, ->1-=0.25, ->1-=0.75] (7) to[out=168, in=12, looseness=1] (5);
	\begin{scope}
		\clip (3,0.92) rectangle (4.5,-0.92);
		\draw[fat] (6) to[out=0, in=0, looseness=0.8] (9);	
	\end{scope}
	\node[below] at (6) {${\sss i}$};
	\node[below] at (7) {${\sss i+1}$};
	\end{tikzpicture}
}

\newcommand{\pointObj}[1]{
	\begin{tikzpicture}[scale=1,baseline={([yshift=-.5ex]current bounding box.center)}]
	\ifthenelse{#1 = 1}{
		\node[draw=black,circle,scale=0.5,fill=black!80] at (0,0) {};
	}{}
	\ifthenelse{#1 = 2}{
		\node[draw=black,circle,scale=0.5,fill=gray!60] at (0,0) {};
	}{}
	\ifthenelse{#1 = 3}{
		\node[draw=black,circle,scale=0.5,fill=white] at (0,0) {};
	}{}
	\end{tikzpicture}
}

\newcommand{\stringHom}[1]{
	\begin{tikzpicture}[scale=1,baseline={([yshift=-.5ex]current bounding box.center)}]
	\ifthenelse{#1 = 1}{
		\draw (0,0) -- (2.5,0);
		\node[draw=black,circle,scale=0.5,fill=black!80] at (0,0) {};
		\node[draw=black,circle,scale=0.5,fill=gray!60] at (2.5,0) {};
		\node[draw=black,rectangle,fill=white, inner sep=2pt] at (1.25,0) {${\sss M_{\mc A_\phi \mc B_\psi}}$};
	}{}
	\ifthenelse{#1 = 2}{
		\draw (0,0) -- (5,0);
		\node[draw=black,circle,scale=0.5,fill=black!80] at (0,0) {};
		\node[draw=black,circle,scale=0.5,fill=gray!60] at (2.5,0) {};
		\node[draw=black,circle,scale=0.5,fill=white] at (5,0) {};
		\node[draw=black,rectangle,fill=white, inner sep=2pt] at (1.25,0) {${\sss M_{\mc A_\phi \mc B_\psi}}$};
		\node[draw=black,rectangle,fill=white, inner sep=2pt] at (3.75,0) {${\sss M_{\mc B_\psi \mc C_\varphi}}$};
	}{}
	\ifthenelse{#1 = 3}{
		\draw (0,0) -- (5,0);
		\node[draw=black,circle,scale=0.5,fill=black!80] at (0,0) {};
		\node[draw=black,circle,scale=0.5,fill=white] at (5,0) {};
		\node[draw=black,rectangle,fill=white, inner sep=2pt] at (2.5,0) {${\sss M_{\mc A_\phi \mc B_\psi} \otimes_{\mc B_\psi} M_{\mc B_\psi \mc C_\varphi}}$};
	}{}
	\ifthenelse{#1 = 4}{
		\draw (0,0) -- (2.5,0);
		\node[draw=black,circle,scale=0.5,fill=black!80] at (0,0) {};
		\node[draw=black,circle,scale=0.5,fill=white] at (2.5,0) {};
		\node[draw=black,rectangle,fill=white, inner sep=2pt] at (1.25,0) {${\sss M_{\mc A_\phi \mc C_\varphi}}$};
	}{}
	\ifthenelse{#1 = 5}{
		\draw (0,0) -- (2.5,0);
		\node[draw=black,circle,scale=0.5,fill=black!80] at (0,0) {};
		\node[draw=black,circle,scale=0.5,fill=black!80] at (2.5,0) {};
		\node[draw=black,rectangle,fill=white, inner sep=2pt] at (1.25,0) {${\sss (V,R_{V,-})}$};
	}{}
	\ifthenelse{#1 = 6}{
		\draw (0,0) -- (5,0);
		\node[draw=black,circle,scale=0.5,fill=black!80] at (0,0) {};
		\node[draw=black,circle,scale=0.5,fill=black!80] at (5,0) {};
		\node[draw=black,rectangle,fill=white, inner sep=2pt] at (2.5,0) {${\sss (V,R_{V,-}) \otimes (W,R_{W,-})}$};
	}{}
	\end{tikzpicture}
}

\newcommand{\loopHom}[2]{
	\begin{tikzpicture}[scale=#1,baseline={([yshift=-.5ex]current bounding box.center)}]
	\ifthenelse{#2 = 1}{
		\draw (1.25,-0.9) to[out=180, in=-180, looseness=0.8] (1.25,0.9);
		\draw[fat] (0,0) -- (2.5,0);
		\begin{scope}
			\clip (1.25,-0.91) rectangle (1.8,0.91);
			\draw[fat] (1.25,-0.9) to[out=0, in=0, looseness=0.8] (1.25,0.9);	
		\end{scope}
		\node[draw=black,circle,scale=0.5,fill=black!80] at (0,0) {};
		\node[draw=black,circle,scale=0.5,fill=black!80] at (2.5,0) {};
		\node[above] at (1.25,-0.9) {${\sss V}$};
	}{}
	\ifthenelse{#2 = 2}{
		\draw (1.25,-0.9) to[out=180, in=-180, looseness=0.8] (1.25,0.9);
		\draw[fat] (0,0) -- (2.5,0);
		\begin{scope}
			\clip (1.25,-0.91) rectangle (1.8,0.91);
			\draw[fat] (1.25,-0.9) to[out=0, in=0, looseness=0.8] (1.25,0.9);	
		\end{scope}
		\draw (2.5,0) -- (5,0);
		\node[draw=black,circle,scale=0.5,fill=black!80] at (0,0) {};
		\node[draw=black,circle,scale=0.5,fill=black!80] at (2.5,0) {};
		\node[draw=black,circle,scale=0.5,fill=white] at (5,0) {};
		\node[above] at (1.25,-0.9) {${\sss V}$};
		\node[draw=black,rectangle,fill=white, inner sep=2pt] at (3.75,0) {${\sss M_{\mc A_\phi \mc B_\psi}}$};
	}{}
	\ifthenelse{#2 = 3}{
		\draw (1.25,-0.9) to[out=180, in=-180, looseness=0.8] (1.25,0.9);
		\draw[fat] (0,0) -- (2.5,0);
		\begin{scope}
			\clip (1.25,-0.91) rectangle (1.8,0.91);
			\draw[fat] (1.25,-0.9) to[out=0, in=0, looseness=0.8] (1.25,0.9);	
		\end{scope}
		\draw (-2.5,0) -- (0,0);
		\node[draw=black,circle,scale=0.5,fill=black!80] at (-2.5,0) {};
		\node[draw=black,circle,scale=0.5,fill=white] at (0,0) {};
		\node[draw=black,circle,scale=0.5,fill=white] at (2.5,0) {};
		\node[above] at (1.25,-0.9) {${\sss V}$};
		\node[draw=black,rectangle,fill=white, inner sep=2pt] at (-1.25,0) {${\sss M_{\mc A_\phi \mc B_\psi}}$};
	}{}
	\ifthenelse{#2 = 4}{
		\draw (1.25,-0.9) to[out=180, in=-180, looseness=0.8] (1.25,0.9);
		\draw[fat] (0,0) -- (2.5,0);
		\begin{scope}
			\clip (1.25,-0.91) rectangle (1.8,0.91);
			\draw[fat] (1.25,-0.9) to[out=0, in=0, looseness=0.8] (1.25,0.9);	
		\end{scope}
		\draw (-1.25,-0.9) to[out=180, in=-180, looseness=0.8] (-1.25,0.9);
		\draw[fat] (-2.5,0) -- (0,0);
		\begin{scope}
			\clip (-1.25,-0.91) rectangle (-0.7,0.91);
			\draw[fat] (-1.25,-0.9) to[out=0, in=0, looseness=0.8] (-1.25,0.9);	
		\end{scope}
		\node[draw=black,circle,scale=0.5,fill=black!80] at (-2.5,0) {};
		\node[draw=black,circle,scale=0.5,fill=black!80] at (0,0) {};
		\node[draw=black,circle,scale=0.5,fill=black!80] at (2.5,0) {};
		\node[above] at (1.25,-0.9) {${\sss W}$};
		\node[above] at (-1.25,-0.9) {${\sss V}$};
	}{}
	\ifthenelse{#2 = 5}{
		\draw (1.25,-0.9) to[out=180, in=-180, looseness=0.8] (1.25,0.9);
		\draw[fat] (0,0) -- (2.5,0);
		\begin{scope}
			\clip (1.25,-0.91) rectangle (1.8,0.91);
			\draw[fat] (1.25,-0.9) to[out=0, in=0, looseness=0.8] (1.25,0.9);	
		\end{scope}
		\draw (-1.25,-0.9) to[out=180, in=-180, looseness=0.8] (-1.25,0.9);
		\draw[fat] (-2.5,0) -- (0,0);
		\begin{scope}
			\clip (-1.25,-0.91) rectangle (-0.7,0.91);
			\draw[fat] (-1.25,-0.9) to[out=0, in=0, looseness=0.8] (-1.25,0.9);	
		\end{scope}
		\node[draw=black,circle,scale=0.5,fill=black!80] at (-2.5,0) {};
		\node[draw=black,circle,scale=0.5,fill=black!80] at (0,0) {};
		\node[draw=black,circle,scale=0.5,fill=black!80] at (2.5,0) {};
		\node[above] at (1.25,-0.9) {${\sss V}$};
		\node[above] at (-1.25,-0.9) {${\sss W}$};
	}{}
	\ifthenelse{#2 = 6}{
		\draw (1.25,-0.9) to[out=180, in=-180, looseness=0.8] (1.25,0.9);
		\draw[fat] (0,0) -- (2.5,0);
		\begin{scope}
		\clip (1.25,-0.91) rectangle (1.8,0.91);
		\draw[fat] (1.25,-0.9) to[out=0, in=0, looseness=0.8] (1.25,0.9);	
		\end{scope}
		\node[draw=black,circle,scale=0.5,fill=black!80] at (0,0) {};
		\node[draw=black,circle,scale=0.5,fill=black!80] at (2.5,0) {};
		\node[above] at (2,-0.9) {${\sss V \otimes W}$};
	}{}
	\end{tikzpicture}
}

\makeatletter
\newcommand{\xRrightarrow}[2][]{\ext@arrow 0359\Rrightarrowfill@{#1}{#2}}
\newcommand{\Rrightarrowfill@}{\arrowfill@\equiv\equiv\Rrightarrow}
\newcommand{\xLleftarrow}[2][]{\ext@arrow 3095\Lleftarrowfill@{#1}{#2}}
\newcommand{\Lleftarrowfill@}{\arrowfill@\Lleftarrow\equiv\equiv}
\makeatother

\usepackage{soul}

\hypersetup{%
	bookmarks=true,         % show bookmarks bar first?
	bookmarksopen=true,
	unicode=true,           % to show unicode characters in Acrobat’s bookmarks
	pdftoolbar=true,        % show Acrobat’s toolbar?
	pdfmenubar=true,        % show Acrobat’s menu?
	pdffitwindow=false,     % window fit to page when opened
	pdfstartview={FitH},    % fits the width of the page to the window
	pdftitle={Crossing with the circle in Dijkgraaf-Witten theory and applications to topological phases of matter},
	pdfauthor={},
	pdfsubject={},          % subject of the document
	pdfcreator={},          % creator of the document
	pdfproducer={pdfLaTeX}, % producer of the document
	pdfkeywords= {topological phases} {category theory} {anyons},
	pdfnewwindow=true,      % links in new window
	colorlinks,
	citecolor=BrickRed,
	filecolor=BrickRed,
	linkcolor=BlueViolet,
	urlcolor=BrickRed,
	linktocpage=true
}

\title{\boldmath Crossing with the circle in Dijkgraaf-Witten theory and applications to topological phases of matter}

\author[\Square]{Alex Bullivant,}
\author[\pentagon, \hexagon]{Clement Delcamp}

\affiliation[\Square]{Department of Theoretical Physics, \\ University of Maynooth, Ireland}

\affiliation[\pentagon]{Max-Planck-Institut f{\"u}r Quantenoptik, \\ Hans-Kopfermann-Str. 1, 85748 Garching, Germany}
\affiliation[\hexagon]{Munich Center for Quantum Science and Technology (MCQST), \\ Schellingstr. 4, D-80799 M{\"u}nchen, Germany}

\emailAdd{bullivant.alex@gmail.com}
\emailAdd{clement.delcamp@mpq.mpg.de}

\abstract{\\~\\ {Given a fully extended topological quantum field theory, the `crossing with the circle' conditions establish that the dimension, or categorification thereof, of the quantum invariant assigned to a closed $k$-manifold $\Sigma$ is equivalent to that assigned to the ($k$+1)-manifold $\Sigma \times \mathbb S^1$. We compute in this manuscript these conditions for the 4-3-2-1 Dijkgraaf-Witten theory. In the context of the lattice Hamiltonian realisation of the theory, the quantum invariants assigned to the circle and the torus encode the defect open string-like and bulk loop-like excitations, respectively. The corresponding `crossing with the circle' condition thus formalises the process by which loop-like excitations are formed out of string-like ones. Exploiting this result, we revisit the statement that loop-like excitations define representations of the linear necklace group as well as the loop braid group.}}

\begin{document} 
	\vspace*{-2em}
	\maketitle
	\flushbottom
	\newpage
	
\section{Introduction}

\noindent
The axiomatic formulation of \emph{topological quantum field theories} (TQFTs) pioneered by Atiyah bears a strong category theoretical flavour. Indeed, given a ring $\Bbbk$, Atiyah defined in \cite{Atiyah:1989vu} a $d$-dimensional TQFT as a \emph{symmetric monoidal functor} $\mc Z : \msf{Cob}(d) \to \msf{Vec}(\Bbbk)$, where $\msf{Cob}(d)$ is the category whose objects are closed oriented ($d$$-$1)-manifolds and morphisms are equivalence classes of bordisms, while $\msf{Vec}(\Bbbk)$ is the category whose objects are vector spaces over $\Bbbk$ and morphisms are linear maps. Unpacking this definition, we obtain that a TQFT $\mathcal Z$ is determined by a choice of finite dimensional $\Bbbk$-vector space $\mc{Z}(\Sigma)$ for every closed oriented ($d$$-$1)-manifold $\Sigma$, a choice of linear maps $\mc{Z}(\Sigma \to \Sigma'): \mc Z(\Sigma) \to \mc Z(\Sigma')$ for every diffeomorphism class of bordism $\Sigma \to \Sigma'$, as well as isomorphisms $\mc Z(\Sigma \sqcup \Sigma') \simeq \mc Z(\Sigma) \otimes \mc Z(\Sigma')$ and $\mc Z(\varnothing) \simeq \Bbbk$, where $\varnothing$ refers to the empty manifold considered as a closed, oriented ($d$$-$1)-manifold. Furthermore, the functoriality conditions translate into the statements that $\mc Z(\Sigma \times \mathbb I) = {\rm id}_{{\mc Z}(\Sigma)} $ and $\mc Z(\Sigma \to \Sigma' \cup_{\Sigma'} \Sigma' \to \Sigma'') = \mc Z(\Sigma \to \Sigma') \circ \mc Z(\Sigma' \to \Sigma'')$. More precisely, given a surface $\Sigma$, the manifold $\Sigma \times \mathbb I$ can be interpreted as either one of the following bordisms: $\Sigma \to \overline \Sigma$, $\overline \Sigma \to \Sigma$, $\overline \Sigma \sqcup \Sigma \to \varnothing$, or $\varnothing \to  \Sigma \sqcup \overline \Sigma$ where $\overline \Sigma$ refers to the manifold with opposite orientation. The corresponding morphisms in $\msf{Cob}(d)$ are the identity maps ${\rm id}_\Sigma$ and ${\rm id}_{\overline \Sigma}$, the evaluation map ${\rm ev}_\Sigma$ and the coevaluation map ${\rm coev}_{\Sigma}$, respectively. Applying the functor $\mc Z$ to ${\rm ev}_\Sigma$ yields the following canonical pairing
\begin{equation*}
	\la - , - \ra : \mc Z(\overline \Sigma) \otimes_{\Bbbk} \mc Z(\Sigma) \simeq \mc Z(\overline \Sigma \sqcup \Sigma) \xrightarrow{\mc Z({\rm ev}_\Sigma)} \mc Z(\varnothing) \simeq \Bbbk \; .
\end{equation*}
Similarly, applying $\mc Z$ to ${\rm coev}_{\Sigma}$ yields the map
\begin{equation*}
	\Bbbk \simeq \mc Z(\varnothing) \xrightarrow{\mc Z({\rm coev}_\Sigma)} \mc Z(\Sigma \sqcup \overline \Sigma) \simeq \mc Z(\Sigma) \otimes_\Bbbk \mc Z(\overline \Sigma) \; .
\end{equation*}
Denoting by $\mc Z(\Sigma)^*$ the vector space dual to $\mc Z(\Sigma)$, it follows from the defining axioms of $\mc Z$ that $v \mapsto \la v , - \ra$ is an isomorphism so that $\mc Z(\Sigma)^* \simeq \mc Z(\overline \Sigma)$. The identification between ${\rm End}(V)$ and $V \otimes_\Bbbk V^*$ for any vector space $V$ further implies that the map ${\rm ev}_\Sigma$ corresponds to the trace operation in ${\rm End}(\mc Z(\Sigma))$, while ${\rm coev}_\Sigma$ is given by the inclusion of the identity map. Since the manifold $\Sigma \times \mathbb S^1$ is diffeomorphic to the composition of ${\rm coev}_\Sigma$, $\Sigma \sqcup \overline \Sigma \to \overline \Sigma \sqcup \Sigma$ and ${\rm ev}_\Sigma$, we obtain that $\mc Z(\Sigma \times \mathbb S^1)$ can be expressed as the composition of maps
\begin{equation*}
	\Bbbk \simeq \mc Z(\varnothing) \xrightarrow{\mc Z({\rm coev}_\Sigma)} {\rm End}(\mc Z(\Sigma)) \xrightarrow{\mc Z({\rm ev}_\Sigma)} \mc Z(\varnothing) \simeq \Bbbk \; ,
\end{equation*}
which corresponds to the scalar multiplication by the dimension of the vector space $\mc Z(\Sigma)$, and thus ${\rm Dim}_{\Bbbk}(\mc Z(\Sigma)) = \mc Z(\Sigma \times \mathbb S^1)$. Henceforth, we shall refer to this equation as the first `crossing with the circle' condition. 

In virtue of $\mc Z(\varnothing) \simeq \Bbbk$, we can state that a $d$-dimensional TQFT assigns an element of $\Bbbk$ to any closed  oriented $d$-manifold, a $\Bbbk$-vector space to every closed oriented ($d$$-$1)-manifold and a vector in the vector space associated to its boundary to every open oriented $d$-manifold. We might also want to consider ($d$$-$1)-manifolds with non-empty boundaries and extend the theory so that is assigns something to the corresponding ($d$$-$2)-manifold. The result is a so-called $2$-extended TQFT, which, in addition to the data described above, assigns a finite semisimple $\Bbbk$-linear abelian category to every ($d$$-$2)-manifold such that the hom-set between two objects of this category are finite $\Bbbk$-vector spaces, and to an open ($d$$-$1)-manifold an object in the category associated with its boundary. At least formally, we can further extend this definition by assigning an appropriate $\Bbbk$-linear 2-category to a closed ($d$$-$3)-manifold, and so on and so forth. A TQFT that assigns non-trivial information all the way down to the point is referred to as a \emph{fully-extended} TQFT. Crucially, such fully extended TQFTs (possibly with the addition of framing data) can be concisely described via the so-called \emph{corbordism hypothesis} put forward by Baez and Dolan in \cite{Baez:1995xq} and proven by Lurie in \cite{lurie2009higher,Lurie:2009keu}. Instead of stating this result here, we shall merely quote one of its byproducts, namely that a fully extended TQFT is fully characterized by what it assigns to the \emph{point}. This implies that given the data assigned to the point, there must be a mechanism allowing us to recover the data assigned to higher-dimensional closed manifolds. One instance of such a mechanism is the first crossing with the circle condition, which establishes, let us recall, that the dimension of the vector space associated with a closed ($d$$-$1)-manifold $\Sigma$ equals $\mc Z(\Sigma \times \mathbb S^1)$. More generally, the following relation is expected to hold \cite{Baez:1995xq,bartlett2009unitary}:
\begin{equation}
	\label{eq:crossingCircle}
	\tag{$*$}
	\msf{Dim} \, \mc Z(\Sigma^{d-n}) \cong \mc Z (\Sigma^{d-n} \times \mathbb S^1) \, ,
\end{equation}
where $\msf{Dim}$ refers to a \emph{categorification} of the notion of dimension of a vector space that is suitable to the type of data assigned to the manifold $\Sigma^{d-n}$ \cite{Baez:1995xq,ganter2008representation,bartlett2009unitary}. Henceforth, we shall refer to such a relation as the $n$-th crossing with the circle condition. Given that a fully extended TQFT assigns an ($n$$-$1)-category to a closed ($d$$-$$n$)-manifold and an ($n$$-$2)-category to a closed ($d$$-$$n$$+$$1$)-manifold, any crossing with the circle condition amounts to a \emph{decategorification} process. Generally speaking, decategorification refers to a collection of techniques whereby statements about categories are reduced to statements about sets.\footnote{Note that decategorification is usually not the inverse to \emph{categorification}, the process of replacing statements about sets by statements about categories.} In practice, this is typically done by discarding morphisms so that only equivalence classes of objects remain. The operation that consists in computing the dimension of vector spaces is one example of such procedure since isomorphic vector spaces do have the same dimension.  

\bigskip \noindent
The purpose of the present manuscript is to compute the crossing with the circle conditions for a special type of fully extended TQFTs that have a lattice gauge theory interpretation, and elucidate their physical interpretations. 
In $d$ dimensions, the input data of such a theory is a finite group $G$ as well as a cohomological class $[\omega] \in H^d(G, \rU(1))$, and the corresponding state-sum invariant for a triangulated manifold was explicitly constructed by Dijkgraaf and Witten in \cite{dijkgraaf1990topological}. In (2+1)d, the theory is equivalent to the \emph{Turaev-Viro-Barrett-Westbury} theory \cite{Turaev:1992hq, Barrett:1993ab}, with input \emph{spherical fusion category} the category $\msf{Vec}^\alpha_G$ of $G$-graded vector spaces whose monoidal structure is twisted by a cohomology class in $[\alpha] \in H^3(G,\rU(1))$. The refinement of the theory to a fully extended (2+1)d TQFT was presented in  
\cite{Freed:1991bn, Freed:1994ad, Freed:1995fn, Freed:2009qp}. Given that it assigns non-trivial information to three-, two-, one- and zero-dimensional manifolds, such a refinement is usually referred to as the 3-2-1-0 Dijkgraaf-Witten theory. By analogy, we can also consider the 4-3-2-1 theory, which is a (3+1)d TQFT that assigns non-trivial information all the way down to the circle.\footnote{The 4-3-2-1 theory can be further extended to the point, making it a fully extended TQFT, but this is beyond the scope of this manuscript. We conjecture that the data assigns to the point is the tricategory of module bicategories for the fusion bicategory $\msf{2Vec}_G^\pi$ of graded 2-vector spaces.}  Henceforth, we shall denote the corresponding functor by $\mc Z^\pi_G$, where $\pi$ is a normalised representative in a cohomology class $[\pi]\in H^4(G,\rU(1))$. The goal of our work is to compute and interpret the crossing with the circle conditions \eqref{eq:crossingCircle} for this specific theory. For simplicity, we shall focus on the scenario where $\Sigma^{4-n}$, with $n=1,\ldots,3$, is homeomorphic to the ($4$$-$$n$)-torus, although we could treat the more general case analogously. 

Choosing the simplest triangulation of the four-torus, the complex number $\mc Z^\pi_G(\mathbb T^4)$ the theory assigns to it can be straightforwardly computed. There is a particularly concise way to quote the result, which we review in sec.~\ref{sec:introDW}, invoking the notion of \emph{loop groupoid}. Given a finite groupoid $\mc G$, its loop groupoid is equivalent to the functor category $\msf{Fun}(\overline{\mathbb Z}, \mc G)$, where $\overline{\mathbb Z}$ is the group $\mathbb Z$ treated as a one-object groupoid \cite{willerton2008twisted}. Applying this definition to the one-object groupoid $\overline G$ yields the loop groupoid denoted by $\Lambda G$. This procedure can be iterated so as to define the four-fold loop groupoid $\Lambda^4 G$. Similarly, one can define a map $\msf t$ that sends a given groupoid cocycle to a loop groupoid cocycle, i.e. $\mathsf{t} : Z^{n}(\mc G,\rU(1)) \to Z^{n - 1}(\Lambda \mc  G, \rU(1))$ so that given the group 4-cocycle $\pi$ one can construct a groupoid 0-cocycle $\msf t^4(\pi)$ in $H^0(\Lambda ^4G, \rU(1))$. Given the above, the number $\mc Z^\pi_G(\mathbb T^4)$ reads
\begin{equation*}
	\mc Z^\pi_G(\mathbb T^4) = \frac{1}{|G|}\sum_{X \in {\rm Ob}(\Lambda^4 G)}\msf t^4(\pi)(X) =: \int_{\Lambda^4 G}\msf t^4(\pi) \, .
\end{equation*}
The data the Dijkgraaf-Witten theory assigns to the manifolds $\mathbb T^3$, $\mathbb T^2$ and $\mathbb S^1$ was recently computed in the context of \emph{topological phases of matter} within the \emph{lattice Hamiltonian} formalism of the theory \cite{Wan:2014woa,Wang:2014oya,Bullivant:2019fmk,Bullivant:2020xhy}. Generally speaking, given a $d$-dimensional state-sum invariant, one can define in a canonical way a lattice Hamiltonian whose ground state subspace on a closed ($d$$-$1)-manifold is isomorphic to the vector space the TQFT assigns to it. The resulting model is a concrete realisation of a topological phase whose low-energy effective description realises the TQFT \cite{wen2004quantum,Levin:2004mi,Chen:2010gda}. The ground state subspace on $\mathbb T^3$ of the lattice Hamiltonian realisation of the 4-3-2-1 Dijkgraaf-Witten theory was computed explicitly in \cite{Wan:2014woa}, which in the loop groupoid terminology reads \cite{willerton2008twisted}
\begin{equation*}
	\mc Z^\pi_G(\mathbb T^3) = 
	{\rm Span}_\mathbb C \Big\{ \msf s: {\rm Ob}(\Lambda^3 G) \to \mathbb C \, \big| \, \msf s(Y) = 
	\msf t^{3}(\pi)(\fr g) \,  \msf s( X) \; \forall \, \fr g \in {\rm Hom}_{\Lambda^3 G}(X,Y) \Big\} =: \mc V_{\Lambda^3 G}(\msf t^3(\pi)) \, . 
\end{equation*}
The remaining data that the theory assigns to $\mathbb T^2$ and $\mathbb S^1$ can be conveniently found as the category theoretical structures encoding the \emph{defects} and \emph{excitations} hosted by the lattice Hamiltonian realisation. Models of topological phases of matter in (2+1)d famously host point-like excitations with \emph{anyonic} statistics. The exchange statistics of these \emph{anyons} is governed by representations of the \emph{braid group}, where the braids are formed by the worldlines of the point-like particles upon exchanges. Given a lattice Hamiltonian realisation of the Turaev-Viro-Barrett-Westbury theory, it is well-known that anyons are encoded into the \emph{Drinfel'd centre} of the input category \cite{Levin:2004mi, alex2011stringnet}. Importantly, braids can always be untangled in (3+1)d so that anyons only exist in two-dimensional systems. Nevertheless, higher-dimensional topological models yield higher-dimensional excitations, which also possess statistics beyond the bosonic and fermionic ones, and as such behave like \emph{spatially extended} anyons. In particular, three-dimensional models, such as the one we are interested in, host bulk loop-like excitations. Using a generalisation of the \emph{tube algebra} approach \cite{ocneanu1994chirality,ocneanu2001operator}, the authors found in \cite{Delcamp:2017pcw,Bullivant:2019fmk} that the bulk loop-like excitations and their statistics were encoded into the braided monoidal category of modules over the \emph{twisted groupoid algebra} $\mathbb C[\Lambda^2 G]^{\msf t^2(\pi)}$. As we briefly review in sec.~\ref{sec:introDW}, this is precisely the category that the TQFT assigns to the two-torus, i.e.
\begin{equation*}
	\mc Z^{\pi}_G(\mathbb T^2) = \Mod(\mathbb C[\Lambda^2 G]^{\msf t^2{\pi}}) \, .
\end{equation*}
In addition to bulk loop-like excitations, the Hamiltonian yields (open) string-like excitations that terminate at zero-dimensional defects. The authors showed in \cite{Bullivant:2020xhy}---using a categorified version of the tube algebra approach---that these string-like objects were encoded into the bicategory of module categories over the category $\VectGr$ of loop-groupoid-graded vector spaces (see \cite{kong2020defects} for a detailed treatment of the case $G=\mathbb Z / 2 \mathbb Z$). This bicategory corresponds to the 2-category the TQFT assigns to the circle. In symbols, we have
\begin{equation*}
	\mc Z^{\pi}_G(\mathbb S^1) = \MOD(\VectGr) \, .
\end{equation*}
In sec.~\ref{sec:crossing}, we compute the dimension, and categorifications thereof, of the above data to find
\begin{align*}
	{\rm Dim}_\mathbb{C} (\mc V_{\Lambda^3 G}(\mathsf{t}^3(\pi))) &= \int_{\Lambda^4G} \msf t^4(\pi) \, , 
	\\
	\msf{Dim} \, \Mod(\GrAlg) &\simeq Z(\GrAlg) \, ,
	\\[0.1em]
	\msf{Dim} \, \MOD(\VectGr) &\cong \ms Z(\VectGr) \, ,
\end{align*}
where $Z(-)$ and $\ms Z(-)$ refers to the centre of an algebra and the Drinfel'd centre of a monoidal category, respectively.
In order to check the second and third crossing with the circle conditions, it remains to establish that the mathematical objects on the right-hand-side of the equations above are equivalent to the data introduced previously, namely that $\mc V_{\Lambda^3 G}(\msf t^3(\pi)) \simeq Z(\GrAlg)$ and $\Mod(\GrAlg) \cong \ms Z(\VectGr)$. Showing these equivalences is the purpose of sec.~\ref{sec:centres}. Although we do not require it for our derivation, it is interesting to note that the bicategory $\MOD(\VectGr)$ also admits an equivalent definition as the higher-categorical centre $\ms Z(\msf{2 Vec}_G^\pi)$ of the bicategory $\msf{2 Vec}_G^\pi$ of $G$-graded 2-vector spaces whose pentagon identity is weakened by a pentagonator 2-isomorphism characterised by the group 4-cocycle $\pi$ \cite{Kong:2019brm}. Our findings can be conveniently summarized in the following table:

\setlength{\tabcolsep}{2.2em}
\renewcommand{\arraystretch}{1.5}
\begin{center}
	\begin{tabularx}{1\columnwidth}{c@{\hskip 40pt} c@{\hskip 5pt} c@{\hskip 5pt} c@{\hskip 10pt} |@{\hskip 8pt} Y}
		Manifold & Quantum invariant& & Equivalent structure & Physical interpretation 
		\\ \cline{1-5}
		\tikzmark{a}{$\mathbb T^4$} 
		& \tikzmark{e}{$\int_{\Lambda^4 G} \mathsf{t}^4(\pi)$} 
		& \tikzmark{w}{$=$}
		& \tikzmark{i}{${\rm Dim}_\mathbb{C} (\mc V_{\Lambda^3 G}(\mathsf{t}^3(\pi)))$} 
		& {\small Ground state}
		\\[-1.2em]
		& & & & {\small degeneracy on $\mathbb T^3$}
		\\
		\tikzmark{b}{$\mathbb T^3$} 
		& \tikzmark{f}{$\mc V_{\Lambda^3 G}(\mathsf{t}^3(\pi))$} 
		& \tikzmark{x}{$\simeq$}
		& \tikzmark{j}{$Z(\mathbb C[\Lambda^2 G]^{\mathsf{t}^2(\pi)})$} 
		& {\small Ground state}
		\\[-1.2em]
		& & & & {\small subspace on $\mathbb T^3$}
		\\
		\tikzmark{c}{$\mathbb T^2$} 
		& \tikzmark{g}{$\msf{Mod}(\mathbb C[\Lambda^2 G]^{\mathsf{t}^2(\pi)})$} 
		& \tikzmark{y}{$\cong$}
		& \tikzmark{k}{$\ms Z(\msf{Vec}_{\Lambda G}^{\mathsf{t}(\pi)})$}
		& {\small Category of loop-like}
		\\[-1.3em]
		& & & &  {\small bulk excitations}
		\\
		\tikzmark{d}{${\mathbb S^1}$} 
		& \tikzmark{h}{$\msf{MOD}(\msf{Vec}_{\Lambda G}^{\mathsf{t}(\pi)})$}
		& \tikzmark{z}{$\cong$}
		& \tikzmark{l}{$\ms Z(\mathsf{2Vec}^\pi_G)$} 
		& {\small Bicategory of defects and}
		\\[-1.3em]
		& & & & {\small  string-like excitations} 
	\end{tabularx} 
\end{center}
\curvTabL{b}{a}
\curvTabR{c}{b}
\curvTabL{d}{c}
\linkTab{f}{i}
\linkTab{g}{j}
\linkTab{h}{k}
\mapTab{a}{e}
\mapTab{b}{f}
\mapTab{c}{g}
\mapTab{d}{h}

\noindent
that makes transparent the subtle interplay between the $n$-fold loop-groupoid of the input group, iterative categorifications of the centre construction, and higher-categorical analogues of the notion of module over an algebra. 

From a physical standpoint, the results summarised in the table above formalise among other things the relation between string- and loop-like excitations hosted by the lattice Hamiltonian realisation of the theory, such that loop-like excitations can be thought as descending from string-like ones via a tracing mechanism. Exploiting this result, we revisit in sec.~\ref{sec:braiding} the statement that loop-like excitations provide representations of the \emph{linear necklace group} \cite{bellingeri2016braid}, which is isomorphic to the braid group, hence confirming that these excitations constitute extended anyon-like objects \cite{Wang:2014xba, Wang:2014oya, putrov2016braiding, Bullivant:2018djw, Bullivant:2019fmk}. Furthermore, we explain that a subset of such loop-like excitations yield representations of the \emph{loop braid group} \cite{Baez:2006un,Bullivant:2018pju}.

\bigskip \noindent
In a subsequent work we aim to verify that the necklace and loop braid group representations described within this manuscript can alternatively be derived utilising the duality structures for objects in $\ms Z(\msf{2Vect}^{\pi}_{G})$. In particular, we wish to refine the pseudo-graphical calculus introduced in sec.~\ref{sec:braiding} by drawing an analogy with the graphical presentation of the category of \emph{2-tangles} \cite{baez19982}, which describes a combinatorial presentation of knotted surfaces embedded in the 4-disk \cite{carter1997combinatorial}. 

Although this work focuses on the example of the (3+1)d Dijkgraaf-Witten theory, we conjecture that all results apply with minor modifications to every (3+1)d TQFT whose input data is a \emph{spherical fusion bicategory} \cite{douglas2018fusion}. In particular, given such an input data, one can construct the corresponding categorified tube algebra such that the bicategory of module categories over it defines the invariant the theory assigns to the circle \cite{highertubedraft}. This bicategory can be further shown to be equivalent, as a braided monoidal bicategory, to the higher-categorical centre of the input bicategory. We then conjecture that the crossing with the circle conditions we establish in the present manuscript apply analogously, yielding among other things new representations of the linear necklace group and the loop braid group.

\bigskip\noindent

\begin{center}
	\textbf{Organisation of the paper}
\end{center}
\noindent
We begin in sec.~\ref{sec:introDW} by introducing relevant notions of category theory and by briefly reviewing what the 4-3-2-1 Dijkgraaf-Witten theory assigns to the manifolds $\mathbb T^4$, $\mathbb T^3$, $\mathbb T^2$ and $\mathbb S^1$. In sec.~\ref{sec:centres}, we present an alternative description of the data introduced in the preceding section in terms of the notion of centre of an algebra, and categorifications thereof. The crossing with the circle conditions are computed in sec.~\ref{sec:crossing} in terms of categorifications of the notion of dimension of a vector space. Finally, in sec.~\ref{sec:braiding}, the crossing with the circle conditions are exploited in order to recover the fact that loop-like excitations hosted by the lattice Hamiltonian realisation of the theory yield representations of the linear necklace and loop braid groups.

\newpage
\section{Dijkgraaf-Witten theory as a 4-3-2-1 extended TQFT\label{sec:introDW}}

\noindent
\emph{In this section, we review some relevant categorical notions and tabulate what data the (3+1)d Dijkgraaf-Witten state-sum invariant assigns to the manifolds $\mathbb T^4$, $\mathbb T^3$, $\mathbb T^2$ and $\mathbb S^1$ using the language of loop groupoids. We shall motivate the relevant data from the lattice Hamiltonian perspective.}

\subsection{Preliminaries\label{sec:preliminaries}}

\noindent
Let us begin by fixing our conventions for categories. Given a category $\mc C$, we notate via $\Ob(\mc C)$ and $\Hom(\mc C)$ the sets of objects and morphisms in $\mc C$, respectively. For each pair $X,Y \in \Ob(\mc C)$ of objects, the set of morphisms (hom-set) from $X$ to $Y$ is denoted by $\Hom_\mc C(X,Y)$. For each $f:X\to Y\in\Hom_\mc{C}(X,Y)$, we define ${\rm s}(f):= X$ and $ {\rm t}(f):=Y$ to be the \emph{source} and \emph{target} objects of $f$, respectively. For each triple $X,Y,Z \in \Ob(\mc C)$, the composition rule is written as $\circ : \Hom_\mc C(X,Y) \times \Hom_\mc C(Y,Z) \to \Hom_\mc C(X,Z)$ and the identity morphism associated with an object $X \in \Ob(\mc C)$ is denoted by ${\rm id}_X \in {\rm End}_\mc C(X):=\Hom_{\mc{C}}(X,X)$. Finally, given three composable morphisms $f,g,h \in \Hom(\mc C)$, this data is subject to the relations ${\rm id}_{{\rm s}(f)} \circ f =f = f \circ {\rm id}_{{\rm t}(f)}$ and $(f \circ g)\circ h = f \circ (g \circ h)$.  A functor between two categories $\mc C$ and $\mc C'$ is a map that sends objects $X \in \Ob(\mc C)$ to $F(X) \in \Ob(\mc C')$ and morphisms $f: X \to Y \in \Hom(\mc C)$ to $F(f):F(X)\to F(Y) \in \Hom(\mc C')$ such that composition is preserved. Maps between functors then lead to the notion of natural transformations, which will play an important role later:
\begin{definition}[Natural transformation]
	Let $F,F': \mc C \to \mc C'$ be two functors between two categories $\mc C$ and $\mc C'$. A natural transformation $\eta : F \Rightarrow F'$ between $F$ and $F'$ is an assignment of a morphism $\eta_X:F(X)\to F'(X) \in \Hom(\mc C')$ to every $X \in \Ob(\mc C)$ such that the diagram
	\begin{equation}
	 	\begin{tikzcd}[ampersand replacement=\&, column sep=1.8em, row sep=1.3em]
		 	|[alias=A]|F(X)
		 	\&\&
		 	|[alias=B]|F(Y)
		 	\\
		 	\\
		 	|[alias=AA]|F'(X)
		 	\&\&
		 	|[alias=BB]|F'(Y)
		 	\arrow[from=A,to=B,"F(f)"]
		 	\arrow[from=B,to=BB,"\eta_{Y}"]
		 	\arrow[from=A,to=AA,"\eta_{X}"']
		 	\arrow[from=AA,to=BB,"F'(f)"']
	 	\end{tikzcd}
	\end{equation}
	commutes for all $f:X \to Y \in \Hom(\mc C)$.
\end{definition}

\noindent
In the following, we shall often consider a special class of categories:
\begin{definition}[Groupoid]
	A groupoid $\mc G$ is a category whose morphisms are all invertible, i.e., there exists a function $^{-1}:\Hom_\mc G(X,Y)\to \Hom_\mc G(Y,X)$ satisfying the relations $\fr g\circ \fr g^{-1}={\rm id}_{{\rm s}(\fr g)}$ and  $\fr g^{-1}\circ \fr g={\rm id}_{{\rm t}(\fr g)}$ for all $\fr g \in \Hom(\mc G)$. A groupoid $\mc{G}$ is called `finite' if the collection of objects and the hom-sets are finite.
\end{definition}
\noindent 
Henceforth, we shall denote the composition $\fr g \circ \fr h$ of two composable groupoid morphisms as $\fr g \fr h$ in analogy with group theory. The following concept of connected component of a groupoid is often very useful:
\begin{definition}[Connected component]
	Let $\mc G$ be a finite groupoid. The morphisms of $\mc{G}$ define an equivalence relation $\sim_{\mc{G}}$ on the set ${\rm Ob}(\mc{G})$ of objects given by the relation $X\sim_{\mc{G}} Y$ if there exists a $\mathfrak{g}:X\rightarrow Y\in\Hom(\mc{G})$. We refer to the equivalence classes of $\sim_\mc G$ as `connected components' and utilise the notation $\uppi_{0}(\mc{G}):={\rm Ob}(\mc{G})/\sim_{\mc{G}}$.
\end{definition}
\noindent
The groupoid cohomology $H^n(\mc G, \rU(1))$ of a finite groupoid $\mc G$ is defined as the simplicial cohomology of its classifying space $B \mc G$, where $B \mc G$ is defined as a simplicial set resulting from the gluing of abstract $n$-simplices that are identified with strings $X_0 \xrightarrow{\fr g_1} X_1 \xrightarrow{\fr g_2} \cdots \xrightarrow{\fr g_n} X_n$ of $n$ composable morphisms in $\mc G$. Given a cohomology class $[\omega_n] \in H^n(\mc G, \rU(1))$, a normalised representative $\omega_n \in [\omega_n]$ is a groupoid $n$-cocycle for which $\omega_n(\fr g_1,\ldots,\fr g_n)=1$, whenever any of the arguments is an identity morphism. 

Given a finite groupoid, we can define another groupoid following a recipe that is ubiquitous in our construction:
\begin{definition}[Loop groupoid]
	Let $\mc G$ be a finite groupoid. We define the loop groupoid $\Lambda \mc G$ as the groupoid whose objects are endomorphisms $\fr g \in {\rm End}_{\mc G}(X)$, for every object $X \in {\rm Ob}(\mc G)$, and morphisms are of the form $\fr h: \fr g \to \fr h ^{-1} \fr g \fr h$, for all $\fr h \in {\rm Hom}_{\mc G}(X,Y)$ and $\fr g \in {\rm End}_{\mc G}(X)$, such that the composition is inherited from the one in $\mc G$. 
\end{definition}
\noindent
Furthermore, there is the so-called $\mathbb S^1$-transgression map that sends a given groupoid cocycle to a loop groupoid cocycle, i.e. $\mathsf{t} : Z^{\bul}(\mc G,\rU(1)) \to Z^{\bul - 1}(\Lambda \mc  G, \rU(1))$, such that
\begin{align}
	\nn
	&\mathsf{t}(\omega)\big(\fr x \xrightarrow{\fr g_1} \, , \fr g_1^{-1}\fr x \fr g_1 \xrightarrow{\fr g_2} \, , \ldots, (\fr g_1\cdots \fr g_{n-1})^{-1}\fr x (\fr g_1 \cdots \fr g_n) \xrightarrow{\fr g_n}\big) 
	\\
	& \q :=
	\prod_{i=0}^{n}\omega(\fr g_1, \ldots,\fr g_i, (\fr g_1 \cdots \fr g_i)^{-1}\fr x (\fr g_1 \cdots \fr g_i), \fr g_{i+1}, \ldots, \fr g_n)^{(-1)^{n-i}} \, ,
\end{align}
where we used the shorthand notation $\fr g \xrightarrow{\fr h} \;  \equiv \fr g \xrightarrow{\fr h}\fr h^{-1} \fr g \fr h$.
The loop groupoid of a finite groupoid being a finite groupoid itself, the procedure described above can be iterated so as to define the $n$-fold loop groupoid $\Lambda^n \mc G := \Lambda (\Lambda^{n-1}\mc G)$ and the corresponding $\mathbb T^n$-transgression map $\mathsf{t}^n : Z^{\bul}(\mc G,\rU(1)) \to Z^{\bul - n}(\Lambda \mc  G, \rU(1))$.
 
\bigskip \noindent
In addition to categories, in the following we will make use of higher categorical structures which we will model algebraically in terms of \emph{bicategories}.
 \begin{definition}[\emph{Bicategory}\label{def:bicategory}]
 	A bicategory $\mc{B}$ consists of:\\[-1.8em]
 	\begin{enumerate}[itemsep=0.3em,parsep=0pt,leftmargin=3em]
 		\item[${\ssss \bullet}$] A set of objects ${\rm Ob}(\mc{B})$.
 		\item[${\ssss \bullet}$] A category $\HOM_{\mc{B}}(X,Y)$, for every pair of objects $X,Y\in {\rm Ob}(\mc{B})$, such that objects and morphisms in $\HOM_{\mc B}(X,Y)$ are referred to as 1- and 2-morphisms, respectively. 
 		\item[${\ssss \bullet}$] A binary functor $\otimes: \mathsf{Hom}_{\mc{B}}(X,Y) \times \mathsf{Hom}_{\mc{B}}(Y,Z) \to \mathsf{Hom}_{\mc{B}}(X,Z)$, for every triple of objects $X,Y,Z\in{\rm Ob}(\mc{B})$.
 		\item[${\ssss \bullet}$] A natural isomorphism $\alpha_{f,g,h}:(f\otimes g)\otimes h\Rightarrow f\otimes (g\otimes h)$ called the `1-associator', for every triple of composable 1-morphisms $f$, $g$ and $h$, satisfying the `pentagon' axiom.
 		\item[${\ssss \bullet}$] A 1-morphism $\mathbbm{1}_{X}\in {\rm Ob}(\mathsf{Hom}_{\mc{B}}(X,X))$, for every object $X\in{\rm Ob}(\mc{B})$, and a pair of natural isomorphisms $\ell_{f}:\mathbbm{1}_{X}\otimes f\Rightarrow f$ and $r_f:f\otimes \mathbbm{1}_{Y}\Rightarrow f$ called the `left' and `right unitors', respectively, for every 1-morphism $f : X \to Y$. This data must satisfy the `triangle' axiom.
 	\end{enumerate}
 \end{definition}

\noindent 
A particularly important class of bicategories for the following discussion is provided by \emph{monoidal categories}:
\begin{definition}[Monoidal category]
	Let $\mc{B}$ be a bicategory with a unique object $\{\bul \} =\Ob(\mc{B})$. The hom-category $\HOM_{\mc{B}}(\bul,\bul) \equiv \mc{C}$ equipped with the composition functor $\otimes$ and the natural isomorphisms $\alpha$, $\ell$ and $r$ is called a `monoidal' category.
\end{definition}
\noindent
Endowing a monoidal category with additional properties, whose precise definitions can be found for instance in chap. 4 of \cite{etingof2016tensor},  yields the concept of \emph{multi-fusion category}:
 
\begin{definition}[Multi-fusion category]
	A `multi-fusion' category $\mc C$ is a rigid monoidal category that is $\mathbb{C}$-linear, abelian, semi-simple and such that the monoidal structure is given via a bifunctor $\otimes:\mc{C}\btimes\mc{C}\rightarrow\mc{C}$, where $\btimes$ denotes the Deligne tensor product of abelian categories. In addition, if  ${\rm Hom}_{\mc{C}}(\mathbbm{1},\mathbbm{1})\simeq \mathbb C$, then we call $\mc{C}$ a `fusion' category.
\end{definition}
\noindent
We shall now review the essential definitions of \emph{module categories} over a multi-fusion category $\mc{C}$, \emph{module category functors} and \emph{module category natural transformations}. Note that for the sake of conciseness, we shall omit to reproduce some of the relevant coherence relations. These can be found for instance in \cite{etingof2016tensor}, or in \cite{Bullivant:2020xhy}, where the notation is the same as here. 
\begin{definition}[Module category] 
	Let $\mc{C}\equiv (\mc C, \otimes , \mathbbm 1_\mc C, \ell, r, \alpha)$ be a multi-fusion category. We define a (left) $\mc{C}$-module category as a triple $(\mc M , \odot, \dot{\alpha})$ that consists of a category $\mc M$, an action bifunctor $\odot:\mc{C}\btimes \mc{M}\rightarrow \mc{M}$ and a natural isomorphism
	\begin{align}
	\dot{\alpha}_{X,Y,M}:(X\otimes Y)\odot M\xrightarrow{\sim}X\odot (Y\odot M) \, ,
	\q \forall \,  X,Y\in {\rm Ob}(\mc{C})\;\;  {\rm and}\;\; M\in {\rm Ob}(\mc{M}) \, .
	\end{align}
	The isomorphism $\dot{\alpha}$, which is referred to as the module associator, is subject to a `pentagon' axiom involving the monoidal associator $\alpha$. 
	In addition, there is a unit isomorphism
	$\ell_{M}:\mathbbm{1}_\mc C\odot M\xrightarrow{\sim} M$, that is subject to a `triangle' axiom involving the right unitor $r$.
\end{definition}
\noindent
\begin{definition}[Module category functor]
	Let $\mc C \equiv (\mc C, \otimes, \mathbbm 1_\mc C, \ell , r , \alpha)$ be a multi-fusion category and $(\mc M_1, \mc M_2)$ a pair of left $\mc C$-module categories with module associators $\dot{\alpha}$ and $\ddot{\alpha}$, respectively. We define a $\mc C$-module functor as a pair $(F,s)$, where $F: \mc M_1 \to \mc M_2$ is a functor, and $s$ a natural isomorphism such that
	\begin{align}
		s_{X,M}:F(X\odot M)\rightarrow X\odot F(M) \, , \q \forall \, X \in {\rm Ob}(\mc C) \; {\rm and} \; M \in {\rm Ob}(\mc M_1) \, .
	\end{align}
	Together, they satisfy a `pentagon' axiom involving $\dot{\alpha}$ and $\ddot{\alpha}$.
\end{definition}
\noindent 
\begin{definition}[Module category natural transformation]
	Let $\mc C \equiv (\mc C, \otimes, \mathbbm 1_\mc C, \ell , r, \alpha)$ be a multi-fusion category and $(F,s)$, $(F',s')$ a pair of $\mc C$-module functors. We define a $\mc C$-module natural transformation (or morphism of $\mc C$-module functors) between $F$ and $F$' as a natural transformation  $\eta:F \to  F'$ such that the following diagram commutes:
	\begin{align}
			\begin{tikzcd}[ampersand replacement=\&, column sep=1.8em, row sep=1.3em]
			|[alias=A]|F(X\odot M)
			\&\&
			|[alias=B]|X\odot F(M)
			\\\\
			|[alias=AA]|F'(X\odot M)
			\&\&
			|[alias=BB]|X\odot F'(M)
			\arrow[from=A,to=B,"s_{X,M}"]
			\arrow[from=B,to=BB,"{\rm id}_{X}\odot\eta_{M}"]
			\arrow[from=A,to=AA,"\eta_{X\odot M}"']
			\arrow[from=AA,to=BB,"s'_{X,M}"]
		\end{tikzcd} \, , 
	\end{align}
	for every $X \in {\rm Ob}(\mc C)$ and $M \in {\rm Ob}(\mc M)$.
\end{definition}
\noindent
Similarly, one can define \emph{right} module categories as well as the corresponding module category functors and morphisms of module category of functors.
Given the above definitions we can naturally form a bicategory of $\mc{C}$-module categories $\MOD(\mc{C})$ in analogy with the category of modules over a ring:
\begin{definition}[Bicategory of module categories\label{def:MOD}] Let $\mc C \equiv (\mc C, \otimes, \mathbbm 1_\mc C, \ell , r, \alpha)$ be a multi-fusion category.
	We define the bicategory $\mathsf{MOD}(\mc{C})$ as the bicategory with objects, $\mc{C}$-module categories, 1-morphisms, $\mc{C}$-module functors, and 2-morphisms, $\mc{C}$-module natural transformations.
\end{definition}
\noindent
Given a pair $(\mc M_1, \mc M_2)$ of $\mc C$-module categories, we will sometimes refer to the category $\msf{Fun}_\mc C(\mc M_1, \mc M_2)$ of $\mc C$-module functors $(F,s) : \mc M_1 \to \mc M_2$ and $\mc C$-module natural transformations.
An important instance of def.~\ref{def:MOD} is the bicategory of \emph{finite dimensional 2-vector spaces} $\TVect:=\MOD(\Vect)$, where $\Vect$ is the fusion category of finite dimensional complex vector spaces and linear maps. The motivation for such a nomenclature is that we should think of $\TVect$ as a categorification of $\Vect$. Indeed, by definition a (complex) vector space is a $\mathbb{C}$-module. Considering $\Vect$ as a possible categorification of $\mathbb{C}$, a 2-vector space (an object in $\TVect$) is defined analogously as a $\Vect$-module category. Moreover, finite dimensional 2-vector spaces are given by finite, $\mathbb{C}$-linear abelian, semi-simple categories.

\subsection{Algebras and higher algebras\label{sec:alg}}

\noindent
In the following, we shall present the data the Dijkgraaf-Witten theory assigns to certain closed manifolds of different dimensions in terms of \emph{algebras} and \emph{higher algebras}. Given a multi-fusion category, there is a notion of algebra internal to it:
\begin{definition}[Algebra object]\label{def:algObject}
	Let $\mc{C}\equiv (\mc C, \otimes , \mathbbm 1, \ell, r, \alpha)$ be a multi-fusion category. We define an (associative) algebra object in $\mc{C}$ as a triple $(A,m,u)$ that consists of an object $A$ and morphisms $m:A\otimes A\rightarrow A$ and $u:\mathbbm{1}\rightarrow A$ in $\mc{C}$ referred to as the multiplication and the unit, respectively. The morphisms $m$ and $u$ are subject to an associativity and unitality conditions involving the monoidal structure $\mc C$.
\end{definition}
\noindent
Using the previous definitions, we can define an important class of algebra objects in $\mathsf{Vec}$, namely
\emph{twisted groupoid algebras}, which generalise straightforwardly the concept of twisted group algebra: 
\begin{definition}[Twisted groupoid algebra]
	Given a groupoid $\mc G$ and a normalised groupoid 2-cocycle in $[\beta] \in H^2(\mc G, \rU(1))$, we define the twisted groupoid algebra $\mathbb C[\mc G]^\beta$ as the associative algebra with defining vector space ${\rm Span}_\mathbb{C}\{| \fr g \ra \, | \, \forall \, \fr g \in {\rm Hom}(\mc G)\}$ and algebra product
	\begin{equation}
		| \fr g \ra \star | \fr g' \ra := \delta_{{\rm t}(\fr g),{\rm s}(\fr g')} \, \beta(\fr g , \fr g') \, | \fr g \fr g' \ra \, ,
	\end{equation}
	for all $\fr g, \fr g' \in \Hom(\mc G)$.
\end{definition}
\noindent
Following the theory of twisted group algebras, we can show that twisted groupoid algebras are \emph{semi-simple} algebras, i.e. every module is isomorphic to a direct sum of simple modules.

The notion of algebra object admits a natural categorification in terms of \emph{pseudo-algebra objects}, whose definition can be found in \cite{Bullivant:2020xhy}. Guided by the observation that $\TVect$ defines a \emph{fusion 2-category} in the sense of \cite{douglas2018fusion}, we would like to consider an appropriate categorification of the notion of twisted groupoid algebras, referred to as twisted groupoid 2-algebras, which are examples of \emph{pseudo-algebra} objects internal to $\TVect$. Akin to the categorification of $\Vect$ to $\TVect$, groupoid 2-algebras are obtained by promoting the field $\mathbb C$ to $\Vect$, yielding the notion of groupoid-graded vector spaces:

\begin{example}[Category of groupoid-graded vector spaces\label{def:twistedgroupoid2alg}]
	Let $\mc{G}$ be a finite groupoid and $\alpha$ a normalised groupoid 3-cocycle in $H^{3}(\mc{G},\rU(1))$ for the trivial $\mc{G}$-module $\rU(1)$. The twisted groupoid 2-algebra $\Vect^{\alpha}_{\mc{G}}$ is the category, whose objects are $\Hom(\mc{G})$-graded vector spaces and morphisms, grading preserving linear maps. There are $|\Hom(\mc G)|$-many simple objects notated via $\mathbb C_\fr g$, $\forall \, \fr g \in \Hom(\mc G)$. Including the zero vector space $\varnothing$ as the vector space with no grading, the monoidal structure is given by a bifunctor $\otimes:\Vect^{\alpha}_{\mc{G}}\btimes \Vect^{\alpha}_{\mc{G}}\rightarrow \Vect^{\alpha}_{\mc{G}}$
	which acts on objects via
	\begin{align}
		\otimes:V_{\fr g}\times W_{\fr h}\mapsto
		\begin{cases}
			(V\otimes W)_{\fr g \fr h}\q &\text{if} \;\; {\rm t}(\fr g)={\rm s}(\fr h)
			\\
			\q\q\; \varnothing &\text{else}
		\end{cases}\, ,
	\end{align}
	for all $V,W \in {\rm Ob}(\Vect^\alpha_\mc G)$ and $\fr g, \fr h\in \Hom(\mc G)$, and on morphisms $f_{\fr g}:V_{\fr g}\to V'_{\fr g}$ and $f'_{\fr h}:W_{\fr h}\to W'_{\fr h}$ via
	\begin{align}
		\otimes:f_{\fr g}\otimes f'_{\fr h}\mapsto
		\begin{cases}
			(f\otimes f')_{\fr g \fr h}\q &\text{if} \;\; {\rm t}(\fr g)={\rm s}(\fr h)
			\\
			\q\q \; 0 &\text{else}
		\end{cases} \, ,
	\end{align}
	for all $\fr g,\fr h\in \Hom(\mc{G})$. The monoidal associator is given by the natural transformation
	\begin{align}
		\alpha_{U_{\fr g},V_{\fr h},W_{\fr k}} = 
		\alpha(\fr g , \fr h, \fr k )
		\cdot (\eta_{U,V,W})_{\fr g \fr h \fr k} : 
		(U_{\fr g}\otimes V_{\fr h})\otimes W_{\fr k}
		\xrightarrow{\sim} U_{\fr g}\otimes (V_{\fr h}\otimes W_{\fr k}) \, ,
	\end{align}
	for all composable morphisms $\fr g, \fr h, \fr k \in \Hom(\mc G)$, where $\eta_{U,V,W}$
	is the canonical isomorphism of vector spaces $(V\otimes W)\otimes Z\xrightarrow{\sim} V\otimes (W\otimes Z)$. The monoidal unit is given by $\mathbbm{1}:=\bigoplus_{ X\in\Ob(\mc{G})}\mathbb{C}_{{\rm id}_X}$ and the unitality conditions are
	\begin{align}
		\ell_{V_{\fr g}}:V_{\fr g}\otimes\mathbbm{1}\xrightarrow{\sim} V_{\fr g} \; , \q 
		r_{V_{\fr g}}:\mathbbm{1} \otimes V_{\fr g}\xrightarrow{\sim} V_{\fr g} \; ,
	\end{align}
	which are defined by the canonical isomorphisms $(V\otimes \mathbb{C})_{\fr g}\simeq V_{\fr g}$ and $(\mathbb C \otimes V)_{\fr g} \simeq V_\fr g$, respectively
	for all vector spaces $V$ and $\fr g\in\Hom(\mc{G})$. Subsequently we will assume $\Vect^{\alpha}_{\mc{G}}$ is equipped with the following rigid structure, which in particular ensures $\Vect^{\alpha}_{\mc{G}}$ is multi-fusion. The dual of an object $V^*$ is given by $\bigoplus_{\fr g \in \Hom(\mc G)}(V^*)_\fr g$ with $(V^*)_\fr g :=\Hom_{\Vect}(V_{\fr g^{-1}},\mathbb{C})$, and the evaluations ${\rm ev}_{V_\fr g}:(V^*)_\fr g \otimes V_{\fr g^{-1}} \rightarrow \mathbbm{1}$ and $\widetilde{{\rm ev}}_{V_\fr g}: V_{\fr g^{-1}} \otimes (V^*)_\fr g \rightarrow \mathbbm{1}$ are defined by the linear maps
	\begin{align}
		{\rm ev}_{V_\fr g}: (f\otimes v)_{{\rm id}_{{\rm s}(\fr g)}}\mapsto f(v)_{{\rm id}_{{\rm s}(\fr g)}}
		\; , \q 
		\widetilde{\rm ev}_{V_\fr g}: (v\otimes f)_{{\rm id}_{{\rm s}(\fr g)}}\mapsto \alpha(\fr g, \fr g^{-1}, \fr g) \, f(v)_{{\rm id}_{{\rm s}(\fr g)}} \; , 
	\end{align}
	respectively, for all $f\in(V^*)_\fr g$ and $v\in V_{\fr g^{-1}}$. The corresponding coevaluation maps are provided by the grading preserving linear maps
	\begin{align}
		{\rm coev}_{V_\fr g}:{\rm id}_{{\rm id}_{\fr g}}\mapsto \alpha(\fr g, \fr g^{-1}, \fr g)^{-1}\bigoplus^{{\rm dim}(V)}_{i=1} (v_{i}\otimes f_{i})_{{\rm id}_{\fr g}}
		\; , \q
		\widetilde{\rm coev}_{V_\fr g}:{\rm id}_{{\rm id}_{\fr g}}\mapsto\bigoplus^{{\rm dim}(V)}_{i=1} (f_{i}\otimes v_{i})_{{\rm id}_{\fr g}}	
	\end{align}
	where $\{v_{i}\}^{{\rm dim}(V)}_{i=1}$ denotes a basis for $V_\fr g$ and $\{f_{i}\}^{{\rm dim}(V)}_{i=1}$ denotes a basis for $(V^*)_\fr g$ such that $f_{j}(v_{i})=\delta_{i,j}$ for all $i,j\in 1,\ldots,{\rm dim}(V)$.
\end{example}

\noindent
As we shall evoke below, this procedure can be iterated by promoting the category $\Vect$ to $\TVect$ so as to define a notion of \emph{3-algebra}.

Crucially, algebra objects in the category $\Vect^{\alpha}_{\mc{G}}$ of groupoid-graded vector spaces admit a simple characterisation in terms of $(\mc G, \alpha)$-subgroupoids:
\begin{definition}
	Let $\mc G$ be a finite groupoid and $\alpha$ a normalised groupoid 3-cocycle in $H^3(\mc G, \rU(1))$. We define a $(\mc G, \alpha)$-subgroupoid as a pair $(\mc A,\phi)$ that consists of a subgroupoid $\mc A \subseteq \mc G$ and a cochain $\phi \in C^2(\mc A, \rU(1))$ satisfying $d^{(2)}\phi(\fr a, \fr a' , \fr a '') = \alpha^{-1}(\fr a, \fr a', \fr a '')$ for any triple $(\fr a, \fr a', \fr a'')$ of composable morphisms in $\Hom(\mc A)$.
\end{definition}

\noindent
Given a $(\mc G, \alpha)$-subgroupoid $(\mc A, \phi)$, an algebra object in $\Vect^{\alpha}_{\mc{G}}$  is  defined as $\mc A_\phi \equiv (\bigoplus_{\fr a \in \Hom(\mc A)} \mathbb C_\fr a, m ,u)$ where
\begin{equation}
	\begin{array}{ccccl}
	m & : & \mc{A}_\phi \otimes \mc{A}_\phi & \to & \mc{A}_\phi
	\\
	& : & {\fr a} \otimes {\fr a'} &\mapsto & \delta_{{\rm t}(\fr a),{\rm s}(\fr a')} \, \phi(\fr a, \fr a')\,  {\fr a \fr a '}
	\end{array} 
	\q {\rm and} \q 
	u(\mathbbm{1}_{\mathsf{Vec}^{\alpha}_{\mc{G}}}):=\sum_{X\in {\rm Ob}(\mc{A}_\phi)}{\rm id}_{X} \, .
\end{equation}
Finally, we shall require the notion of module objects:
\begin{definition}[Right module object\label{def:algObjMod}]
	Let $\mc C \equiv (\mc C, \otimes, \mathbbm 1, \ell, r, \alpha)$ be a multi-fusion category and $a \equiv (A,m,u)$ an algebra object in $\mc C$. We define a right module object over $A$ as a pair $(M,p)$ consisting of an object $M \in \Ob(\mc C)$ and an action morphism $p: M \otimes A \to M \in \Hom(\mc C)$ that satisfies a compatibility condition involving the multiplication $m$ and the associator $\alpha$, as well as a unit constraint involving $r$ and $u$.
\end{definition}

\noindent
Given two module objects, a module object homomorphism between them is a morphism between the corresponding objects in the underlying category that satisfies a compatibility condition involving the action morphisms of both algebra objects. Left module objects and the corresponding homomorphisms are defined in a similar fashion.
Furthermore, given an algebra object, the subspace of module object homomorphisms is stable under composition and as such we can define the following category:
\begin{definition}[Category of module objects] 
	Let $\mc C$ be a multi-fusion category and $A \equiv (A,m,u)$ and algebra object in $\mc C$. We define the category $\Mod_\mc C(A)$ as the category with objects $A$-module objects and morphisms $A$-module homomorphisms.
\end{definition}

\noindent
Henceforth, given an algebra object $A$ in $\Vect$, we shall notate the category $\Mod_{\Vect}(A)$ as $\Mod(A)$, which is equivalent to the category $\msf{Rep}(A)$ of representations and intertwiners of $A$.

\subsection{Dijkgraaf-Witten theory in a nutshell\label{sec:DW}}

\noindent
We shall now review from a physical perspective what the (3+1)d Dijkgraaf-Witten theory assigns to the manifolds $\mathbb T^4$, $\mathbb T^3$, $\mathbb T^2$ and $\mathbb S^1$, in terms of the category-theoretical notions presented above. Since we shall only briefly motivate and quote the results, we encourage the reader to consult \cite{dijkgraaf1990topological, Wan:2014woa, Wang:2014oya, Bullivant:2019fmk, Bullivant:2020xhy} for details.

The input of the theory is a finite group $G$ and a normalised representative in $[\pi] \in H^4(G,\rU(1))$. Given a four-manifold $\mc M$, we endow it with a triangulation $\mc M_\triangle$ equipped with a total ordering $v_0 < v_1 < \ldots < v_{|\mc M_\triangle|^0}$ of its 0-simplices. This total ordering induces an orientation $\epsilon(\triangle^{(n)}) = \pm 1 $ for every $n$-simplex $\triangle^{(n)}$. We define a $G$-colouring of such a triangulation as an assignment of group variables to every 1-simplex such that for every 2-simplex $(v_i v_j v_k)$ with $v_i < v_j < v_k$ the \emph{flatness} condition $g_{v_iv_j}g_{v_j v_k} = g_{v_i v_k}$ is verified. The set of $G$-colourings is denoted by ${\rm Col}(\mc M_\triangle, G)$. Writing the restriction of a $G$-colouring $g$ to a $4$-simplex $\triangle^{(4)} = (v_0 v_1v_2v_3v_4)$ as $g[v_0 \ldots v_4] \equiv (g_{v_0v_1}, \ldots, g_{v_{3}v_4})$, the evaluation of the 4-cocycle $\omega$ on $\triangle^{(4)}$ is given by $\omega(g[v_0 \ldots v_{4}]) \equiv \omega(g_{v_0 v_1}, \ldots, g_{v_3 v_{4}})$. Given the above conventions, the state-sum assigns to $\mc M_\triangle$ the following complex number
\begin{equation}
	\mc Z_G^\omega(\mc M_\triangle)= \frac{1}{|G|^{|\mc M_\triangle|^0}}\sum_{g \in {\rm Col}(\mc M_\triangle,G)}\prod_{\triangle^{(n)} \subset \mc M_\triangle}
	\!\!\! \la \omega(g), \triangle^{(n)}\ra \, ,
\end{equation} 
where we introduced the topological action $\la \omega(g),\triangle^{(n)}\ra := \omega(g[\triangle^{(n)}])^{\epsilon(\triangle^{(n)})}$. Let us now specialize to the four-torus $\mathbb T^4$. It can be triangulated as a 4-cube with opposite 3-cubes identified that is decomposed into twenty-four 4-simplices. In the spirit of \cite{willerton2008twisted}, we shall write the complex number $\mc Z_G^\omega(\mathbb T^4_\triangle)$ using the language of loop groupoids. Let $\overline G$ be the delooping of the finite group $G$, i.e. the group treated as a one object groupoid. The loop groupoid $\Lambda \overline G$ is the groupoid such that ${\rm Ob}(\Lambda \overline G) = G$ and ${\rm Hom}(\Lambda \overline G) = \{g \xrightarrow{a}a^{-1}ga \equiv g \xrightarrow{a} \, | \, \forall \, g,a \in G\}$. Henceforth, we shall use the shorthand notation $\Lambda G \equiv \Lambda \overline G$, whenever no confusion is possible. A groupoid cocycle in $H^3(\Lambda G, \rU(1))$ can then be obtained by applying the $\mathbb S^1$-transgression map to the group $4$-cocycle $\pi$ such that
\begin{equation}
	\mathsf t (\pi)(a \xrightarrow{b} \, ,  a^b \xrightarrow{c} \, , a^{bc} \xrightarrow{d}) = 
	\frac{\omega(b,a^b,c,d) \, \omega(b,c,d,a^{bcd})}{\omega(a,b,c,d) \, \omega(b,c,a^{bc},d)} \, ,
\end{equation}
where we introduced the shorthand notation $x^y := y^{-1}xy$.
Iterating this process, we can construct the 4-fold loop groupoid $\Lambda ^4  G$ together with the $\mathbb T^4$-transgression map such that $\mathsf t^4(\pi) \in H^0(\Lambda ^4 G, \rU(1))$. By definition, objects in ${\rm Ob}(\Lambda^4 G)$ are characterised by quadruples $\{a,b,c,d\}$ of group variables in $G$ such that $[x,y] := xyx^{-1}y^{-1}=\mathbbm 1_G$, for every $(x,y) \in \{a,b,c,d\}$. Moreover, given a groupoid $\mc G$, a groupoid 0-cocycle can be interpreted as a $\rU(1)$-valued function over the objects of $\mc G$, which only depend on connected components of $\mc G$. Given the following notation
\begin{equation}
	\label{eq:DimGr}
	\int_{\mc G} \gamma := \sum_{[X] \in \uppi_0(\mc G)}\frac{\gamma(X)}{|\mathtt{Aut}(X)|} \, ,
\end{equation}
where $\mathtt{Aut}(X)$ refers to the group of morphisms in ${\rm End}_\mc G(X)$ and $\uppi_0(\mc G)$ is the set of connected components,
the complex number the TQFT assigns to $\mathbb T^4$ can finally be expressed in the following concise way \cite{willerton2008twisted}
\begin{equation}
	\label{eq:ZTfour}
	\mc Z^\pi_G(\mathbb T^4) = \int_{\Lambda^4  G}\mathsf{t}^4(\pi) \, .
\end{equation}

\bigskip
\noindent
By definition, the Dijkgraaf-Witten theory assigns a vector space to every closed three-manifold. Specialising to the case of the three-torus, let us now compute this vector space. We shall find this vector space as the ground state subspace of the Hamiltonian realisation of the theory on $\mathbb T^3$. This lattice Hamiltonian, whose explicit definition can be found in \cite{Wan:2014woa,Bullivant:2019fmk}, is such that its ground state projector on $\mathbb T^3$ is equal to the linear map the theory assigns to the bordism $\mathbb T^3 \times \mathbb I$. Since the theory maps the manifold $\mathbb T^3 \times \mathbb I$ to the identity map ${\rm id}_{\mc Z^\pi_G(\mathbb T^3)               }$, we have $\mc Z^\pi_G(\mathbb T^3) = {\rm Im}\, \mc Z^\pi_G(\mathbb T^3 \times \mathbb I)$. Triangulating the three-torus as a cube with opposite plaquettes identified that is decomposed into six 3-simplices, we define a microscopic state on this triangulation as a state $|a,b,c\ra \in \mathbb C[G]^{\otimes 3}$, where it is understood that $a$, $b$ and $c$ label the 1-simplices going along each non-contractible 1-cycle, respectively. Acting with the ground state projector $\mc Z^\pi_G(\mathbb T^3 \times \mathbb I)$ yields
\begin{align}
	\mc Z^\pi_G(\mathbb T^3) 
	&\simeq {\rm Span}_\mathbb{C}\Big\{\frac{1}{|G|}\sum_{\fr x \in \Hom_{\Lambda ^3 G}(\fr a,-)}
	\!\!\! \msf{t}^3(\pi)(\fr a\xrightarrow{ \fr x}) \, 
	|\fr x^{-1} \fr a \fr x \ra \Big\}_{\fr a \in \Ob(\Lambda ^3  G)} 
	\, ,
\end{align}
where objects $\fr a \in {\rm Ob}(\Lambda^3  G)$ are characterised by triples $\{a,b,c\}$ of $G$-variables such that $[a,b] = [b,c] = [a,c] = \mathbbm 1_G$. Crucially, it follows from the triangulation invariance of the state-sum that any two choices of triangulation for $\mathbb T^3$ yields isomorphic vector spaces. More generally, given a groupoid $\mc G$ and groupoid 1-cocycle $\epsilon$, we define the following vector space \cite{willerton2008twisted}:
\begin{equation}
	\label{eq:defVSpace}
	\mc V_{{\mc G}}(\epsilon) := {\rm Span}_\mathbb C \Big\{ \msf s: {\rm Ob}(\mc G) \to \mathbb C \, \big| \, \msf s(Y) = 
	\epsilon(\fr g) \,  \msf s( X) \; \forall \, \fr g \in {\rm Hom}_\mc{G}(X,Y) \Big\} \ .
\end{equation}	
Using the 1-cocycle condition of $\msf t^3(\pi)$, we obtain that the vector space the TQFT assigns to $\mathbb T^3$ is isomorphic to
\begin{equation}
	\label{eq:ZTthree}
	\mc Z_G^\pi(\mathbb T^3) \simeq \mc V_{\Lambda^3  G}(\mathsf{t}^3(\pi)) 
	\, .
\end{equation}

\bigskip \noindent
Let us now motivate from the lattice Hamiltonian point of view what the theory assigns to the two-torus. By the definition of an extended TQFT, we expect the theory to assign to every closed two-manifold a finite dimensional 2-vector space, and to an open three-manifold an object in the 2-vector space associated with its boundary. We shall argue that the quantum invariant the Dijkgraaf-Witten theory assigns to the two-torus can be interpreted as the category of loop-like bulk anyonic excitations hosted by its lattice Hamiltonian realisation.

As mentioned above, the lattice Hamiltonian realisation of the theory on a three-manifold is such that its ground state subspace is isomorphic to the vector space the theory assigns to the closed three-manifold. More specifically, it is an exactly solvable model obtained as a sum of mutually commuting projectors, which act on neighbourhoods of the $0$-simplices that are in the interior of the manifold \cite{Wan:2014woa,Bullivant:2019fmk}. It follows that the Hamiltonian has \emph{open boundary conditions}. Generically, such boundary conditions can be interpreted as excitations that are linear superpositions of electric \emph{charges} and magnetic \emph{fluxes}. Indeed, given an excitation, which is by definition a subcomplex whose energy is higher than that of the ground state, the equivalence class of topological excitations up to the insertion of a local excitation are encoded onto the boundary conditions of the manifold that results from removing this subcomplex from the three-manifold \cite{Lan:2013wia,Bullivant:2019fmk}. Given that a loop-like excitation is an excitation whose topology is that of the circle $\mathbb S^1$, and that a regular neighbourhood of $\mathbb S^1$ has the topology of a solid torus $\mathbb D^2 \times \mathbb S^1$, we find that loop-like excitations are classified by boundary conditions on the two-torus $\mathbb T^2 = \partial (\mathbb D^2 \times \mathbb S^1)$---the solid torus providing here the aforementioned subcomplex, for which the energy is above the ground state one.

Let $\Sigma^{\rm o}$ be the open manifold obtained by removing a solid torus from a closed three-manifold $\Sigma$. Given such a manifold, it is always possible to glue a copy of the manifold $\mathbb T^2 \times \mathbb I$ along $\partial \Sigma^{\rm o}$ without altering its topology. As detailed in \cite{Bullivant:2019fmk}, this gluing operation can be extended to an action of the ground states on $\mathbb T^2 \times \mathbb I$ onto those on $\Sigma^{\rm o}$. Similarly, the operation that consists in gluing two copies of $\mathbb T^2 \times \mathbb I$ and apply an orientation-preserving diffeomorphism from $\mathbb T^2 \times [0,2]$ to $\mathbb T^2 \times [0,1]$ can be extended so as to endow the ground state subspace on $\mathbb T^2 \times \mathbb I$ with the structure of an associative semi-simple $*$-algebra. This algebra was shown in \cite{Bullivant:2019fmk} to be Morita equivalent to the groupoid algebra $\mathbb C[\Lambda^2 G]^{\mathsf t^2(\pi)}$ of the 2-fold loop-groupoid twisted by the $\mathbb T^2$-transgression map $\mathsf t^2(\pi)$ in $H^2(\Lambda ^2  G, \rU(1))$. Applying the definition of sec.~\ref{sec:preliminaries}, this is the algebra with underlying vector space
\begin{equation}
	{\rm Span}_\mathbb C \big\{\big| \fr g \xrightarrow{\fr a} \big\ra \; \big| \; 	\fr g \xrightarrow{\fr a} \; \in {\rm Hom}(\Lambda^2  G)\big\} \, ,
\end{equation}
and product rule
\begin{equation}
	\label{eq:twistedQTriple}
	| \fr g \xrightarrow{\fr a} \ra \star | \fr g' \xrightarrow{\fr a'} \ra = \delta_{\fr g', \fr a^{-1}\fr g \fr a}
	\, \msf t^2(\pi)(\fr g \xrightarrow{\fr a}, \fr a^{-1 }\fr g \fr a\xrightarrow{\fr a'}) \, | \fr g \xrightarrow{\fr a \fr a'} \ra
	\, .
\end{equation}
It follows that the ground state subspace on $\Sigma^{\rm o}$ has the structure of a module over $\mathbb C[\Lambda^2 G]^{\mathsf t^2(\pi)}$, which can be decomposed over isomorphism classes of \emph{simple} modules. Noticing that we can always find a collar neighbourhood of the torus-like boundary that is diffeomorphic to $\mathbb T^2 \times \mathbb I$, the task of finding such simple modules reduces to classifying the simple modules of the \emph{regular} module of $\mathbb C[\Lambda^2  G]^{\mathsf t^2(\pi)}$, which in turn boils down to computing the irreducible representations of the algebra. Putting everything together, we obtain that the loop-like bulk excitations of the lattice Hamiltonian realisation are encoded into the category $\mathsf{Mod}(\mathbb C[\Lambda^2 G]^{\msf t^2(\pi)})$ of $\GrAlg$-modules and module homomorphisms, which is the quantum invariant that the TQFT assigns to $\mathbb T^2$, i.e.
\begin{equation}
	\label{eq:ZTtwo}
	\mc Z^\pi_G(\mathbb T^2) = \mathsf{Mod}(\mathbb C[\Lambda^2 G]^{\msf t^2(\pi)}) \, .
\end{equation}
Note that this category can be further endowed with a braided monoidal structure that encode the \emph{fusion} and the \emph{braiding} of the loop-like excitations \cite{Bullivant:2019fmk}. We shall comment further on these aspects in the following sections.

\bigskip \noindent
Finally, let us present the 2-category that the theory assigns to the circle $\mathbb S^1$. We shall introduce this 2-category as the bicategory of open string-like excitations hosted by the lattice Hamiltonian realisation of the theory \cite{Bullivant:2020xhy}. In the vein of the discussion above, we shall reveal this bicategory via a categorification of the tube algebra approach. 

Given a three manifold $\Sigma$ with a non-empty boundary $\partial \Sigma$, we consider the lattice Hamiltonian realisation of the theory on $\Sigma$. We are interested in string-like excitations that terminate at the spatial boundary, a regular neighbourhood of which has the topology of a solid cylinder, i.e. $\mathbb D^2 \times \mathbb I$. Removing such a regular neighbourhood leaves a cylinder-like boundary component referred to as the \emph{excitation boundary}, which is incident on $\partial \Sigma$. We notate the resulting manifold via $\Sigma^{\rm o}$ and the excitation boundary via $\partial \Sigma^{\rm o}{\sss |}_{\rm ex.}$.  Similarly to the loop-like excitations, we would like to classify these string-like excitations via a classification of the boundary conditions along the excitation boundary $\partial \Sigma^{\rm o}{\sss |}_{\rm ex.}$. Following the tube algebra approach, we begin by noticing that we can always glue a copy of the \emph{pinched interval bordism} $(\mathbb S^1 \times \mathbb I) \times_{\rm p} \mathbb I$ along $\partial \Sigma^{\rm o}{\sss |}_{\rm ex.}$, where $\Xi \times_{\rm p} \mathbb I$ is defined as $\Xi \times \mathbb I / \sim$ such that $(x,i)\sim (x,i')$ for $(x,i) \in \partial \Xi \times \mathbb I$. Similarly, the gluing of two copies of the manifold $(\mathbb S^1 \times \mathbb I) \times_{\rm p} \mathbb I$ along $\mathbb S^1 \times \mathbb I$ is diffeomorphic to $(\mathbb S^1 \times \mathbb I) \times_{\rm p} \mathbb I$. We thus enrich the vector space of ground states on $(\mathbb S^1 \times \mathbb I) \times_{\rm p} \mathbb I$ with the structure of an associative semi-simple $*$-algebra, so that the ground state subspace on $\Sigma^{\rm o}$ has the structure of a module over it. This algebra can be checked to be isomorphic to the algebra $\mathbb C^{\Lambda  G}$ of functions over the loop groupoid $\Lambda  G$. The tube algebra approach thus prescribes  that the theory assigns the category $\msf{Mod}(\mathbb C^{\Lambda  G})$ to $\mathbb S^1 \times \mathbb I$, which happens to be equivalent to the category of loop-groupoid-graded vector spaces $\Vect_{\Lambda  G}$.\footnote{Recall that for a finite group $G$, we have the equivalences $\msf{Rep}(G) \cong \Mod(\mathbb C[G])$ and $\Vect_G \cong \Mod(\mathbb C^G)$.} Alternatively, we can think of $\Mod(\mathbb C^{\Lambda G})$ as the category of functors from the groupoid, with object-set $\Lambda G$ and trivial morphisms, to $\msf{Vec}$.

In sharp contract to the case of the torus, we can define another gluing operation for the manifold $(\mathbb S^1 \times \mathbb I) \times_{\rm p} \mathbb I$, namely along the circle boundary components. This gluing operation can be lifted to a product rule $\otimes : \msf{Vec}_{\Lambda G} \btimes \msf{Vec}_{\Lambda  G} \to \msf{Vec}_{\Lambda G}$, which equips $\msf{Vec}_{\Lambda G}$ with a monoidal structure that is the one discussed in def.~\ref{def:twistedgroupoid2alg}. In particular, it follows from the triangulation invariance of the theory that the monoidal associator is characterised by $\msf{t}(\pi)$, as expected, so that the manifold $(\mathbb S^1 \times \mathbb I) \times_{\rm p} \mathbb I$ is endowed with the structure of the 2-algebra $\Vect_{\Lambda  G}^{\msf{t}(\pi)}$. We shall now find what the theory assigns to the circle by computing the `representations' of this categorified tube algebra, which in this context are provided by the module categories over the 2-algebra $\Vect_{\Lambda  G}^{\msf{t}(\pi)}$. Putting everything together, we can argue that the theory assigns to the circle the following bicategory (see def.~\ref{def:MOD}):
\begin{equation}
	\label{eq:ZSone}
	\mc Z^\pi_G(\mathbb S^1) = \msf{MOD}(\VectGr)
	 \; ,
\end{equation}
such that objects in this bicategory are interpreted as defect boundary conditions for the endpoints of a string-like excitation---isomorphism classes of which specifying in particular allowed magnetic fluxes for the string---the 1-morphisms as dyonic quantum numbers associated with string-like topological excitations that are constrained by a choice of boundary conditions at the endpoints, and 2-morphisms as implementing the renormalisation of string-like excitations that are glued along their endpoints  \cite{Bullivant:2020xhy}. The 2-algebra $\VectGr$ can further be equipped with the structure of a \emph{quasi-triangular quasi-Hopf category}, categorifying the notion of \emph{quasi-triangular quasi-Hopf algebra}\footnote{The twisted quantum double, which is the tube algebra associated with the point-like excitations hosted by the lattice Hamiltonian realisation of the (2+1)-dimensional theory, is an example of quasi-triangular quasi-Hopf algebra \cite{Dijkgraaf1991}.} and serving as a non-strict example of the construction in \cite{neuchl1997representation}. Such an extension in turn equips $\MOD(\VectGr)$ with a braided monoidal bicategorical structure.

\section{Centres\label{sec:centres}}

\noindent
\emph{In preparation for the following, we will reformulate the category-theoretical data that the theory assigns to the manifolds $\mathbb T^3$, $\mathbb T^2$ and $\mathbb S^1$ in terms of the notion of `centre' and categorifications thereof.}

\subsection{Centre of a (1-)algebra}

In the previous section, we briefly summarised the result that the Dijkgraaf-Witten theory assigns the vector space $\mc V_{\Lambda ^3 G}(\msf{t}^3(\pi))$ to the three-torus $\mathbb T^3$. We now would like to express this vector space in terms of the algebra $\mathbb C[\Lambda ^2  G]^{\msf t^2(\pi)}$, the modules of which label the bulk loop-like excitations of the Hamiltonian realisation. 

Given a groupoid $\mc G$ and normalised groupoid 2-cocycle in $[\beta] \in H^2(\mc G, \rU(1))$, let us consider the centre $Z(\mathbb C[\mc G]^\beta)$ of the twisted groupoid algebra $\mathbb C[\mc G]^\beta$. Recall that the centre is given by the commutative subalgebra consisting of all elements $|\psi \ra \in \mathbb C[\mc G]^\beta$ that satisfy the relation
\begin{equation}
	| \psi \ra \star | \fr g \ra = | \fr g \ra \star | \psi \ra \, , \q \forall \, |\fr g\ra \in \mathbb C[\mc G]^\beta \, .
\end{equation}
The fact that it is indeed an algebra follows from the associativity of $\mathbb C[\mc G]^\beta$ and the observation that
\begin{equation}
	(| \fr \psi \ra \star | \phi \ra) \star | \fr g \ra 
	= | \psi \ra \star | \fr g \ra \star | \phi \ra 
	= | \fr g \ra \star (|\psi \ra \star | \phi \ra) \, ,
\end{equation}
for every $|\psi \ra,|\phi \ra \in Z(\mathbb C[\mc G]^\beta)$ and $| \fr g \ra \in \mathbb C[\mc G]^\beta$.
We shall now establish the fact that as a vector space, this centre is isomorphic to $\mc V_{\Lambda \mc G}(\msf{t}(\beta))$, as defined in \eqref{eq:defVSpace}. Given a function $\msf s \in \mc V_{\Lambda \mc G}(\msf t(\beta))$, we consider the groupoid algebra element $| \psi \ra = \sum_{\fr g \in {\rm Ob}(\Lambda \mc G)}\overline{\msf s(\fr g)}| \fr g \ra$. It satisfies
\begin{align}
	\nn
	| \fr x \ra \star | \psi \ra &= 
	\sum_{\substack{Z \in {\rm Ob}(\mc G) \\ \fr g \in {\rm End}_\mc{G}(Z)}}
	\!\!\! \overline{{\msf{s}(\fr g)}} \, | \fr x \ra \star | \fr g \ra 
	= 
	\sum_{\fr g \in {\rm End}_\mc{G}(Y)} \overline{\msf s(\fr g)} \, \beta(\fr x , \fr g) \, | \fr x \fr g \ra 
	\\ 
	\nn
	& =
	\sum_{\fr g \in {\rm End}_\mc{G}(X)} \overline{\msf s(\fr x^{-1} \fr g \fr x)} \, 
	\beta(\fr x, \fr x^{-1} \fr g \fr x) \, | \fr g \fr x \ra
	=
	\sum_{\fr g \in {\rm End}_\mc G(X)} \overline{\msf s(\fr x^{-1} \fr g \fr x)} \, \msf t(\beta)(\fr g \xrightarrow{\fr x}) \, \beta(\fr g, \fr x) \, | \fr g \fr x \ra
	\\ &=
	\sum_{\fr g \in {\rm End}_\mc{G}(X)} \overline{\msf s(\fr g)} \, \beta(\fr g, \fr x) \, |\fr g \fr x \ra = 
	\sum_{\substack{Z \in {\rm Ob}(\mc G) \\ \fr g \in {\rm End}_\mc{G}(Z)}}
	\!\!\! \overline{\msf s(\fr g)} \, | \fr g \ra \star | \fr x \ra = | \psi \ra \star | \fr x \ra
\end{align}
for every $\fr x \in {\rm Hom}_{\mc G}(X,Y)$, where we used the defining property of $\msf s$ as well as the explicit expression of $\msf t(\beta)$. This proves that $| \psi \ra \in Z(\mathbb C[\mc G]^\beta)$. Conversely, given an element $\sum_{\fr g \in {\rm Hom}(\mc G)}\overline{\msf s(\fr g)}| \fr g \ra$ in the centre, we can check that $\msf s \in \mc V_{\Lambda \mc G}(\msf t(\beta))$, proving the isomorphism $\mc V_{\Lambda \mc G}(\msf t(\beta)) \simeq Z(\mathbb C[\mc G]^\beta)$. Specialising to $\mc G = \Lambda ^2 G$ finally yields
\begin{equation}
	\label{eq:isoVect}
	\mc V_{\Lambda ^3  G}(\msf t^3(\pi)) \simeq Z(\mathbb C[\Lambda^2 G]^{\msf t^2(\pi)}) \, .
\end{equation}
Notice that the twisted groupoid algebra $\mathbb C[\Lambda ^2  G]^{\msf t^2(\pi)}$ enters the definition of the category that the theory assigns to $\mathbb T^2$. This is premonitory of the relation $\msf{Dim}: \mc Z^\pi_G(\mathbb T^2) \mapsto \mc Z^\pi_G(\mathbb T^3)$ that we shall establish in the following section.

\subsection{Centre of a 2-algebra\label{sec:centre2Algebra}}

We showed above that the vector space assigned by $\mc Z_G^\pi$ to $\mathbb T^3$ is isomorphic to that spanned by the central elements of the algebra $\mathbb C[\Lambda^2 G]^{\msf t(\pi)}$. Similarly, we shall now demonstrate that the category assigned by the theory to $\mathbb T^2$ is equivalent to the categorified centre of a 2-algebra, namely $\VectGr$. More specifically, we shall employ a categorification of the notion of centre of an algebra that is suitable for any multi-fusion category.

From the earlier discussions, a multi-fusion category can be viewed as a natural categorification of the notion of semi-simple algebra, where the multiplication rule is replaced by the tensor product bifunctor $\otimes : \mc C \btimes \mc C \to \mc C$. We could then naively define the centre of a multi-fusion category as the category whose objects commute with all other objects in $\mc C$ with respect to $\otimes$. However, in the spirit of categorification, we ask $X \otimes A$ and $A \otimes X$ to be isomorphic for all $A \in {\rm Ob}(\mc C)$ as opposed to equal. In this manner, objects in the centre should be provided by pairs $(X,R_{X,-})$, where $X$ is an object of $\mc{C}$ and $R_{X,-}: X \otimes - \xrightarrow{\sim} - \otimes X$ a collection of isomorphisms. Due to the weak associativity of $\otimes$, it is natural to further require that the isomorphisms $R_{X,-}$ compose weakly, relating $R_{X,A}$ and $R_{X,B}$ to $R_{X,A \otimes B}$ for every $A,B \in {\rm Ob}(\mc C)$. This yields the definition of the monoidal (or Drinfel'd) centre of a multi-fusion category \cite{Drinfeld:1989st,majid2000foundations}
\begin{definition}[Centre of a multi-fusion category] 
	Let $\mc C \equiv (\mc C, \otimes, \mathbbm 1_\mc C, \ell, r, \alpha)$ be a multi-fusion category. The centre $\ms Z(\mc C)$ of $\mc C$ is a category defined as follows: Objects in $\ms Z(\mc C)$ consist of pairs $(X,R_{X,-})$ with $X \in {\rm Ob}(\mc C)$ and $R_{X,-}: X \otimes - \xrightarrow{\sim} - \otimes X$ a collection of natural isomorphisms such that the diagram
	\begin{align}
		\label{eq:compoPent}
		\begin{tikzcd}[ampersand replacement=\&, column sep=1.8em, row sep=1.3em]
			{} \& |[alias=A]|(A \otimes X) \otimes B
			\&\&
			|[alias=D]|A \otimes ( X\otimes B)
			\& {}
			\\\\
			|[alias=AA]|(X\otimes A)\otimes B
			\&\&\&\&
			|[alias=DD]|A\otimes (B\otimes X)
			\\\\
			\&
			|[alias=BBB]|X\otimes (A\otimes B)
			\&\&
			|[alias=CCC]|(A\otimes B)\otimes X
			\arrow[from=A,to=D,"\alpha_{A,X,B}"]
			\arrow[from=AA,to=A,"R_{X,A}\otimes {\rm id}_B"]
			\arrow[from=BBB,to=AA,"\alpha^{-1}_{X,A,B}"]
			\arrow[from=BBB,to=CCC,"R_{X,A \otimes B}"']
			\arrow[from=DD,to=CCC,"\alpha^{-1}_{A,B,X}"]
			\arrow[from=D,to=DD,"{\rm id}_A \otimes R_{X,B}"]
		\end{tikzcd}
	\end{align}
	commutes for every $A,B \in {\rm Ob}(\mc C)$. The naturality of $R_{X,-}$ further implies the commutativity of the following diagram commutes
	\begin{align}
		\begin{tikzcd}[ampersand replacement=\&, column sep=1.8em, row sep=1.3em]
			|[alias=A]|X\otimes A
			\&\&
			|[alias=B]|X\otimes B
			\\
			\\
			|[alias=AA]|A\otimes X
			\&\&
			|[alias=BB]|B\otimes X
			\arrow[from=A,to=B,"{\rm id}_{X}\otimes f"]
			\arrow[from=B,to=BB,"R_{X,B}"]
			\arrow[from=A,to=AA,"R_{X,A}"']
			\arrow[from=AA,to=BB,"f\otimes {\rm id}_{X}"']
		\end{tikzcd} \, ,
	\end{align}
	for every $f \in \Hom_\mc C(A,B)$. 
	Given a pair objects $(X, R_{X,-}), (Y,R_{Y,-}) \in {\rm Ob}(\ms Z(\mc C))$, a morphism from $(X,R_{X,-})$ to $(Y,R_{Y,-})$ is a morphism $f \in \Hom_\mc C(X,Y)$ such that for every $A \in {\rm Ob}(\mc C)$ the square below commutes:
	\begin{align}
		\label{eq:squareHomCentre}
		\begin{tikzcd}[ampersand replacement=\&, column sep=1.8em, row sep=1.3em]
			|[alias=A]|X\otimes A
			\&\&
			|[alias=B]|Y\otimes A
			\\
			\\
			|[alias=AA]|A\otimes X
			\&\&
			|[alias=BB]|A\otimes Y
			\arrow[from=A,to=B,"f\otimes {\rm id}_{A}"]
			\arrow[from=B,to=BB,"R_{Y,A}"]
			\arrow[from=A,to=AA,"R_{X,A}"']
			\arrow[from=AA,to=BB,	"{\rm id}_{A}\otimes f"']
		\end{tikzcd} \, .
	\end{align}
	Composition of the morphisms in $\ms Z(\mc C)$ is induced from that in $\mc C$ and the identify morphism associated with the object $(X,R_{X,-})$ is ${\rm id}_X$.
\end{definition}
\noindent
In analogy with the centre of an algebra being closed under multiplication, the centre of a multi-fusion category is furthermore a monoidal category:
\begin{property} 
	Let $\mc C \equiv (\mc C, \otimes, \mathbbm 1_\mc C, \ell, r, \alpha)$ be a multi-fusion category. The centre $\ms Z(\mc C)$ is a monoidal category such that the unit is given by the pair $(\mathbbm 1_\mc C, r^{-1}\ell)$, the tensor product of two objects $(X,R_{X,-}),(Y,R_{Y,-})\in{\rm Ob}(\ms Z(\mc C))$ is provided by $(X,R_{X,-})\otimes (Y,R_{Y,-}) := (X \otimes Y,R_{X \otimes Y,-})$ where the isomorphism $R_{X\otimes Y,-}$ is defined via the following commutative diagram:
	\begin{align}
		\label{eq:monoCentre}
		\begin{tikzcd}[ampersand replacement=\&, column sep=1.8em, row sep=1.3em]
			{} \& |[alias=A]|X \otimes (A \otimes Y)
			\&\&
			|[alias=D]|(X \otimes A)\otimes Y
			\& {}
			\\\\
			|[alias=AA]|X \otimes (Y \otimes A)
			\&\&\&\&
			|[alias=DD]|(A \otimes X)\otimes Y
			\\\\
			\&
			|[alias=BBB]|(X \otimes Y) \otimes A
			\&\&
			|[alias=CCC]|A \otimes (X \otimes Y)
			\arrow[from=A,to=D,"\alpha^{-1}_{X,A,Y}"]
			\arrow[from=AA,to=A,"{\rm id}_X \otimes R_{Y,A}"]
			\arrow[from=BBB,to=AA,"\alpha_{X,Y,A}"]
			\arrow[from=BBB,to=CCC,"R_{X \otimes Y,A}"']
			\arrow[from=DD,to=CCC,"\alpha_{A,X,B}"]
			\arrow[from=D,to=DD," R_{X,A}\otimes {\rm id}_Y"]
		\end{tikzcd}
	\end{align}
	for every $A \in {\rm Ob}(\mc C)$. The monoidal associator is inherited from that in $\mc C$.
\end{property}
\noindent
In the same vein as the definition of the centre of a multi-fusion category, we can categorify the notion of commutative algebra. This yields the notion of braided monoidal category:
\begin{definition}[Braided monoidal category] 
	A braided monoidal category is a monoidal category $\mc C \equiv (\mc C, \otimes, \mathbbm 1_\mc C, \ell , r,\alpha)$ equipped with a natural isomorphism $R_{X,Y}: X \otimes Y \xrightarrow{\sim} Y \otimes X$ for every $X,Y \in {\mc C}$ such that coherence relations
	\begin{equation}
	\begin{split}
		\begin{tikzcd}[ampersand replacement=\&, column sep=2.2em, row sep=1.3em]
			|[alias=B]|(Y \otimes X) \otimes Z
			\&\& 	|[alias=A]|(X \otimes Y) \otimes Z \&\&
			|[alias=BB]|X \otimes (Y \otimes Z)
			\\\\
			|[alias=C]|Y \otimes (X \otimes Z)
			\&\& |[alias=D]|Y \otimes (Z \otimes X)\&\&
			|[alias=CC]|(Y \otimes Z) \otimes X
			\arrow[from=A,to=B,"R_{X,Y} \otimes {\rm id}_Z"']
			\arrow[from=A,to=BB,"\alpha_{X,Y,Z}"]
			\arrow[from=B,to=C,"\alpha_{Y,X,Z}"']
			\arrow[from=BB,to=CC,"R_{X,Y \otimes Z}"]
			\arrow[from=C,to=D,"{\rm id}_{Y} \otimes R_{X,Z}"']
			\arrow[from=CC,to=D,"\alpha_{Y,Z,X}"]
		\end{tikzcd}
		\\[0.3em]
		\begin{tikzcd}[ampersand replacement=\&, column sep=2.2em, row sep=1.3em]
			|[alias=B]|(Y \otimes Z) \otimes X
			\&\& 	|[alias=A]|Y \otimes (Z \otimes X) \&\&
			|[alias=BB]|Y \otimes (X \otimes Z)
			\\\\
			|[alias=C]|X \otimes (Y \otimes Z)
			\&\& |[alias=D]|(X \otimes Y) \otimes Z \&\&
			|[alias=CC]|(Y \otimes X) \otimes Z
			\arrow[from=A,to=B,"\alpha^{-1}_{Y,Z,X}"']
			\arrow[from=A,to=BB,"{\rm id}_Y \otimes R_{Z,X}"]
			\arrow[from=B,to=C,"R_{Y \otimes Z}\otimes X"']
			\arrow[from=BB,to=CC,"\alpha^{-1}_{Y,X,Z}"]
			\arrow[from=C,to=D,"\alpha^{-1}_{X,Y,Z}"']
			\arrow[from=CC,to=D,"R_{Y,X} \otimes {\rm id}_Z"]
		\end{tikzcd}
	\end{split}
	\end{equation}
	are satisfied for all $X,Y,Z \in {\rm Ob}(\mc C)$.
\end{definition}
\noindent
The fact that the centre of an algebra is a commutative algebra gets naturally categorified into the following property \cite{majid2000foundations,etingof2016tensor}: 
\begin{property}
	The centre $\ms Z(\mc C)$ of a multi-fusion category $\mc C \equiv (\mc C, \otimes, \mathbbm 1_\mc C, \ell , r ,\alpha)$ is a braided monoidal category with braiding isomorphism
	$R_{(X,R_{X,-}),(Y,R_{Y,-})}:= R_{X,Y}$
	for every pair of objects $(X,R_{X,-}),(Y,R_{Y,-})\in \Ob(\ms Z(\mc C))$.
\end{property}

\bigskip \noindent
Henceforth, we shall assume that the algebra $\mathbb C[\Lambda^2 G]^{\msf t(\pi)}$ is further equipped with the quasi-coassociative comultiplication map and the compatible $R$-matrix, as described explicitly in \cite{Bullivant:2019fmk} and recalled below, such that $\Mod(\mathbb C[\Lambda^2 G]^{\msf t(\pi)})$ is a braided monoidal category.
We are now ready to state the main result of this section, namely that the category the Dijkgraaf-Witten theory assigns to the two-torus $\mathbb T^2$ can be expressed as the centre of a 2-algebra:\footnote{This is a generalisation of the familiar equivalence $\Mod(\mathbb C[\Lambda G]^{\msf t(\alpha)}) \cong \ms Z(\Vect_G^\alpha)$ as braided monoidal categories \cite{majid2000foundations,gruen2018computing}, where $\mathbb C[\Lambda G]^{\msf t(\alpha)}$ is isomorphic to the so-called \emph{quantum double} of $G$ \cite{Dijkgraaf1991,Drinfeld:1989st}.}
\begin{theorem}\label{thm:Centre}
	There is a braided monoidal equivalence between the category of modules over the twisted groupoid algebra $\GrAlg$ and the centre of the twisted groupoid 2-algebra $\VectGr$. In symbols,
	\begin{equation}
		\label{eq:congCat}
		\Mod(\mathbb C[\Lambda^2 G]^{\msf t^2(\pi)}) \cong \ms Z(\Vect_{\Lambda G}^{\msf t(\pi)}) \, .
	\end{equation}
\end{theorem}
\noindent
In order to prove this statement, we shall proceed incrementally by first proving the equivalence of the categories, and then extend it to a braided monoidal equivalence:
\begin{lemma}
	There is an equivalence between the categories $\Mod(\mathbb C[\Lambda^2 G]^{\msf t^2(\pi)})$ and $\ms Z(\Vect_{\Lambda G}^{\msf t(\pi)})$.
\end{lemma} 
\noindent
\begin{proof}
	The semi-simplicity of $\VectGr$ implies that every object $V = \bigoplus_{\fr g \in \Hom(\Lambda G)}V_\fr g$ in $\VectGr$ decomposes as a direct sum of finitely many simple objects of the form $\{\mathbb C_\fr g\}_{\forall \, \fr g \in \Hom(\Lambda G)}$. Given this observation, let us construct explicitly the objects in the centre. By definition, we know that an object in $\ms Z(\VectGr)$ is provided by a pair $(V,R_{V,-})$ with $V$ a $\Hom(\Lambda G)$-graded vector space and $R_{V,-}$ a collection of isomorphisms defined by
	\begin{equation}
		R_{V,\mathbb C_\fr a}: V \otimes \mathbb C_\fr a \xrightarrow{\sim} \mathbb C_{\fr a} \otimes V \, , \q \forall \, \fr a \in \Hom(\Lambda G) \, ,
	\end{equation}
	which satisfy the coherence relation \eqref{eq:compoPent}. 
	Provided such a family of isomorphisms and a morphism $\fr g \in \Hom(\Lambda G)$, we have
	\begin{equation}
		(R_{V,\mathbb C_\fr a})_{\fr g \fr a}:
		\begin{cases}
			V_{\fr g} \xrightarrow{\sim} V_{\fr a^{-1}\fr g \fr a} \q &\text{if} \;\; {\rm s}(\fr g)={\rm t}(\fr g) = {\rm s}(\fr a)
			\\
			 V_ \fr g \xrightarrow{\sim} \varnothing &\text{otherwise}
		\end{cases} \, ,
	\end{equation}
	for every $\fr a \in \Hom(\Lambda G)$, so that the existence of a non-trivial isomorphism $R_{V,\mathbb C_{\fr a}}$ implicitly constrains $V$ to be an ${\rm End}(\Lambda G)$-graded vector space. Noticing that $\Ob(\Lambda^2 G) = {\rm End}(\Lambda G)$, it follows that such a vector space decomposes as
	\begin{equation}
		V = \bigoplus_{[\fr g'] \in \uppi_0(\Lambda^2 G)}V_{[\fr g']} 
		\q {\rm with} \q
		V_{[\fr g']}:= \bigoplus_{\fr g \in [\fr g']}V_\fr g \, ,
	\end{equation}
	such that all the vector spaces appearing in the decomposition of $V_{[\fr g']}$ are equal when neglecting the grading.
	Given the above, we find that the isomorphisms $R_{V,-}$ induce a family of endomorphisms
	\begin{equation}
		\begin{array}{ccccl}
			\rho_{\mathbb C_\fr a} & : & V_{[\fr g']} & \to & V_{[\fr g']}
			\\
			& : & V_{\fr g \in [\fr g']} &\mapsto & 	
			\delta_{{\rm s}(\fr a), {\rm t}(\fr g)}\, [V_\fr g \triangleleft \rho( v_\fr g,\fr a)] \in V_{\fr a^{-1}\fr g \fr a}
		\end{array} \; , 
	\end{equation}
	where $\rho(v_\fr g, \fr a): V_\fr g \xrightarrow{\sim} V_{\fr a^{-1}\fr g \fr a}$ for non-zero basis vectors $(\fr a, v_\fr g) \in \mathbb C_\fr a \times V_{\fr g \in [\fr g']}$ with ${\rm s}(\fr a)={\rm t}(\fr g)$. This family of endomorphisms is such that
	\begin{equation}
		\label{eq:braidingBasisElts}
		R_{v_\fr g, \fr a} : v_\fr g \otimes \fr a \mapsto \fr a \otimes [v_\fr g \triangleleft \rho(v_\fr g,\fr a)] \; ,
	\end{equation}
	and more generally\footnote{Recall that we use the convention $f(g(-)) \equiv g \circ f$ for the composition of morphisms.}
	\begin{equation}
		\label{eq:braidingV}
		R_{V_{[\fr g']},\mathbb C_\fr a}= {\rm swap} \circ \Big({\rm id}\otimes \rho \big(\sum_{\fr g \in [\fr g']}v_\fr g, \fr a \big) \Big) \, ,
	\end{equation}
	where `swap' is the transposition map that permutes the order of vector spaces in the tensor product.
	Classifying the objects of $\ms Z(\VectGr)$ thus amounts to classifying pairs
	\begin{equation}
		(V_{[\fr g']\in \uppi_0(\Lambda^2 G)} \, , \{ \rho_{\mathbb C_\fr a}\}_{\forall \fr a \in \Hom(\Lambda G)}) \, .	
	\end{equation} 
	Utilising \eqref{eq:compoPent}, we find that 
	\begin{align*}
		R_{V_{[\fr g']},\mathbb C_{\fr a\fr a'}}
		=
		\msf t(\pi)^{-1}_{V_{[\fr g']},\mathbb C_\fr a,\mathbb C_{\fr a'}} \circ
		(R_{V_{[\fr g']},\mathbb C_\fr a}\otimes{\rm id}_{\mathbb{C}_\fr a}) \circ
		\msf t(\pi)_{\mathbb C_\fr a, V_{[\fr g']}, \mathbb C_{\fr a'}} \circ
		({\rm id}_{\mathbb C_\fr a} \otimes R_{V_{[\fr g']},\mathbb C_\fr a'}) \circ
		\msf t(\pi)^{-1}_{\mathbb C_\fr a, \mathbb C_{\fr a'},V_{[\fr g']}}
		\, , 
	\end{align*}
	where $\msf t(\pi)$ here refers to the associator isomorphism in $\VectGr$.
	In virtue of \eqref{eq:braidingBasisElts}, this implies that $\rho(-,-)$ satisfies the algebra
	\begin{equation}
		\label{eq:algebraModule}
		\rho(v_\fr g, \fr a) \triangleleft
		\rho(v_{\fr g'}, \fr a') = \delta_{{\rm t}(\fr a), {\rm s}(\fr a')}
		\, \msf t^2(\pi)(\fr g \xrightarrow{\fr a}, \fr a^{-1 }\fr g \fr a\xrightarrow{\fr a'}) \, \rho(v_\fr g, \fr a \fr a') 
		\, ,
	\end{equation}
	for all $\fr g \in \Ob(\Lambda ^2 G)$ and $\fr a \in \Hom_{\Lambda G}({\rm s}(\fr g),-)$.
	We recognize this algebra as the twisted groupoid algebra of the two-fold loop groupoid of $\overline G$.
	It follows that objects $(V,R_{V,-})$ can be conveniently described via weak functors
	\begin{equation}
		\label{eq:functorCentre}
		\begin{array}{ccccl}
			F_{\rho,V} & : & \Lambda^2 G & \to & \Vect
			\\
			& : & \fr g \in \Ob(\Lambda^2 G) &\mapsto & 	V_\fr g \subset V
			\\
			& : & \fr g \xrightarrow{\fr a} \, \in \Hom(\Lambda^2 G) & \mapsto & \rho(v_\fr g, \fr a): V_\fr g \to V_{\fr a^{-1}\fr g \fr a}
		\end{array} \; , 
	\end{equation}
	such that every isomorphism $\rho(v_\fr g,\fr a)$ satisfies the weak composition rule \eqref{eq:algebraModule}. Exploiting the well-known equivalence between representations and modules, we can interpret $F_{\rho,V}$, or the pair $(V,\rho)$, as a module over $\mathbb C[\Lambda^2\overline G]^{\msf t^2(\pi)}$. Furthermore, morphisms in $\ms Z(\VectGr)$ can be defined by natural transformations $F_{\rho,V} \to F_{\rho',V'}$, or equivalently, as intertwiners between representations of the twisted groupoid algebra. Putting everything together, this establishes the equivalence $\Mod(\mathbb C[\Lambda^2 G]^{\msf t^2(\pi)}) \cong \ms Z(\VectGr)$.
\end{proof}
\medskip \noindent

\begin{lemma}
	The equivalence $\Mod(\mathbb C[\Lambda^2 G]^{\msf t^2(\pi)}) \cong \ms Z(\VectGr)$ can be extended to a monoidal equivalence.
\end{lemma}
\noindent
\begin{proof}
	In light of the equivalence at the level of the categories, showing the monoidal equivalence amounts to proving that the monoidal product in the centre corresponds to the tensor product over $\mathbb C$ of representations of $\mathbb C[\Lambda^2 G]^{\msf t^2(\pi)}$ encoded by a comultiplication map that is quasi-coassociative with respect to a quasi-invertible algebra element characterized by $\msf t(\pi)^{-1}$.
	Given a pair of objects $(V,R_{V,-}),(W,R_{W,-})\in {\rm Ob}(\ms{Z}(\VectGr)$, the monoidal structure is provided by $(V,R_{V,-})\otimes (W,R_{W,-})=(V\otimes W,R_{V\otimes W,-})$ together with \eqref{eq:monoCentre} such that
	\begin{align}
		R_{V\otimes W,A}
		=\msf t(\pi)_{V,W,A} \circ 
		({\rm id}_{V}\otimes R_{W,A}) 
		\circ \msf t(\pi)^{-1}_{V,A,W}
		\circ
		(R_{V,A}\otimes {\rm id}_{W})
		\circ \msf t(\pi)_{A,V,W} \,
	\end{align}
	for every $A \in \Ob(\VectGr)$. The monoidal associator is the one of $\VectGr$. Provided non-zero basis vectors $(\fr a, v_{\fr g_1}, w_{\fr g_2}) \in \mathbb C_\fr a \times V_{\fr g_1 \in [\fr g_1']} \otimes W_{\fr g_2 \in [\fr g_2']}$, we denote by $(v \otimes w)_{\fr g \in [\fr g']}$ the basis vector of $(V \otimes W)_\fr g$ defined as
	\begin{equation}
		(v \otimes w)_\fr g = \sum_{\substack{\fr g_1, \fr g_2 \in \End(\Lambda G) \\ \fr g_1 \fr g_2 = \fr g}}v_{\fr g_1} \otimes w_{\fr g_2} \, .
	\end{equation}
	Considering the pair of modules $(V,\rho)$ and $(W,\sigma)$ associated with $(V,R_{V,-})$ and $(W,R_{W,-})$, respectively, we find that
	\begin{equation}
		R_{(v \otimes w)_{\fr g}, \fr a} : 
		(v \otimes w)_{\fr g} \otimes \fr a 
		\mapsto 
		\sum_{\substack{\fr g_1, \fr g_2 \in \End(\Lambda G) \\ \fr g_1 \fr g_2 = \fr g}}\!
		\gamma_\fr a (\fr g_1, \fr g_2) \;
		\fr a \otimes ([v_{\fr g_1} \triangleleft\rho(v_{\fr g_1},\fr a)] \otimes [w_{\fr g_2} \triangleleft \sigma(w_{\fr g_2},\fr a)]) \; ,
	\end{equation}
	where
	\begin{align}
		\gamma_\fr a(\fr g_1, \fr g_2):=
		\frac{
			\msf t(\pi)(\fr g_1, \fr g_2, \fr a) \,
			\msf t(\pi)(\fr a , \fr a^{-1}\fr g_1 \fr a, \fr a ^{-1}\fr g_2 \fr a)
		}{
			\msf t(\pi)(\fr g_1, \fr a, \fr a^{-1}\fr g_2 \fr a)
		}\,.
	\end{align}
 	Following the line of argument of the previous proof, we find that the pair $(V \otimes W, R_{V \otimes W,-})$ can be equivalently described by the module $(V \otimes W, (\rho \otimes \sigma)\circ \Delta)$ in terms of the weak functor
	\begin{equation}
			\begin{array}{ccccl}
			F_{(\rho \otimes \sigma) \circ \Delta,V \otimes W} & : & \Lambda^2 G & \to & \Vect
			\\
			& : & \fr g \in \Ob(\Lambda^2 G) &\mapsto & (V \otimes W)_\fr g \subset V \otimes W
			\\
			& : & \fr g \xrightarrow{\fr a} \, \in \Hom(\Lambda^2 G) & \mapsto & (\rho \otimes \sigma) \circ \Delta(\fr g \xrightarrow{\fr a})
			\end{array} \; , 
	\end{equation}
	such that the map $\Delta$ is identified with the \emph{comultiplication} map of the twisted groupoid algebra \cite{Bullivant:2019fmk}
	\begin{align}
	\label{eq:comultiplication}
	\begin{array}{ccccl}
		\Delta & : & \GrAlg & \to & \GrAlg \otimes \GrAlg
		\\
		& : & | \fr g \xrightarrow{\fr a} \ra &\mapsto & 		\sum_{\substack{\fr g_1, \fr g_2 \in \End(\Lambda G) \\ \fr g_1 \fr g_2 = \fr g}}
		\gamma_\fr a (\fr g_1, \fr g_2)\,
		|\fr g_1 \xrightarrow{\fr a} \ra
		\otimes 
		|\fr g_2 \xrightarrow{\fr a}\ra 
	\end{array} \, .
	\end{align}
	It follows from the cocycle relation
	\begin{equation}
		\frac{\gamma_\fr a(\fr g_2, \fr g_3) \, \gamma_\fr a(\fr g_1, \fr g_2 \fr g_3)}{\gamma_\fr a (\fr g_1 \fr g_2, \fr g_3) \, \gamma_\fr a(\fr g_1, \fr g_2)}
		=
		\frac{\msf t(\pi)(\fr g_1, \fr g_2, \fr g_3)}{\msf t(\pi)(\fr a^{-1}\fr g_1\fr a, \fr a^{-1}\fr g_2 \fr a, \fr a^{-1}\fr g_3 \fr a)}
	\end{equation}
	that this comultiplication map, which can be verified to be an algebra homomorphism, satisfies the following \emph{quasi-coassociativity} condition
	\begin{align}
		(\Delta\otimes {\rm id})\circ \Delta(|\fr a \xrightarrow{\fr a} \ra)
		=
		\Phi \star [({\rm id} \otimes \Delta)\circ \Delta(| \fr g \xrightarrow{\fr a} \ra)] \star \Phi^{-1} \, ,
	\end{align}
	with $\Phi$ the invertible element of $\GrAlg \otimes \GrAlg \otimes \GrAlg$ given by
	\begin{align}
		\Phi:=\sum_{\fr g_1, \fr g_2, \fr g_3 \in \Ob(\Lambda^2 G)}
		\msf t(\pi)^{-1}(\fr g_1, \fr g_2, \fr g_3) \,
		|\fr g_1 \xrightarrow{{\rm id}_{{\rm s}(\fr g_1)}} \ra
		\otimes 
		|\fr g_2 \xrightarrow{{\rm id}_{{\rm s}(\fr g_2)}} \ra
		\otimes 
		|\fr g_3 \xrightarrow{{\rm id}_{{\rm s}(\fr g_3)}} \ra
		\; ,
	\end{align}
	where the loop groupoid cocycle $\msf t(\pi)$ characterises the monoidal associator in the centre. This quasi-coassociativity condition ensures that the $\GrAlg$-modules defined according to
	\begin{align}
		(\rho \otimes \sigma \otimes \varrho) \circ (\Delta \otimes {\rm id})\Delta 
		\q {\rm and} \q
		(\rho \otimes \sigma \otimes \varrho) \circ ({\rm id} \otimes \Delta)\Delta 
	\end{align}
	are isomorphic, from which follows the associativity condition in $\Mod(\GrAlg)$.
\end{proof}

\medskip \noindent
\begin{lemma}
	The monoidal equivalence $\Mod(\mathbb C[\Lambda^2 G]^{\msf t^2(\pi)}) \cong \ms Z(\VectGr)$ can be extended to a braided monoidal equivalence.
\end{lemma}

\noindent
\begin{proof}
	Let us consider a pair of objects $(V,R_{V,-}),(W,R_{W,-})\in \ms Z(\VectGr)$ described by the modules $(V,\rho)$ and $(W,\sigma)$, respectively. On the one hand, the braiding isomorphism with respect to the monoidal structure of the centre is given by
	\begin{align}
		R_{(V,R_{V,-}),(W,R_{W,-})} = R_{V,W}:(V,R_{V,-})\otimes (W,R_{W,-})\xrightarrow{\sim} (W,R_{W,-})\otimes (V,R_{V,-}) \, .
	\end{align}
	On the other hand, the braiding isomorphism on $\Mod(\GrAlg)$ is given by $ (\rho \otimes \sigma)(\msf R) \circ {\rm swap}$, where $\msf R$ is an invertible algebra element defined as \cite{Bullivant:2019fmk}
	\begin{align}
		\label{eq:Rmatrix}
		\msf R:=
		\sum_{\substack{
				\fr g\in {\rm Ob}(\Lambda^{2} G)
				\\
				\fr g' \in {\rm End}_{\Lambda G}({\rm s}(\fr g))
		}} \!\!\!\!
		| \fr g'\xrightarrow{\fr g} \ra
			\otimes
		|\fr g \xrightarrow{{\rm id}_{{\rm s}(\fr g)}} \ra
		\; \in 
		\GrAlg \otimes \GrAlg \, ,
	\end{align}
	such that it satisfies
	\begin{equation}
		\msf R \star \Delta(| \fr g \xrightarrow{\fr a} \ra)\star \msf R^{-1} =  \Delta(|\fr g \xrightarrow{\fr a}\ra) \circ {\rm swap}
		\, , \q \forall \, |\fr g \xrightarrow{\fr a} \ra \in \GrAlg \, .
	\end{equation}
	This algebra element is referred to as the $R$-matrix in the context of the study of quasi-triangular quasi-Hopf algebra. Equivalence between the two descriptions is then ensured by the following derivation:
	\begin{align}
		\nn
		R_{V,W} &= 
		\sum_{\fr g \in \Ob(\Lambda^2 G)}\big({\rm id} \otimes \sigma(w_\fr g,{\rm id}_{\rm s(\fr g)}) \big)\circ R_{V,W_\fr g}
		\\
		\nn
		&=
		{\rm swap} \circ \!\!\!\!\! \sum_{\substack{\fr g \in \Ob(\Lambda^2 G)}} \!\!\!\!\! \sigma(w_\fr g,{\rm id}_{\rm s(\fr g)}) \otimes \rho \Big(\sum_{\fr g' \in {\rm End}_{\Lambda G}(\rm s(\fr g))} \!\!\!\!\!\! v_{\fr g'},\fr g \Big) 
		\\
		&= \sum_{\substack{\fr g \in \Ob(\Lambda^2 G) \\ \fr g' \in {\rm End}_{\Lambda G}(\rm s(\fr g))}} \!\!\! \Big(\rho(v_{\fr g'},\fr g) \otimes \sigma(w_\fr g,{\rm id}_{\rm s(\fr g)}) \Big) \circ {\rm swap} = (\rho \otimes \sigma)(\msf R) \circ {\rm swap} \, ,
	\end{align}
where we used the fact that $\sigma(w_\fr g, {\rm id}_{\rm s(\fr g)})$ acts as a projector onto $W_\fr g \subset W$, as well as \eqref{eq:braidingV}.	 
\end{proof}	

\bigskip
\noindent
Putting the previous three lemmas together yields th.~\ref{thm:Centre}. It follows in particular that the simple objects of the centre $\ms Z(\VectGr)$ can be conveniently obtained via the irreducible representations of the twisted groupoid algebra, which were computed explicitly in \cite{Bullivant:2019fmk}.

\subsection{Centre of a 3-algebra}
We mentioned in the previous section that the Dijkgraaf-Witten theory assigns to the circle $\mathbb S^1$ the braided monoidal bicategory $\MOD(\VectGr)$. In close analogy with what the theory assigns to $\mathbb T^2$ and $\mathbb T^3$, this bicategory can be equivalently presented as the categorified centre of a 3-algebra, namely $\TVectGr$. The same way the centre defined above is a categorification of the notion of centre of an algebra suitable to multi-fusion categories, the categorified centre required here is a natural categorification of the notion of centre of a multi-fusion category suitable to monoidal bicategories. Since showing this equivalence of bicategories was the purpose of the manuscript \cite{Kong:2019brm} by Kong et al., we shall merely state the result in this section.

First we need to enrich our notion of bicategory with a monoidal structure whose consistency conditions are weakened in an appropriate way according to the ethos of categorification \cite{kapranov19942, baez1996higher, gurski2011loop}:
\begin{definition}[Monoidal bicategory]
	A monoidal bicategory is defined as a decuple $\mc B \equiv (\mc B, \otimes , \mathbbm 1, \alpha, r, \ell, \pi,\tau_1,\tau_2,\tau_3)$ that consists of a bicategory $\mc B$ together with a monoidal structure $(\otimes, \mathbbm 1, \alpha, \ell, r)$ such that the coherence diagrams of the pseudo-natural equivalences $\alpha: (- \otimes -) \otimes  - \to - \otimes (- \otimes -)$, $\ell: \mathbbm 1 \otimes - \to -$ and $r: - \otimes \mathbbm 1 \to -$ commute up to invertible modifications $\pi$, $\tau_1$, $\tau_2$ and $\tau_2$, which fulfil various coherence relations.
\end{definition}

\noindent
An example of monoidal bicategory is provided by twisted group 3-algebras. These are obtained as a categorification of the concept of 2-algebras by promoting the category $\Vect$ to $\TVect$:

\begin{example}[Bicategory of group-graded 2-vector spaces]
	Let $G$ be a finite group and $\pi$ a normalised group 4-cocycle in $H^4(G,\rU(1))$. The twisted group 3-algebra $\TVectGr$ is a monoidal bicategory whose objects are $G$-graded 2-vector spaces of the form $V = \bigbplus_{g \in G}V_g$, 1-morphisms, grading preserving $\Vect$-module functors, and 2-morphisms, $\Vect$-module natural transformations. There are $|G|$-many simple objects notated via $\Vect_g$, $\forall \, g \in G$. The monoidal structure  is defined on homogeneous components via $\btimes: \Vect_g \times \Vect_{h}\to \Vect_{gh}$ for all $g,h \in G$ together with the pseudo-natural equivalences
	\begin{equation}
	\begin{split}
		(\Vect_g \btimes \Vect_h) \btimes \Vect_k \xrightarrow{\alpha_{\Vect_g,\Vect_h,\Vect_k}} \Vect_g \btimes (\Vect_h \btimes \Vect_k) \; ,
		\\
		\Vect_g \btimes \Vect_{\mathbbm 1_G} \xrightarrow{r_{\Vect_g}}\Vect_g \q {\rm and} \q
		\Vect_{\mathbbm 1_G} \btimes \Vect_g \xrightarrow{\ell_{\Vect_g}} \Vect_g \, ,
	\end{split}
	\end{equation}
	which are identity 1-morphisms. The invertible modifications $\tau_1$, $\tau_2$ and $\tau_3$ are trivial, whereas the `pentagonator' $\pi$ is defined by
	\begin{align}
		\pi_{\Vect_g, \Vect_h, \Vect_k, \Vect_l} = \;&\pi(g,h,k,l) \cdot {\rm id}_{\Vect_{ghkl}}
		\\ \nn
		: \;
		&(\alpha_{\Vect_g, \Vect_h, \Vect_k} \btimes {\rm id}_{\Vect_l})
		\circ
		\alpha_{\Vect_g, \Vect_h \btimesFt \Vect_k, \Vect_l} 
		\circ
		({\rm id}_{\Vect_g} \btimesFt \alpha_{\Vect_h, \Vect_k, \Vect_l})
		\\ \nn
		&\Rightarrow		
		\alpha_{\Vect_{gh}, \Vect_k, \Vect_l} \circ \alpha_{\Vect_g, \Vect_h, \Vect_{kl}} \, ,
	\end{align}
	for all $g,h,k,l \in G$. 
\end{example}
\noindent
Monoidal bicategories can be further equipped with a braiding structure, whose definition we omit here (see e.g. \cite{gurski2011loop}). A categorification of the centre of a multi-fusion category suitable to monoidal bicategories goes as follows:

\begin{definition}[Centre of a monoidal bicategory]
		Let $\mc B \equiv (\mc B, \otimes, \mathbbm 1,\alpha, \ell, r, \pi, \tau_1, \tau_2, \tau_3)$ be a monoidal bicategory. The centre $\ms Z(\mc B)$ of $\mc B$ is a bicategory such that:\\[-1.8em]
 	\begin{enumerate}[itemsep=0.3em,parsep=1pt,leftmargin=3em]
		\item[${\ssss \bullet}$] Objects are triple $(X,R_{X,-},R_{X|-,-})$ that consists of an object $X \in \mc B$, a pseudo-natural equivalence $R_{X,-}: X \otimes - \to - \otimes X$ and an invertible modification $R_{X|-,-}$ weakening the `hexagon' coherence relation \eqref{eq:compoPent}, which satisfy a `permuto-associahedron' axiom involving the pentagonator $\pi$.
		\item[${\ssss \bullet}$] 1-morphisms between two objects $(X,R_{X,-},R_{X|-,-})$ and $(Y,R_{Y,-},R_{Y|-,-})$ are tuples $(f,R_{f,-})$ that consists of a morphism $f:X \to Y$ and an invertible modification $R_{f,-}$ weakening the coherence relation \eqref{eq:squareHomCentre}, which satisfy a `prism' axiom involving $R_{X|-,-}$ and $R_{Y|-,-}$. Composition of morphisms in $\ms Z(\mc B)$ is of the form $(f,R_{f,-})\circ (g,R_{g,-}) = (f \circ g, ({\rm id}_{f \otimes -} \circ R_{g,-})\cdot(R_{f,-}\circ {\rm id}_{- \otimes g}))$, where $f$ and $g$ are composable morphisms in $\mc B$.
		\item[${\ssss \bullet}$] 2-morphisms between 1-morphisms $(f,R_{f,-})$ and $(g,R_{g,-})$ are 2-morphisms $f \Rightarrow g$ in $\mc B$ subject to a `prism' axiom involving $R_{f,-}$ and $R_{g,-}$.
	\end{enumerate}
\end{definition}

\noindent
Similarly to the centre of a multi-fusion category, the centre of a monoidal bicategory can be verified to be a braided monoidal bicategory. In \cite{Kong:2019brm}, they computed explicitly the centre of the 3-algebra $\TVectGr$ and showed that it satisfies an equivalence of bicategories, which in our terminology reads
\begin{equation}
	\label{eq:MODCentre}
	\MOD(\VectGr) \cong \ms Z(\TVectGr) \, .
\end{equation}
This equivalence can be further lifted to a braided monoidal equivalence of bicategories by noting there exists a quasi-triangular quasi-Hopf category structure on $\TVectGr$ making $\MOD(\VectGr)$ braided monoidal \cite{neuchl1997representation}. Although we do not require \eqref{eq:MODCentre} for our exposition per se, it brings the content of this section together and sheds light on universal features of the theory as a fully extended TQFT.

\bigskip
\section{Dimension and crossing with the circle\label{sec:crossing}}

\noindent
\emph{In this section, we establish the `crossing with the circle' conditions for the manifolds $\mathbb T^3$, $\mathbb T^2$ and $\mathbb S^1$ obtained by computing the dimension, and categorifications thereof, of the data the theory assigns to these manifolds.}

\subsection{Dimension of the vector space $\mc Z^\pi_G(\mathbb T^3)$}

We explained in the introduction that if a fully extended topological quantum field theory is fully characterised by the data it assigns to the point, we must be able to recover from this data what the theory assigns to higher-dimensional manifolds. In particular, the `crossing with the circle' condition is the statement that the dimension of the quantum invariant assigned to a ($d$$-$$n$)-manifold is equivalent to that assigned to the Cartesian product of this manifold with the circle, where `dimension' here refers to a suitable categorification of the notion of dimension of a vector space.  In symbols, we expect
\begin{equation*}
	\msf{Dim} \,  \mc Z(\Sigma^{d-n}) = \mc Z(\Sigma^{d-n} \times \mathbb S^1) \; .
\end{equation*}
The first---and simplest---instance of this equation states that the complex number assigned to a manifold of the form $\Sigma^{3} \times \mathbb S^1$ equals the dimension of the vector space assigned to $\Sigma^{3}$. Invoking general arguments, we have already established in the introduction that it is indeed true, but we shall now confirm it in the case of the three-torus by computing explicitly the dimension of the vector space $\mc V_{\Lambda^3  G}(\msf t^3(\pi))$.

Let us consider a finite groupoid $\mc G$ and a groupoid 1-cocycle $\epsilon$ in $H^1(\mc G, \rU(1))$. On the one hand, we have the following relation \cite{willerton2008twisted}:
\begin{equation*}
	\int_{\Lambda \mc G}\msf t(\epsilon)
	\stackrel{\eqref{eq:DimGr}}{=} 
	\sum_{[\fr g]\in \uppi_0(\Lambda \mc G)}\frac{\msf t(\epsilon)(\fr g)}{|\mathtt{Aut}(\fr g)|} 
	= \sum_{\substack{[X]\in \uppi_0(\mc G) \\ \fr g \in {\rm End}_{\mc G}(X)}}\frac{\epsilon(\fr g)}{|\mathtt{Aut}(X)|}
	=
	\sum_{[X] \in \uppi_0(\mc G)}
	\begin{cases}
		1 \q {\rm if} \; \; \epsilon{\sss |}_{\mathtt{Aut}(X)} \equiv 1 
		\\ 0 \q {\rm otherwise}
	\end{cases}
\end{equation*}
where we used the fact that, since $\epsilon$ is a groupoid 1-cocycle,  $\epsilon{\sss |}_{\mathtt{Aut}(X)}$ defines a one-dimensional representation of $\mathtt{Aut}(X)$ for every $[X] \in \uppi_0(\mc G)$, which satisfies the usual \emph{orthogonality} condition.
On the other hand, since any groupoid $\mc G$ can be decomposed over its connected components as 
\begin{equation}
	\mc G \cong \bigsqcup_{[X] \in \uppi_0(\mc G)}\overline{\mathtt{Aut}(X)} \, ,
\end{equation}
we find that
\begin{equation}
	{\rm Dim}_\mathbb C \mc V_\mc G(\epsilon) = \sum_{[X]\in \uppi_0(\mc G)} {\rm Dim}_\mathbb C
	\mc V_{\, \overline{\!\mathtt{Aut}(X)\!}\,}(\epsilon {\sss |}_{\mathtt{Aut}(X)})	
	= \sum_{[X] \in \uppi_0(\mc G)}	
	\begin{cases}
		1 \q {\rm if} \; \; \epsilon{\sss |}_{\mathtt{Aut}(X)} \equiv 1 
		\\ 0 \q {\rm otherwise} 
	\end{cases} \, ,
\end{equation}
hence the equality $\int_{\Lambda \mc G} {\msf t}(\epsilon) = {\rm Dim}_\mathbb C \mc V_\mc G(\epsilon)$.
Applying this formula to $\mc G = \Lambda^2  G$ finally yields 
\begin{equation}
	{\rm Dim}_\mathbb C \, \mc Z^\pi_G(\mathbb T^3) 
	\stackrel{\eqref{eq:ZTthree}}{=} 
	{\rm Dim}_\mathbb C \, \mc V_{\Lambda^3  G}(\msf t^3(\pi)) 
	=
	\int_{\Lambda^4 \overline G}{\msf t}^4(\pi) 
	\stackrel{\eqref{eq:ZTfour}}{=} \mc Z^\pi_G(\mathbb T^4) \, ,
\end{equation}
as expected. Exploiting the isomorphism \eqref{eq:isoVect}, we know that a complete and orthogonal basis for $\mc V_{\Lambda^3 G}(\msf t^3(\pi))$ is labelled by the characters of the twisted groupoid algebra $\mathbb C[\Lambda^2  G]^{\msf t^2(\pi)}$. It follows that the number $\mc Z^\pi_G(\mathbb T^4)$ that the theory assigns to $\mathbb T^4$ equals the number of irreducible representations of the algebra. Physically, this is the statement that the ground state degeneracy of  the lattice Hamiltonian realisation of the theory on $\mathbb T^3$ equals the number of irreducible loop-like excitation types in the model.

Note finally that computing the dimension of a vector space is a trivial example of a \emph{decategorification} process, upon which the vector space thought as a 0-category is reduced to a number thought as a (-1)-category. In the following, we consider higher-categorical analogues of this process.

\subsection{Dimension of the category $\mc Z^\pi_G(\mathbb T^2)$}

Let us now consider the crossing with the circle condition for the two-torus $\mathbb T^2$. As previously, we want to recover what the theory assigns to $\mathbb T^3$ as the dimension of the category $\mc Z^\pi_G(\mathbb T^2)$, where by dimension we mean a categorification of the usual notion that is suitable to categories. Our approach mimics Bartlett's who performs in \cite{bartlett2009unitary} analogous computations for the fully extended (2+1)-dimensional theory. More specifically, we consider a categorification of the well-known statement that the dimension of a vector space equals the trace of the identity linear map:

\begin{definition}[Dimension of a category]
	Let $\mc{C}$ be a category and ${\rm id}_{\mc{C}}:\mc{C}\rightarrow \mc{C}$ the identity functor on $\mc C$. We define the dimension $\msf{Dim}(\mc{C})$ of $\mc C$ as the commutative monoid $\msf{Nat}({\rm id}_\mc C, {\rm id}_\mc C)$ of natural transformations $\eta: {\rm id}_\mc C \Rightarrow {\rm id}_\mc C$. If $\mc{C}$ is a $\Bbbk$-linear category, then the dimension is defined as a commutative $\Bbbk$-algebra.
\end{definition}
\noindent
Expanding this definition, a natural transformation $\eta:{\rm id}_\mc C \Rightarrow {\rm id}_\mc C$ of the identity functor is an assignment of a morphism $\eta_{X}:X\to X \in \Hom(\mc C)$ to each object $X\in{\rm Ob}(\mc{C})$ subject to the condition that the diagram
\begin{align}
	\label{eq:monoidNat}
	\begin{tikzcd}[ampersand replacement=\&, column sep=1.8em, row sep=1.3em]
		|[alias=A]|X
		\&\&
		|[alias=B]|Y
		\\
		\\
		|[alias=AA]|X
		\&\&
		|[alias=BB]|Y
		\arrow[from=A,to=B,"f"]
		\arrow[from=B,to=BB,"\eta_{Y}"]
		\arrow[from=A,to=AA,"\eta_{X}"']
		\arrow[from=AA,to=BB,"f"']
	\end{tikzcd}
\end{align}
commutes for all morphisms $f:X \to Y\in \Hom(\mc C)$.
Given a pair of natural transformations $\eta,\mu:{\rm id}_\mc C \Rightarrow {\rm id}_\mc C$, these can be composed so as to yield another natural transformation $\eta \circ \mu: {\rm id}_\mc C \Rightarrow {\rm id}_\mc C$, which assigns to every $X \in \Ob(\mc C)$ the morphism $(\eta \circ \mu)_X := \eta_X \circ \mu_X$. Applying the relation \eqref{eq:monoidNat} to the morphisms $f \equiv \mu_X : X \to X$ for every $X \in \Ob(\mc C)$ yields $\eta_X \circ \mu_X = \mu_X \circ \eta_X$, hence the commutativity $\eta \circ \mu = \mu \circ \eta$. Given this definition, establishing the crossing with the circle condition for $\mathbb T^2$ amounts to proving the following theorem:

\begin{theorem}\label{thm:DimRep}
	The dimension of the category $\Mod(\GrAlg)$ is isomorphic, as a commutative $\mathbb C$-algebra, to the centre $Z(\GrAlg)$ of the twisted groupoid algebra $\GrAlg$, i.e.
	\begin{equation}
		\label{eq:DimRep}
		\msf{Dim} \, \Mod(\GrAlg)
		\simeq
		Z(\GrAlg) \, .
	\end{equation}
\end{theorem}

\noindent
We shall prove this theorem by considering a series of lemmas. Letting $\Mod(\GrAlg){\sss |}_{\rm reg.}$ denote the full subcategory of $\Mod(\GrAlg)$ consisting of a single object, which is the \emph{right} regular representation of the algebra, and endomorphic intertwiners of this unique representation, we have the following property:
\begin{lemma}\label{lem:regRep}
	A natural transformation of the identity functor on $\Mod(\GrAlg)$ is completely determined by a natural transformation of the identify functor on $\Mod(\GrAlg){\sss |}_{\rm reg.}$.
\end{lemma}

\noindent
\begin{proof}
	Let us begin with some observations. Recall that the right regular representation of an algebra $A$ is the representation whose vector space is given by the underlying vector space of $A$ with action given by multiplication in $A$. As such, the single object in $\Mod(\GrAlg){\sss |}_{\rm reg.}$ is $\GrAlg$ itself. Given a representation $(V,\rho)\in {\rm Ob}(\Mod(\GrAlg ))$, an intertwiner $f: \GrAlg \to V$ between the regular representation and $V$ satisfies by definition
	\begin{align}
		(- \star | \fr g \ra) \circ f = f \circ \rho(| \fr g\ra) \;, \q \forall \, | \fr g \ra \in \GrAlg \, . 
	\end{align}
	Applying this defining formula to the identity algebra element $|\mathbbm 1 \ra$ in $\GrAlg$ yields $f(|\fr g\ra) = \rho(|\fr g \ra)(v)$, for  every $| \fr g \ra \in \GrAlg$, where $v := f (| \mathbbm 1\ra)$. Conversely, given $v \in V$, we can define an intertwiner $f_v$ via $f_v(| \fr g \ra) := \rho(| \fr g \ra)(v)$. It follows that a choice of intertwiner $f: \GrAlg \to V$ uniquely specifies a vector $v \in V$. 
	As a corollary, we obtain that an endomorphic intertwiner of the regular representation is defined by an element $| \fr g \ra\in \GrAlg$, which we write $f_{| \fr g\ra}$.
	
	Let us now consider a natural transformation $\eta$ of the identity functor on $\Mod(\GrAlg)$ and notate via $f_{| \eta \ra}$ the intertwiner $\eta_{\GrAlg}: \GrAlg \to \GrAlg$, emphasizing that a choice of $\eta$ specifies in particular a choice of algebra element $| \eta \ra$. The morphism $\eta_V: V \to V$ assigned by $\eta$ to any representation $(V,\rho)$ in $\Mod(\GrAlg)$ is such that the diagram
	\begin{align}
		\begin{tikzcd}[ampersand replacement=\&, column sep=1.8em, row sep=1.3em]
		|[alias=A]|\GrAlg
		\&\&
		|[alias=B]|V
		\\
		\\
		|[alias=AA]|\GrAlg
		\&\&
		|[alias=BB]|V
		\arrow[from=A,to=B,"f_v"]
		\arrow[from=B,to=BB,"\eta_{V}"]
		\arrow[from=A,to=AA,"f_{| \eta \ra}"']
		\arrow[from=AA,to=BB,"f_v"']
		\end{tikzcd} 
	\end{align}
	commutes for every $v \in V$. This condition stipulates that $\eta_V: v \mapsto \rho(|\eta \ra)(v)$. The commutativity of the diagram
	\begin{align}
		\begin{tikzcd}[ampersand replacement=\&, column sep=1.8em, row sep=1.3em]
		|[alias=X]|\GrAlg
		\&\&
		|[alias=A]|V
		\&\&
		|[alias=B]|W
		\\
		\\
		|[alias=XX]|\GrAlg
		\&\&
		|[alias=AA]|V
		\&\&
		|[alias=BB]|W
		\arrow[from=A,to=B,"f"]
		\arrow[from=B,to=BB,"\eta_{W}"]
		\arrow[from=A,to=AA,"\eta_{V}"']
		\arrow[from=AA,to=BB,"f"']
		%%%%%
		\arrow[from=X,to=A,"f_{v}"]
		\arrow[from=X,to=XX,"f_{| \eta \ra}"']
		\arrow[from=XX,to=AA,"f_{v}"']
		\end{tikzcd} 
	\end{align}
	for every pair of representations $(V,\rho)$ and $(W,\sigma)$ and intertwiner $f: V \to W$,
	which follows from the definition property of the intertwiner $f: V \to W$ so that
	\begin{align}
		(\eta_V \circ f)(v) 
		=
		\big(\rho(| \eta \ra) \circ f \big)(v)
		=
		\big(f \circ \sigma(| \eta \ra) \big)(v)
		=
		(f \circ \eta_W)(v) \, ,
	\end{align}
	then confirms that the map $\eta$ as so defined is indeed a natural transformation of the identity functor. The natural transformation $\eta$ being solely defined in terms of its component associated with the regular representation, this concludes the proof of the lemma.
\end{proof}

\medskip \noindent
Let us now proceed to showing the following property:
\begin{lemma}\label{lem:DimRegRep}
	The dimension of the subcategory $\Mod(\GrAlg){\sss |}_{\rm reg.}$ is isomorphic, as a commutative $\mathbb C$-algebra, to the centre $Z(\GrAlg)$ of the twisted groupoid algebra $\GrAlg$.
\end{lemma}

\noindent
\begin{proof}
	Let $\eta \in \msf{Dim} \, \Mod(\GrAlg){\sss |}_{\rm reg.}$ be a natural transformation of the identity functor, whose unique component is the endomorphic intertwiner $f_{| \eta \ra} \equiv \eta_{\GrAlg}$. By definition, $f_{| \eta \ra}$ must commute with all intertwiners in the category. But we established in the proof of the previous lemma that endomorphic intertwiners of the regular representation are of the form $f_{| \fr g \ra}(-) = - \star | \fr g \ra$, for every $| \fr g \ra \in \GrAlg$. Therefore, the diagram
	\begin{align}
		\begin{tikzcd}[ampersand replacement=\&, column sep=1.8em, row sep=1.3em]
		|[alias=A]|\GrAlg
		\&\&
		|[alias=B]|\GrAlg
		\\
		\\
		|[alias=AA]|\GrAlg
		\&\&
		|[alias=BB]|\GrAlg
		\arrow[from=A,to=B,"f_{| \fr g \ra}"]
		\arrow[from=B,to=BB,"f_{| \eta \ra}"]
		\arrow[from=A,to=AA,"f_{| \eta \ra}"']
		\arrow[from=AA,to=BB,"f_{| \fr g \ra}"']
		\end{tikzcd}
	\end{align}
	must commute for every $| \fr g \ra \in \GrAlg$. It follows that an intertwiner $f_{| \eta \ra}$ defines a natural transformation of the identity functor if and only if
	\begin{equation}
		| \eta\ra \star | \fr g \ra = | \fr g \ra \star | \eta \ra 
		\, , \q \forall \, | \fr g \ra \in \GrAlg \, ,
	\end{equation}
	i.e. $| \eta \ra \in Z(\GrAlg)$. This establishes the following isomorphism of vector space:
	\begin{equation}
		\msf{Dim} \, \Mod(\GrAlg){\sss |}_{\rm reg.} \simeq Z(\GrAlg) \, . 
	\end{equation}
	It follows from the definition of the composition of natural transformations as well as the associativity of the twisted groupoid algebra that the centre element associated with the composition $\eta \circ \mu$ is $| \eta \ra \star | \mu \ra$. Thus, the previous isomorphism lifts in an obvious way to an isomorphism of commutative algebras.
\end{proof}

\medskip \noindent
Putting the previous two lemmas together yields theorem \ref{thm:DimRep}, and thus we have shown that
\begin{equation}
	\msf{Dim} \, \mc Z^\pi_G(\mathbb T^2) 
	\stackrel{\eqref{eq:ZTtwo}}{=}
	\msf{Dim} \, \Mod(\GrAlg)
	\stackrel{\eqref{eq:DimRep}}{\simeq}
	Z(\GrAlg)
	\stackrel{\eqref{eq:isoVect}}{\simeq}
	\mc V_{\Lambda^3 \overline G}(\msf t^3(\pi))
	\stackrel{\eqref{eq:ZTthree}}{=} 
	\mc Z^\pi_G(\mathbb T^3) \, ,
\end{equation}
as required.

\subsection{Dimension of the bicategory $\mc Z^\pi_G(\mathbb S^1)$}

We shall now compute  the crossing with the circle equation for the circle $\mathbb S^1$, which amounts to showing that the dimension of the bicategory $\mc Z^\pi_G(\mathbb S^1)$ is equivalent, as a braided monoidal category, to $\mc Z^\pi_G(\mathbb T^2)$, where by dimension we mean a categorification of the previous notion suitable to bicategories. The derivation will follow in close analogy with the proof of th.~\ref{thm:DimRep} and is related to Bartlett's computations in the lower-dimensional scenario using gerbal representations \cite{bartlett2009unitary}. 

In order to define the dimension of a bicategory, we first need to introduce higher-categorical analogues of the notions of functor and natural transformation, namely \emph{2-functor} and \emph{pseudo-natural transformations}. Since we shall only deal with identity 2-functors in practice, we omit to provide a detailed definition here and simply state that a 2-functor between two bicategories consists of a rule between the object-sets as well as a (1-)functor between the hom-categories such that the structure is preserved up to coherent 2-isomorphisms---all the coherent 2-isomorphisms being trivial in the case of identity 2-functors.  Pseudo-natural transformations between two such 2-functors are then defined as follows:

\begin{definition}[Pseudo-natural transformation]
	Let $F,F': \mc B \to \mc B'$ be two 2-functors between two bicategories $\mc B$ and $\mc B'$. A pseudo-natural transformation $\eta : F \Rightarrow F'$ between $F$ and $F'$ is a rule assigning a 1-morphism $\eta_X:F(X)\to F'(X) \in \Hom(\mc C')$ to every $X \in \Ob(\mc B)$ and an invertible 2-morphism $\eta_f$ to every $f:X \to Y \in \Ob (\msf{Hom}(\mc B))$ defined via\footnote{We part here with the usual definition where the 2-isomorphism $\eta_f$ is defined in regard to the opposite orientation. We choose this backwards convention in order to make the relation with the centre of a multi-fusion category more explicit.}
	\begin{equation}
	\begin{tikzcd}[ampersand replacement=\&, column sep=1.8em, row sep=1.3em]
	|[alias=A]|F(X)
	\&\&
	|[alias=B]|F(Y)
	\\
	\\
	|[alias=AA]|F'(X)
	\&\&
	|[alias=BB]|F'(Y)
	\arrow[from=A,to=B,"F(f)"]
	\arrow[from=B,to=BB,"\eta_{Y}"]
	\arrow[from=A,to=AA,"\eta_{X}"']
	\arrow[from=AA,to=BB,"F'(f)"']
	\arrow[from=AA,to=B,Rightarrow,"\eta_f"] 
	\end{tikzcd} \, .
	\end{equation}
	The 1- and 2-morphisms $\eta_X$ and $\eta_f$ are subject to coherence laws ensuring naturality as well as the preservation of the composition and the units, involving in particular the 1-associators of $\mc B$ and $\mc B'$.
\end{definition}

\noindent
We further require the notion of \emph{modification}, which are maps between pseudo-natural transformations:

\begin{definition}[Modification]
	Let $\eta , \mu : F \Rightarrow F'$ be two pseudo-natural modifications between two 2-functors $F,F' : \mc B \to \mc B'$. A modification $\vartheta : \eta \Rrightarrow \mu$ is a rule assigning a 2-morphism $\vartheta_X : \eta_X \Rightarrow \mu_X$ to every $X \in \Ob(\mc B)$ such that the diagram
	\begin{align}
	\begin{tikzcd}[ampersand replacement=\&, column sep=2.2em, row sep=1.3em]
		|[alias=A]|\eta_X \circ F'(f)
		\&\&
		|[alias=B]|\mu_X \circ F'(f)
		\\
		\\
		|[alias=AA]|F(f) \circ \eta_Y
		\&\&
		|[alias=BB]|F(f) \circ \mu_Y
		\arrow[from=A,to=B,Rightarrow,"\vartheta_X \otimes {\rm id}_{F'(f)}"]
		\arrow[from=B,to=BB,Rightarrow,"\mu_f"]
		\arrow[from=A,to=AA,Rightarrow,"\eta_f"']
		\arrow[from=AA,to=BB,Rightarrow,"{\rm id}_{F(f)}\otimes \vartheta_Y"']
	\end{tikzcd}
	\end{align}
	commutes for every $X,Y \in \Ob(\mc B)$ and $f \in \Ob(\msf{Hom}_\mc B(X,Y))$.
\end{definition}

\noindent
We are now ready to define the dimension of a bicategory \cite{Baez:1995xq,bartlett2009unitary}:
\begin{definition}[Dimension of a bicategory]
	Let $\mc{B}$ be a bicategory and ${\rm id}_{\mc{B}}:\mc{B} \to  \mc{B}$ the identity 2-functor on $\mc B$. We define the dimension $\msf{Dim}(\mc B)$ of $\mc B$ as the braided monoidal category with objects, pseudo-natural transformations $\eta,\mu: {\rm id}_\mc B \Rightarrow {\rm id}_\mc B$ and morphisms, modifications $\vartheta:\eta\Rrightarrow \mu$ between them.
\end{definition}

\noindent
Expanding this definition, a pseudo-natural transformation $\eta : {\rm id}_\mc B \Rightarrow {\rm id}_\mc B$ of the identity 2-functor is a rule assigning a 1-morphism $\eta_X : X \to X$ to every $X \in \Ob(\mc B)$ and an invertible 2-morphism $\eta_f : \eta_X \otimes f \Rightarrow f \otimes \eta_Y$ to every $f \in \Ob(\msf{Hom}_\mc B(X,Y))$ such that the diagram
\begin{align}
	\label{eq:pseudomonoidal}
	\begin{tikzcd}[ampersand replacement=\&, column sep=1.8em, row sep=1.3em]
	|[alias=A]|\eta_X \circ (f \circ g)
	\&\&
	|[alias=B]|(\eta_X \circ f) \circ g
	\&\&
	|[alias=C]| (f \circ \eta_Y ) \circ g
	\\
	\\
	|[alias=AA]| (f \circ g )\circ \eta_Z
	\&\&
	|[alias=BB]| f \circ (g \circ \eta_Z)
	\&\&
	|[alias=CC]| f \circ (\eta_Y \circ g)
	\arrow[from=A,to=B,Rightarrow,"\alpha^{-1}_{\eta_X,f,g}"]
	\arrow[from=B,to=C,Rightarrow,"\eta_f \otimes {\rm id}_g"]
	\arrow[from=C,to=CC,Rightarrow,"\alpha_{f,\eta_Y,g}"]
	\arrow[from=CC,to=BB,Rightarrow,"{\rm id}_f \otimes \eta_g"]
	\arrow[from=BB,to=AA,Rightarrow,"\alpha^{-1}_{f,g,\eta_Z}"]
	\arrow[from=A,to=AA,Rightarrow,"\eta_{f \circ g}"']
	\end{tikzcd}
\end{align}
commutes for every $f: X \to Y$ and $g: Y \to Z$. Moreover, given two pseudo-natural transformations $\eta,\mu : {\rm id}_\mc B \Rightarrow {\rm id}_\mc B$, the composition $\eta \circ \mu$ is defined as the pseudo-natural transformation that assigns to every $X \in \Ob(\mc B)$ the 1-morphism $(\eta_X \circ \mu_X)$ and to every $f \in \Ob(\msf{Hom}_\mc B(X,Y))$ the invertible 2-morphism $(\eta \circ \mu)_f:(\eta \circ \mu)_X \circ g \Rightarrow f \circ (\eta \circ \mu)_Y$ defined via
\begin{align}
	\label{eq:pseudocomposition}
	\begin{tikzcd}[ampersand replacement=\&, column sep=1.9em, row sep=1.3em]
	|[alias=Z]| (\eta \circ \mu)_X \circ f
	\&
	|[alias=A]|(\eta_X \circ \mu_X) \circ f
	\&\&
	|[alias=B]|\eta_X \circ (\mu_X \circ f)
	\&\&
	|[alias=C]| \eta_X \circ (f \circ \mu_Y)
	\\
	\\
	|[alias=ZZ]| f \circ (\eta \circ \mu)_Y
	\&
	|[alias=AA]| f \circ (\eta_Y \otimes \mu_Y)
	\&\&
	|[alias=BB]| (f \circ \eta_Y) \circ \mu_Y
	\&\&
	|[alias=CC]| f \circ (\eta_Y \circ \mu_Y)
	\arrow[from=Z,to=A,equal]
	\arrow[from=ZZ,to=AA,equal]
	\arrow[from=A,to=B,Rightarrow,"\alpha_{\eta_X,\mu_X,f}"]
	\arrow[from=B,to=C,Rightarrow,"{\rm id}_{\eta_X} \otimes \mu_f"]
	\arrow[from=C,to=CC,Rightarrow,"\alpha^{-1}_{\eta_X,f,\mu_Y}"]
	\arrow[from=CC,to=BB,Rightarrow,"\eta_f \otimes {\rm id}_{\mu_Y}"]
	\arrow[from=BB,to=AA,Rightarrow,"\alpha_{f,\eta_Y,\mu_Y}"]
	\arrow[from=Z,to=ZZ,Rightarrow,"(\eta \circ \mu)_f"']
	\end{tikzcd} \, .
\end{align}
Finally, given two pseudo-natural transformations $\eta,\mu : {\rm id}_\mc B \Rightarrow {\rm id}_\mc B$, a modification $\vartheta: \eta \Rrightarrow \mu$ is a rule assigning a 2-morphism $\vartheta_X: \eta_X \Rightarrow \mu_X$ to every $X \in \Ob(\mc B)$ such that $\eta_f \circ ({\rm id}_f \otimes \vartheta_Y) = (\vartheta_X \otimes f) \circ \mu_f$ for every $f \in \Ob(\msf{Hom}_\mc B(X,Y))$. Given this definition, establishing the crossing with the circle equation for $\mathbb S^1$ amounts to proving the following theorem

\begin{theorem}\label{thm:Dim2Rep}
	The dimension of the bicategory $\MOD(\VectGr)$ is equivalent, as a braided monoidal category, to the centre $\ms Z(\VectGr)$ of the twisted groupoid 2-algebra $\VectGr$, i.e.
	\begin{equation}
		\label{eq:Dim2Rep}
		\msf{Dim} \, \MOD(\VectGr) \cong \ms Z(\VectGr) \, .
	\end{equation}
\end{theorem}

\noindent
We shall prove this theorem by considering the higher-categorical analogues of lemmas \ref{lem:regRep} and \ref{lem:DimRegRep}. However, in order to do so, we require an alternative description of the constituents of $\MOD(-)$, which we recall from def.~\ref{def:MOD}  is the bicategory of module categories, module functors and module natural transformations over a multi-fusion category. 

Given a multi-fusion category $\mc C$, every indecomposable $\mc C$-module category is equivalent to the category of module for a \emph{separable} algebra object in $\mc C$. In order to introduce the notion of separable algebra object, we first need to discuss \emph{bimodule} objects. We already introduced the notions of algebra objects and right module objects in def.~\ref{def:algObject} and def.~\ref{def:algObjMod}, respectively. Left module objects are defined analogously, which when combined with right module objects, yield the notion of bimodule objects:
\begin{definition}[Bimodule object]
	Let $\mc C \equiv (\mc C, \otimes, \mathbbm 1, \ell, r, \alpha)$ be a multi-fusion category and $(A,B)$ a pair of algebra objects in $\mc C$. If $(M,p)$ is right $B$-module object, $(M,q)$ is a left $A$-module object, and $(p,q)$ satisfy an obvious coherence relation involving the associator $\alpha$, then the triple $(M,p,q)$ defines an $(A,B)$-bimodule object in $\mc C$. 
\end{definition}

\noindent
Notice that by definition any right $A$-module object $(M,p)$ can be identified with the $(\mathbbm 1,A)$-bimodule object $(M,\ell_M,p)$, and similarly for left $A$-module objects. In the same vein, we can define the concept of bimodule object homomorphisms. A separable object in $\mc C$ is then defined as an algebra object $(A,m,u)$ whose multiplication map admits a \emph{section} $\Delta: A \to A \otimes A$ satisfying 
\begin{equation}
	A \xrightarrow{\Delta} A \otimes A \xrightarrow{m} A = A \xrightarrow{{\rm id}_A} A
\end{equation} 
as an $(A,A)$-bimodule homomorphism. As alluded earlier, we can then show that every indecomposable \emph{left} module category over a multi-fusion category $\mc C$ can be defined as the category of \emph{right} modules over a separable algebra object in $\mc C$ \cite{etingof2016tensor}. Furthermore, the subspace of bimodule object homomorphisms being stable under composition, we can define the following category:
\begin{definition}[Category of bimodule objects] 
	Let $\mc C$ be a multi-fusion category and $(A,B)$ a pair of algebra objects in $\mc C$. We define the category $\msf{Bimod}_\mc C(A,B)$ as the category with objects $(A,B)$-bimodule objects, and morphisms $(A,B)$-bimodule homomorphisms.
\end{definition}

\noindent
Given two $\mc C$-module categories $\Mod_\mc C(A)$ and $\Mod_C(B)$, where $A$ and $B$ are two separable algebra objects in $\mc C$, the hom-category of $\mc C$-module functors between them can be shown to be equivalent to the category of bimodule objects $\msf{Bimod}_\mc C(A,B)$ \cite{etingof2016tensor}.  

\bigskip \noindent
Keeping the preliminary remarks above in mind, let us now proceed with our derivation. Mirroring, the methodology of the previous derivation, we begin by introducing the sub-bicategory $\MOD(\VectGr){\sss |}_{\rm reg.}$ consisting of a single object, namely the input multi-fusion category thought as the module category over itself, and the monoidal hom-category of $\VectGr$-module endofunctors of $\VectGr$. Henceforth, we shall refer to this single object as the \emph{regular} $\VectGr$-module category, by analogy with algebra representation theory. We then have the following:

\begin{lemma}
	Pseudo-natural transformations of the identity 2-functor on $\MOD(\VectGr)$ and modifications are completely determined by their components on the identity 2-functor of the sub-bicategory $\MOD(\VectGr){\sss |}_{\rm reg.}$.
\end{lemma}

\noindent
\begin{proof}
	We begin by remarking that every algebra object in $\VectGr$ can be shown to be separable, and as such we omit the distinction in the following. Moreover, the monoidal identity
	\begin{equation}
		\mathbbm 1_{\VectGr} := \bigoplus_{\fr g \in \Ob(\Lambda G)} \mathbb C_{{\rm id}_\fr g} 
	\end{equation} 
	corresponds to the trivial  algebra object, and the category of modules over it is none other than the regular $\VectGr$-module category. In symbols,
	\begin{equation}	
		\Mod_{\VectGr}(\mathbbm 1_{\VectGr}) = \VectGr \, .
	\end{equation} 
	We know from the preliminaries above that any left module $\VectGr$-module category can be expressed as the category of right modules over an algebra object $\mc A_\phi$, as defined in sec.~\ref{sec:alg}. The hom-category of $\VectGr$-module functors between such a module category and the regular module category is equivalent to the category of bimodule objects between the corresponding algebra objects, i.e.
	\begin{align}
		\msf{Hom}_{\MOD(\VectGr)}(\VectGr, \Mod_{\VectGr}(\mc A_\phi))
		&\, = \,
		\msf{Fun}_{\VectGr}(\VectGr, \Mod_{\VectGr}(\mc A_\phi)) 
		\\
		&\, \cong \,
		\msf{Bimod}_{\VectGr}(\mathbbm 1_{\VectGr}, \mc A_\phi) 
		\, \cong \, 
		\Mod_{\VectGr}(\mc A_\phi) \, .
	\end{align} 
	It follows that a choice of module functors $(F,s)$ of this form is specified by a choice of $\mc A_\phi$-module object $M_{\mc A_\phi}$ such that the functor $F$ is defined via
	\begin{equation}
	\begin{array}{ccccl}
		F = - \otimes M_{\mc A_\phi} & : & \VectGr & \to & \Mod_{\VectGr}(\mc A_\phi)
		\\
		& : & V\in \Ob(\VectGr)  &\mapsto & V \otimes M_{\mc A_\phi} 	
		\\
		& : & f \in \Hom(\VectGr) & \mapsto & f \otimes {\rm id}_{M_{\mc A_\phi}}
	\end{array} 
	\end{equation}
	and the natural isomorphism $s$ is defined on objects $V,V' \in \Ob(\VectGr)$ as
	\begin{align}
		s_{V,V'} : F(V \otimes V')  \to V \otimes F(V') 
	\end{align}
	so that we have the identification $s_{-,-} = \alpha_{-,-,M_{\mc A_\phi}}$. Henceforth, we shall denote such module endofunctors by $(F_{M_{\mc A_\phi}} , s_{M_{\mc A_\phi}})$.
	As a corollary, we obtain that a module endofunctor of the regular module category is fully specified by a choice of $\Lambda G$-graded vector space. 
	
	Let us now consider a pseudo-natural transformation $\eta \in \Ob(\msf{Dim} \, \MOD(\VectGr))$ of the identity 2-functor. By definition, it assigns a module endofunctor to every object and a morphism of module functors to every module functor.
	We notate via $(F_{V^\eta},s_{V^\eta}) \equiv (- \otimes V^\eta, \alpha_{-,-,V^\eta})$ the module endofunctor of the regular $\VectGr$-module category $\eta$ assigned to the object $\VectGr$, emphasizing that a choice of $\eta$ specifies in particular a choice of $\Lambda G$-graded vector space $V^\eta$. Moreover, every module endofunctor of $\VectGr$ is of the form $(F_W,s_W)$ for some $W \in \Ob(\VectGr)$, and we denote by $\eta_{W}$ the isomorphism of module functors $\eta$ assigns to it according to
	\begin{align}
		\label{eq:diagDimReg}
		\begin{tikzcd}[ampersand replacement=\&, column sep=2.8em, row sep=1.3em]
			|[alias=X]|\VectGr
			\&\&
			|[alias=A]|\VectGr
			\\
			\\
			|[alias=XX]|\VectGr
			\&\&
			|[alias=AA]|\VectGr
			\arrow[from=A,to=AA,"(F_{V^\eta}{,}s_{V^\eta})"]
			\arrow[from=X,to=A,"(F_{W} {,} s_{W})"]
			\arrow[from=X,to=XX,"(F_{V^\eta}{,}s_{V^\eta})"']
			\arrow[from=XX,to=AA,"(F_{W} {,} s_{W})"']
			\arrow[Rightarrow,from=XX,to=A,"\eta_W"]
		\end{tikzcd} \, .
	\end{align}
	 Let us now consider any module category $\Mod_{\VectGr} (\mc A_\phi)$ and notate via $\eta_{\mc A_\phi}$ the corresponding module endofunctor. The pseudo-natural transformation $\eta$ must further assign an isomorphism of module functors to every module functor $(F_{M_{\mc A_\phi \mc A_\phi}} {,} s_{M_{\mc A_\phi \mc A_\phi}}): \Mod_{\VectGr}(\mc A_\phi) \to \Mod_{\VectGr}(\mc A_\phi)$ prescribed by a choice of $(\mc A_\phi,\mc A_\phi)$-bimodule object $M_{\mc A_\phi \mc A_\phi}$. This isomorphism, which we shall refer to using the shorthand notation $\eta_{M_{\mc A_\phi \mc A_\phi}}$, is defined via the following diagram:
	\begin{align}
		\label{eq:dimBraidingSpe}
		\begin{tikzcd}[ampersand replacement=\&, column sep=3.5em, row sep=1.3em]
			|[alias=X]|\Mod_{\VectGr}(\mc A_\phi)
			\&\&
			|[alias=A]|\Mod_{\VectGr}(\mc A_\phi)
			\\
			\\
			|[alias=XX]|\Mod_{\VectGr}(\mc A_\phi)
			\&\&
			|[alias=AA]|\Mod_{\VectGr}(\mc A_\phi)
			\arrow[from=A,to=AA,"\eta_{\mc A_\phi}"]
			\arrow[from=X,to=A,"(F_{M_{\mc A_\phi \mc A_\phi}} {,} s_{M_{\mc A_\phi \mc A_\phi}})"]
			\arrow[from=X,to=XX,"\eta_{\mc A_\phi}"']
			\arrow[from=XX,to=AA,"(F_{M_{\mc A_\phi \mc A_\phi}} {,} s_{M_{\mc A_\phi \mc A_\phi}})"']
			\arrow[Rightarrow,from=XX,to=A,"\eta_{M_{\mc A_\phi \mc A_\phi}}"]
		\end{tikzcd} \, .
	\end{align}
	Noticing that $\Mod_{\VectGr}(\mc A_\phi)$ defines in particular a \emph{right} module category over $\msf{Bimod}_{\VectGr}(\mc A_\phi, \mc A_\phi)$ such that the action bifunctor is defined from the functors $F_{M_{\mc A_\phi \mc A_\phi}}$, the diagram above stipulates that $(\eta_{\mc A_\phi}, \eta_{M_{\mc A_\phi \mc A_\phi}})$ defines a $\msf{Bimod}_{\VectGr}(\mc A_\phi, \mc A_\phi)$-module endofunctor of $\Mod_{\VectGr}(\mc A_\phi)$. Let us now consider the defining diagram of the isomorphism of functors $\eta_{M_{\mc A_\phi}}$ that the pseudo-natural transformation $\eta$ assigns to a given module functor $(F_{M_{\mc A_\phi}},s_{M_{\mc A_\phi}}): \VectGr \to \Mod_{\VectGr}(\mc A_\phi)$:
	\begin{align}
		\begin{tikzcd}[ampersand replacement=\&, column sep=2.8em, row sep=1.3em]
			|[alias=X]|\VectGr
			\&\&
			|[alias=A]|\Mod_{\VectGr}(\mc A_\phi)
			\\
			\\
			|[alias=XX]|\VectGr
			\&\&
			|[alias=AA]|\Mod_{\VectGr}(\mc A_\phi)
			\arrow[from=A,to=AA,"\eta_{\mc A_\phi}"]
			\arrow[from=X,to=A,"(F_{M_{\mc A_\phi}} {,} s_{M_{\mc A_\phi}})"]
			\arrow[from=X,to=XX,"(F_{V^\eta}{,}s_{V^\eta})"']
			\arrow[from=XX,to=AA,"(F_{M_{\mc A_\phi}} {,} s_{M_{\mc A_\phi}})"']
			\arrow[Rightarrow,from=XX,to=A,"\eta_{M_{\mc A_\phi}}"]
		\end{tikzcd} \, .
	\end{align}
	This diagram stipulates that the functor component of $\eta_{\mc A_\phi}$ is provided by $ - \otimes V^{\eta}$, such that $M \otimes V^\eta$ with $M \in {\rm Ob}(\Mod_{\VectGr}(\mc A_\phi))$ defines the right module object with action given by
	\begin{equation}
		(M\otimes V^{\eta}) \otimes \mc A_\phi \xrightarrow{\eqref{eq:dimBraidingSpe}}(M\otimes \mc{A}_{\phi})\otimes V^\eta\xrightarrow{p \otimes {\rm id}_{V^\eta}}M\otimes V^\eta \, .
	\end{equation}
	The diagram further stipulates that the isomorphism of module functors $\eta_{M_{ \mc A_\phi}}$ is defined for any $\Lambda G$-graded vector space $V$ as
	\begin{equation}
		\eta_{M_{\mc A_\phi}} : (V \otimes V^{\eta}) \otimes M_{\mc A_\phi}
		\xrightarrow{\sim} (V \otimes M_{\mc A_\phi}) \otimes V^{\eta} \, ,
	\end{equation}
	which is fixed by the family of isomorphisms $\eta_W$ defined according to \eqref{eq:diagDimReg}.
	Note that for a choice of $V^\eta$, it follows from the definition of the monoidal product in $\VectGr$ that $M \otimes V^\eta$ may be the zero vector space for every $M \in \Ob(\Mod_{\VectGr}(\mc A_\phi))$, in which case the $\VectGr$-module endofunctor of $\Mod_{\VectGr}(\mc A_\phi)$ induced by $V^\eta$ is the zero functor.
	Let us now consider two module categories $\Mod_{\VectGr}(\mc A_\phi)$ and $\Mod_{\VectGr}(\mc B_\psi)$. It follows from the previous derivation that $\eta$ assigns a homomorphism $\eta_{M_{ \mc A_\phi \mc B_\psi}}$ to every module functor $(F_{M_{\mc A_\phi \mc B_\psi}}, s_{M_{\mc A_\phi \mc B_\psi}}): \Mod_{\VectGr}(\mc A_\phi) \to \Mod_{\VectGr}(\mc B_\psi)$ via
	\begin{align}
		\label{eq:dimBraiding}
		\begin{tikzcd}[ampersand replacement=\&, column sep=3.5em, row sep=1.3em]
			|[alias=X]| \VectGr
			\&\&
			|[alias=A]|	\Mod_{\VectGr}(\mc A_\phi)
			\&\&
			|[alias=B]| \Mod_{\VectGr}(\mc B_\psi)
			\\
			\\
			|[alias=XX]| \VectGr
			\&\&
			|[alias=AA]| \Mod_{\VectGr}(\mc A_\phi)
			\&\&
			|[alias=BB]|	\Mod_{\VectGr}(\mc B_\psi)
			\arrow[from=A,to=B,"(F_{M_{\mc A_\phi \mc B_\psi}}{,} s_{M_{\mc A_\phi \mc B_\psi}})"]
			\arrow[from=B,to=BB,"\eta_{\mc B_\psi}"]
			\arrow[from=A,to=AA,"\eta_{\mc A_\phi}"']
			\arrow[from=AA,to=BB,"(F_{M_{\mc A_\phi \mc B_\psi}}{,} s_{M_{\mc A_\phi \mc B_\psi}})"']
			%%%%%
			\arrow[from=X,to=A,"(F_{M_{\mc A_\phi}}{,} s_{M_{\mc A_\phi}})"]
			\arrow[from=X,to=XX,"(F_{V^\eta}{,} s_{V^\eta})"']
			\arrow[from=XX,to=AA,"(F_{M_{\mc A_\phi}}{,} s_{M_{\mc A_\phi}})"']
			\arrow[Rightarrow,from=XX,to=A,"\eta_{M_{\mc A_\phi}}"]
			\arrow[Rightarrow,from=AA,to=B,"\eta_{M_{ \mc A_\phi \mc B_\psi}}"]
		\end{tikzcd}
	\end{align}
	such that
	\begin{equation}
		\eta_{M_{\mc A_\phi \mc B_\psi}} :
		F_{M_{\mc A_\phi \mc B_\psi}}(M_{\mc A_\phi} \otimes V^\eta) \xrightarrow{\sim} F_{M_{\mc A_\phi \mc B_\psi}}(M_{\mc A_\phi}) \otimes V^\eta  \, .
	\end{equation}
	Crucially, the assignments $\eta_{\mc A_\phi}$, $\eta_{\mc B_\psi}$ and $\eta_{M_{\mc A_\phi B_\psi}}$ are completely determined by the input data together with the component of $\eta$ associated with the regular module category, namely $(F_{V^\eta}, s_{V^\eta})$ and the collection of isomorphisms $ \eta_{W} \equiv \eta_{(F_W,s_W)}$ for $W \in \Ob(\VectGr)$. It is immediate to check that such assignments are indeed compatible with the coherence relations \eqref{eq:pseudomonoidal} and \eqref{eq:pseudocomposition}. Similarly we can show that any modifications $\vartheta: \eta \Rrightarrow \mu$ are completely determined by their actions with respect to the regular $\VectGr$-module category $\VectGr$. This completes the proof of the lemma.
\end{proof}

\medskip
\noindent
Let us now proceed to showing the following lemma:
\begin{lemma}
	The dimension of the sub-bicategory $\MOD(\VectGr){\sss |}_{\rm reg.}$ is equivalent, as a braided monoidal category, to the centre $\ms Z(\VectGr)$ of the multi-fusion category $\VectGr$.
\end{lemma}

\noindent
\begin{proof}
	Let $\eta \in \msf{Dim} \, \MOD(\VectGr){\sss |}_{\rm reg.}$ be a pseudo-natural transformation of the identity 2-functor. It assigns a unique 1-morphism to the single object of the bicategory, namely the module endofunctor $(F_{V^\eta}, s_{V^\eta}) \equiv (- \otimes V^\eta, \alpha_{-,-,V^\eta})$, where $V^\eta \in \Ob(\VectGr)$. Moreover, $\eta$ assigns to every $\VectGr$-module endofunctor $(F_W,s_W): \VectGr \to \VectGr$, with $W \in \Ob(\VectGr)$, an isomorphism of module functors (see eq.~\eqref{eq:diagDimReg})
	\begin{equation}
		\eta_{(F_W,s_W)} \equiv \eta_W : (V \otimes V^\eta) \otimes W \xrightarrow{\sim} (V \otimes W) \otimes V^\eta \, , \q \forall \, V \in \Ob(\VectGr) \, .
	\end{equation}
	Picking $V  = \mathbb C$, this yields an isomorphism
	\begin{equation}
		R_{V^\eta,-} : V^\eta \otimes - \xrightarrow{\sim} - \otimes V^\eta \, . 
	\end{equation}
	Furthermore, it follows from the coherence relation \eqref{eq:pseudomonoidal} satisfied by the $\VectGr$-module functors assigned by $\eta$ that the collection of natural isomorphisms $R_{V^\eta,-}$ fulfils the hexagon relation \eqref{eq:compoPent}. It follows that objects in $\msf{Dim} \, \MOD(\VectGr){\sss |}_{\rm reg.}$ are in one-to-one correspondence with objects $(V^\eta, R_{V^\eta,-}) \in \Ob(\ms Z(\VectGr))$. 
	
	Given two pseudo-natural transformations $\eta,\mu \in \msf{Dim} \, \MOD(\VectGr){\sss |}_{\rm reg.}$, let us consider a modification $\vartheta: \eta \Rrightarrow \mu$. This modification assigns a morphism of module functor between $(F_{V^\eta},s_{V^\eta})$ and $(F_{V^\mu},s_{V^\mu})$ to the unique object $\VectGr$ of the bicategory. Denoting by $\vartheta_\mathbbm 1$ this morphism---in reference to the fact that $\VectGr = \Mod_{\VectGr}(\mathbbm 1_{\VectGr})$---it is required to satisfy
	\begin{align}
		\begin{tikzcd}[ampersand replacement=\&, column sep=3em, row sep=1.3em]
			|[alias=A]|(F_{V^\eta},s_{V^\eta}) \circ (F_W,s_W)
			\&\&
			|[alias=B]| (F_{V^\eta},s_{V^\eta}) \circ (F_W,s_W)
			\\
			\\
			|[alias=AA]| (F_W,s_W) \circ (F_{V^\eta},s_{V^\eta})
			\&\&
			|[alias=BB]| (F_W,s_W) \circ (F_{V^\mu},s_{V^\mu})
			\arrow[from=A,to=B,Rightarrow,"\vartheta_\mathbbm 1 \otimes {\rm id}_{(F_W,s_W)}"]
			\arrow[from=B,to=BB,Rightarrow,"\mu_W"]
			\arrow[from=A,to=AA,Rightarrow,"\eta_W"']
			\arrow[from=AA,to=BB,Rightarrow,"{\rm id}_{(F_W,s_W)}\otimes \vartheta_\mathbbm 1"']
		\end{tikzcd}
	\end{align}
	for every $\VectGr$-module endofunctor $(F_W,s_W)$ of the regular module category $\VectGr$, with $W \in \Ob(\VectGr)$, $\eta_W \equiv \eta_{(F_W,s_W)}$ and $\mu_W \equiv \mu_{(F_W,s_W)}$. In terms of the objects $(V^\eta, R_{V^\eta,-}), (V^\mu,R_{V^\mu,-}) \in \Ob(\ms Z(\VectGr))$, $\vartheta_\mathbbm 1$ is identified with a morphism $V^\eta \to V^\mu$ satisfying the coherence relation \eqref{eq:squareHomCentre}. Therefore, morphisms of $\msf{Dim} \, \MOD(\VectGr){\sss |}_{\rm reg.}$ are in one-to-one correspondence with those of the centre $\ms Z(\VectGr)$. Putting everything together, we have established the equivalence of categories
	\begin{equation}
		\label{eq:Dim2RepReg}
		\msf{Dim} \, \MOD(\VectGr){\sss |}_{\rm reg.} \cong \ms Z(\VectGr) \, .
	\end{equation}
	It is instructive to rephrase this derivation as follows: By definition, $(F_{V^\eta},s_{V^\eta})$ is a module endofunctor of the left $\VectGr$-module category over itself. But $\VectGr$ is a also a right $\VectGr$-module category with respect to the action bifunctor defined via the set of functors $F_W$ in such a way that $(F_{V^\eta},\eta_W)$ for every $W$ defines a right module endofunctor of $\VectGr$ over itself. It follows that $\eta$ is in one-to-one correspondence with a  $(\VectGr,\VectGr)$-bimodule endofunctor over $\VectGr$, the category of which is known to be equivalent to the centre $\ms Z(\VectGr)$, i.e.
	\begin{equation}
		\msf{Fun}_{\VectGr | \VectGr}(\VectGr, \VectGr) \cong \ms Z(\VectGr) \, .
	\end{equation}
	Let us now extend the equivalence \eqref{eq:Dim2RepReg} to a monoidal equivalence.  Let us consider two objects $\eta, \mu \in \Ob(\msf{Dim} \, \MOD(\VectGr)){\sss |}_{\rm reg.}$. The monoidal structure is provided by the composition $\eta \circ \mu$ of the pseudo-natural transformations in $\MOD(\VectGr)$. By definition, the composite $\eta \circ \mu$ assigns to the regular module category $\VectGr$ the composite $(F_{V^\eta},s_{V^\eta}) \circ (F_{V^\mu}, s_{V^\mu})$, where $F_{V^\eta} \circ F_{V^\mu} \circ \msf t(\pi)_{-,V^\eta,V^\mu}=  F_{V^\eta \otimes V^\mu} $. Given a module endofunctor $(F_W,s_W): \VectGr \to \VectGr$, with $W \in \Ob(\VectGr)$, $(\eta \circ \nu)$ assigns an isomorphism of module functors $(\eta \circ \mu)_W : F_W(V \otimes (V^\eta \otimes V^\mu)) \xrightarrow{\sim} F_W(V) \otimes (V^\eta \otimes V^\mu)$ defined via \eqref{eq:pseudocomposition} for every $V \in \Ob(\VectGr)$. Applied to $V = \mathbb C$, this yields an isomorphism
	\begin{equation}
		R_{V^\eta \otimes V^\mu,-} : (V^\eta \otimes V^\mu) \otimes - \xrightarrow{\sim} - \otimes (V^\eta \otimes V^\mu)
	\end{equation}
	satisfying the relation $\eqref{eq:monoCentre}$. This establishes the monoidal equivalence between $\msf{Dim} \, \MOD(\VectGr){\sss |}_{\rm  reg.}$ and $\ms Z(\VectGr)$. The braided equivalence follows immediately from the same arguments.
\end{proof}

\bigskip \noindent
Putting the previous two lemmas together yields theorem \ref{thm:Dim2Rep}, and thus we have shown that
\begin{equation}
	\msf{Dim} \, \mc Z^\pi_G(\mathbb S^1) 
	\stackrel{\eqref{eq:ZSone}}{=}
	\msf{Dim} \, \MOD(\VectGr)
	\stackrel{\eqref{eq:Dim2Rep}}{\cong}
	\ms Z(\VectGr)
	\stackrel{\eqref{eq:congCat}}{\cong}
	\Mod(\mathbb C[\Lambda^2 G]^{\msf t^2(\pi)})
	\stackrel{\eqref{eq:ZTtwo}}{=} 
	\mc Z^\pi_G(\mathbb T^2) \, ,
\end{equation}
as required. Recall from sec.~\ref{sec:DW} that the bicategory $\MOD(\VectGr)$ encodes boundary conditions for the endpoints of a string-like excitation of the lattice Hamiltonian realisation, quantum numbers associated with string-like topological excitations that are constrained by a choice of endpoints boundary conditions, and the renormalisation of string-like excitations that are glued along their endpoints. On the other hand, the category $\Mod(\mathbb C[\Lambda^2 G]^{\msf t(\pi)})$ encodes the loop-like (bulk) excitations and their statistics. As such, the computation performed in this section formalises the process upon which loop-like excitations are formed out of string-like ones. Heuristically, this process can be interpreted as follows: Consider an open four-manifold whose boundary, which is itself open, contains a two-torus boundary component that we decompose into a \emph{left} and a \emph{right} copy of the manifold $\mathbb S^1 \times [0,1]$. The TQFT functor assigns $(\VectGr,\VectGr)$-bimodule categories to the circles at which the two copies of $\mathbb S^1 \times [0,1]$ meet. Furthermore, it assigns left, respectively right, $\VectGr$-module functors to the copies of $\mathbb S^1 \times [0,1]$. Choosing the regular $(\VectGr,\VectGr)$-bimodule category to be assigned to the copies of $\mathbb S^1$, we find that the TQFT assigns the category $\msf{Fun}_{\VectGr}(\VectGr,\VectGr)\simeq \VectGr$ to the left copy of the manifold $\mathbb S^1 \times [0,1]$ and analogously to the right copy. Within this specific context, the category $\VectGr$ encodes string-like excitations with a special kind of boundary conditions. It follows that the TQFT assigns to $\mathbb T^2$ the category $\msf{Fun}_{\VectGr|\VectGr}(\VectGr, \VectGr)$ of bimodule endofunctors of $\VectGr$ over itself, which is  equivalent to the category of bulk loop-excitations. Loosely speaking, this should be interpreted as two string-like excitations whose endpoints are identified so as to form a loop-like excitation.

\bigskip
\section{Braiding statistics of loop-like excitations\label{sec:braiding}}

\emph{Building upon the previous results, we shall argue in this section that $\VectGr$-module endofunctors induced from objects of $\ms Z(\VectGr)$ describe the salient features of the loop-like excitations hosted by the lattice Hamiltonian realisation of the (3+1)d Dijkgraaf-Witten theory. In particular, we shall revisit the fact that $\ms Z(\VectGr)$ define representations of the linear necklace braid group.}

\subsection{Motion groups}

Given a quantum physical system and a collection of $n \in \mathbb Z^+$ indistinguishable particles, the transformation properties of their joint wavefunction upon exchanging their positions along specified trajectories is known as the \emph{exchange statistics}. For instance, bosons and fermions are characterised by symmetric and antisymmetric joint wave functions, respectively, with respect to the exchange process of two particles. More generally, the exchange process corresponds to the action of a symmetry group $G$ on the joint Hilbert space, so that the wavefunction defines a representation of $G$. For instance, given $n$ indistinguishable particles in $\mathbb D^{d}$ with $d \geq 3$, the permutation of the $n$ entries in the joint wavefunction is governed by the symmetric group $\ms S_n$ so that bosons and fermions are given by the trivial and the sign representations of $\ms S_n$, respectively.  

In (2+1)d, it is well-known that quantum physical systems can host particles with more exotic statistics. Indeed, the exchange statistics of $n$ indistinguishable particles is described by the so-called braid group $\ms B_n$, defined in the following. Anyons are then defined as particles whose exchange statistics define representations of the braid group that are neither the trivial nor sign representations. This phenomenon is specific to two-dimensional systems since the path of a particle exchanging position with another via a half-turn clockwise rotation is not necessarily homotopic to that of exchanging position via a half-turn counter-clockwise rotation.

Although point-like anyons cannot exist in a ($d$+1)-dimensional system with $d \geq 3$ for the topological reason evoked above, extended quasi-particles may have exotic exchange statistics, generalizing the notion of anyons to higher dimensions. As we know, such extended quasi-particles arise as excitations in higher-dimensional topological models. In order to provide a unified description of the symmetry group characterising the exchange statistics of point-like and extended quasi-particles, we shall discuss the notion of \emph{motion groups}, of which $\ms S_n$ and $\ms B_n$ are examples.

Consider an oriented $d$-manifold $\Sigma$ and choose an oriented submanifold $\Xi$. Letting $\mathtt{Homeo}(\Sigma)$ be the orientation-preserving homeomorphism group of $\Sigma$, we notate via $\mathtt{Homeo}(\Sigma,\Xi) \subset \mathtt{Homeo}(\Sigma)$ the subgroup of auto-homeomorphisms that are additionally auto-homeomorphisms of $\Xi$. A \emph{motion} of $\Xi$ is defined as a map $f : [0,1] \times \Sigma \to \Sigma$ such that for all $t \in [0,1]$, $f_t \equiv f(t,-)$ is a \emph{path} in $\mathtt{Homeo}(\Sigma)$ satisfying $f_0 = {\rm id}_{\Sigma}$ and $f_1 \in \mathtt{Homeo}(\Sigma,\Xi)$. Given a pair of motions $(f,g)$, composition is obtained by translating $g$ via multiplication in $\mathtt{Homeo}(\Sigma)$ so that $f_1$ and $g_0$ coincide, and then use the composition of paths, i.e. 
\begin{align}
	\label{eq:motioncomp}
	(f\circ g)_{t}=
	\begin{cases}
		f_{2t}\quad &\text{for}\quad 0\leq t\leq\frac{1}{2}\\
		g_{2t-1}\circ f_{1}\quad &\text{for}\quad \frac{1}{2}\leq t \leq 1
	\end{cases} \, .
\end{align}
The reverse motion $\bar f$ of a motion $f$ is then defined via $\bar f_t = f_{1-t}\circ f_1^{-1}$. Furthermore, two motions $f$ and $g$ are said to be equivalent if the motion $\bar f \circ g$ is smoothly homotopic to a stationary motion, i.e. a motion $h$ such that $h_t \in \mathtt{Homeo}(\Sigma,\Xi)$ for every $t \in [0,1]$. Finally, we define the motion group $\mathtt{Mot}(\Sigma,\Xi)$ as the group of equivalence classes of motions of $\Xi$ in $\Sigma$, with multiplication rule induced from the composition \eqref{eq:motioncomp} of motions, and inverse map the reverse operation \cite{bams/1183535291,10.1307/mmj/1029002454,Baez:2006un}. In the following, we shall consider three examples of motion groups, namely the braid group, the linear necklace group, and the loop braid group.

\subsection{Braid group}

Given $\Sigma = \mathbb D^2$ and the disjoint union $\Xi = \sqcup_{i=1,\ldots,n} \mathbb D^2 \subset \Sigma$ of $n$ copies of $\mathbb D^2$, the $n$-strand braid group $\ms B_n$ is defined as the motion group $\mathtt{Mot}(\mathbb D^2,\sqcup_{i=1,\ldots,n}\mathbb D^2)$. It admits a presentation in terms of generators $\{\sigma_i\}_{i=1,\ldots,n-1}$ subject to the relations
\begin{align*}
	&\text{\small B1.} \;\; \sigma_{i}\sigma_{j}=\sigma_{j}\sigma_{i} \, , \q {\rm for} \; |i-j|>1 ,
	\\
	&\text{\small B2.} \;\; \sigma_{i}\sigma_{i+1}\sigma_{i}=\sigma_{i+1}\sigma_{i}\sigma_{i+1}.
\end{align*}
Labelling each copy of $\mathbb D^2 \subset \Xi$ by a unique integer $i \in \llbracket 1 ,n \rrbracket$ and placing them along a segment embedded in $\mathbb D^2$, each generator $\sigma_i$ can be thought as an operator performing the exchange of the disks $i$ and $i$+1 via a half-turn rotation in the counter-clockwise direction. This can be conveniently visualised in terms of the corresponding worldlines as
\begin{equation}
	\mathbbm 1_n = \braidGroup{1}{1}{1}
	\ldots \braidGroup{1}{2}{i}
	\ldots \braidGroup{1}{1}{n}
	\, , \q
	\sigma_i = \braidGroup{1}{1}{1}
	\ldots \braidGroup{1}{3}{i}
	\ldots \braidGroup{1}{1}{n}
	\, , \q
	\sigma_i^{-1} = \braidGroup{1}{1}{1}
	\ldots \braidGroup{1}{4}{i}
	\ldots \braidGroup{1}{1}{n} \, ,
\end{equation}
where we also represented the identity element and the inverse $\sigma_i^{-1}$ for convenience. This graphical representation can also be used to conveniently visualise the relations above. Doing so would reveal that the second relation can be interpreted as the third Reidemeister move \cite{reidemeister1927elementare}, which plays a crucial role in knot theory \cite{kauffman2001knots}.

An important feature of the braid group is that for any $i \in \llbracket 1 ,n \rrbracket$, $\sigma^2_{i}\neq 1$. Geometrically, this is the statement that the braid resulting from performing twice the exchange in the same direction cannot be continuously deformed in $\mathbb D^2 \times [0,1]$ to the trivial one without cutting the strands. If we were to change the spatial manifold from $\mathbb{D}^2$ to $\mathbb{D}^d$ with $d>2$, and consider $\Xi=\sqcup_{i=1,...,n} \mathbb D^d$ to be the disjoint union of $n$ $d$-balls, the motion group $\mathtt{Mot}(\mathbb{D}^d,\sqcup_{1,...,n} \mathbb D^d)$ would be found to be isomorphic to the symmetric group $\ms S_n$ on $n$ objects. Indeed, the braid associated with the double exchange of two copies of $\mathbb D^d$ along the same trajectory can always be disentangled. It follows from the definition of the braid group that the symmetric group $\ms S_n$ admits a presentation in terms of generators $\{s_i\}_{i=1,\ldots,n-1}$ subject to the relations
\begin{align*}
	&\text{\small S1.} \;\; s_{i}s_{j}=s_{j}s_{i} \, , \q  {\rm for} \; |i-j|>1,
	\\
	&\text{\small S2.} \;\; s_is_{i+1}s_{i} = s_{i+1}s_{i}s_{i+1},
	\\
	&\text{\small S3.} \;\; s_i^2 = 1.
\end{align*}
Each generator $s_i$ can still be thought as an operator exchanging the position of the disks $i$ and $i+1$. As we mentioned earlier, this implies that in order to have anyon-like exchange statistics in spatial dimension higher than three, extended objects must be considered.

\subsection{Linear necklace group}

Let $\Sigma = \mathbb D^3$ and $\Xi \subset \Sigma$ be the $n$-component \emph{linear necklace}. More precisely, given $\Sigma:=[0,n+1]\times[-1,1]\times[-1,1]$ with axes labelled by $x$, $y$ and $z$, respectively, the linear necklace $\Xi$ is defined as $K_{0}\sqcup^{n}_{i=1}K_{i}$, where the `necklace' $K_{0}$ is a local neighbourhood of the $x$-axis, and the $i$-th `loop' $K_{i}$ is a local neighbourhood of a Euclidean unit circle in the $yz$-plane centred around $x=i$. The corresponding motion group $\mathtt{Mot}(\mathbb D^3, \Xi)$ is known as the \emph{linear necklace group} $\ms L \ms N_n$. Interestingly, it was shown in \cite{bellingeri2016braid} that $\ms L \ms N_n$ is isomorphic to the $n$-strand braid group $\ms B_n$, and as such it admits a presentation akin to the one provided above in terms of the generators $\{\sigma_i\}_{i=1,\ldots,n-1}$. In this new context, each generator $\sigma_i$ can now be thought as an operator performing the exchange of the loops $i$ and $i$+1 such that the loops $i$+1 passes through the centre of $i$ as illustrated below:
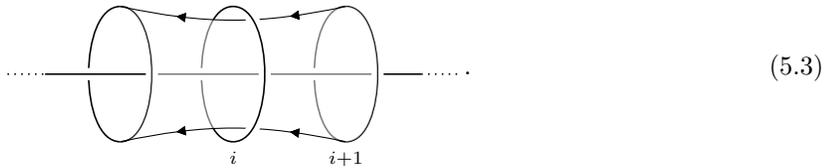
\begin{equation}
	\label{eq:necklaceDrawing}
	\necklace{1} \, .
\end{equation}
In sec.~\ref{sec:DW}, we reviewed that the category $\Mod(\mathbb C[\Lambda^2 G]^{\msf t^2(\pi)})$ encodes the loop-like excitations of the lattice Hamiltonian realisation of Dijkgraaf-Witten theory. Moreover, its braided monoidal structure, which is prescribed by the comultiplication map $\Delta$ and the invertible algebra element $\msf R$ introduced in sec.~\ref{sec:centre2Algebra}, encodes the fusion and the braiding statistics of the excitations. Given the above, we showed in \cite{Bullivant:2019fmk} that the loop-like excitations define representations of the linear necklace group. Let us briefly review the main argument here. Recall from sec.~\ref{sec:centre2Algebra} that objects of $\Mod(\GrAlg)$ are $\GrAlg$-modules of the form $(V,\rho)$, where $V = \bigoplus_{[\fr g'] \in \uppi_0(\Lambda^2G)}V_{[\fr g']}$ with $\uppi_0(\Lambda^2G)$ the set of connected components in the groupoid $\Lambda^2 G$, whose objects are characterized by pairs of commuting group elements in $G$. Given a choice of equivalence class $[\fr g'] \in \uppi_0(\Lambda^2 G)$, the corresponding $\GrAlg$-module further decomposes as $V_{[\fr g']}= \bigoplus_{\fr g \in [\fr g']} V_\fr g$, where every object $\fr g \equiv x \xrightarrow{a} \,  \in {\rm End}(\Lambda G)$ is interpreted as a loop-like excitation with flux $a \in G$,  which is constrained by the presence of a flux $x \in G$ that threads the loop-like excitation, via the requirement $[a,x] = \mathbbm 1_G$. The representation associated with the vector space $V_\fr g$ then corresponds to the charge of the loop-like excitation. Furthermore, it was established in \cite{willerton2008twisted,Bullivant:2019fmk} that we have the following decomposition 
\begin{equation}
	\label{eq:decompMod}
	\Mod(\GrAlg) \cong \bigoplus_{[\fr g]\in \uppi_0(\Lambda^2 G)} \!\!\! \Mod(\mathbb C[\overline{ \mathtt{Aut}(\fr g)}]^{\msf t^2(\pi){\ssss |}_{\fr g}})\cong
	\bigoplus_{[\fr g]\in \uppi_0(\Lambda^2 G)} \!\!\! \msf{Rep}(\mathtt{Aut}(\fr g), \msf t^2(\pi){\sss |}_\fr g) \, ,
\end{equation}
where $\msf{Rep}(\mathtt{Aut}(\fr g), \msf t^2(\pi){\sss |}_\fr g)$ is the category of $\msf t^2(\pi){\sss |_\fr g}$-projective representations of the stabilizer group $\mathtt{Aut}(\fr g) \subseteq G$ with $\msf t^2(\pi){\sss |}_\fr g$ the restriction in $H^2(\mathtt{Aut}(\fr g),\rU(1))$. It follows from this decomposition that the simple objects in $\Mod(\GrAlg)$, which encode the \emph{elementary} loop-like excitations of the model, correspond to pairs $([\fr g] \in \uppi_0(\Lambda^2G), \rho:\mathtt{Aut}(\fr g)\to {\rm End}(V_{[\fr g]}))$ such that $[\fr g]$ and $\rho$ provide the magnetic and electric quantum numbers, respectively.

The definition \eqref{eq:comultiplication} of the map $\Delta$ indicates that two (non necessarily elementary) loop-like excitations whose magnetic components are prescribed by a pair of objects $(\fr g, \fr g')$ can fuse if and only if $\fr g$ and $\fr g'$ are composable as morphisms in ${\rm End}(\Lambda G)$, i.e. they must share the same threading flux. Similarly, the braiding isomorphism that is prescribed by the invertible algebra element $\mathsf R$ introduced in \eqref{eq:Rmatrix} encodes the exchange of two loop-like excitations while being threaded by the same flux. Such a braiding precisely corresponds to the process depicted in \eqref{eq:necklaceDrawing} and thus the braided monoidal category $\Mod(\GrAlg)$, or equivalently $\ms Z(\VectGr)$, define representations of the linear necklace group.

We now would like to revisit this argument using the results established in this manuscript. We begin by introducing a pseudo-graphical calculus for $\MOD(\VectGr)$, where `pseudo' indicates here that this calculus should not be used to perform rigorous computations partly due to the weak associativity of the composition of 1-morphisms, but rather serves as a useful visual support. Recall from the previous section, that any object in $\MOD(\VectGr)$ can be obtained as the category $\Mod_{\VectGr}(\mc A_\phi)$ of right modules over an algebra object $\mc A_\phi$. As we mentioned before, these objects are interpreted as boundary conditions for a string-like excitation, such boundary conditions including in particular the allowed fluxes for the corresponding excitations as measured by a closed holonomy following the non-contractible cycle perpendicular to the length of the string.  
Henceforth, we shall represent such objects as coloured points, e.g.
\begin{equation}
	\label{eq:convPoints}
	\pointObj{1} \equiv \Mod_{\VectGr}(\mc A_\phi) \, , \q 	
	\pointObj{2} \equiv \Mod_{\VectGr}(\mc B_\psi) \, , \q 	
	\pointObj{3} \equiv \Mod_{\VectGr}(\mc C_\varphi) \, .
\end{equation}
Furthermore, recall that we have the equivalence
\begin{equation}
	\HOM_{\MOD(\VectGr)}(\Mod_{\VectGr}(\mc A_\phi), \Mod_{\VectGr}(\mc B_\psi))
	\cong \msf{Bimod}_{\VectGr}(\mc A_\phi, \mc B_\psi)
\end{equation}
for every pair of algebra objects $(\mc A_\phi, \mc B_\psi)$ so that every 1-morphism on the l.h.s., which is by definition a $\VectGr$-module functor, is specified by a choice of $(\mc A_\phi,\mc B_\psi)$-bimodule object $M_{\mc A_\phi \mc B_\psi}$. These 1-morphisms are interpreted as dyonic quantum numbers associated with string-like excitations whose endpoints are given by the corresponding objects. Such a dyonic quantum number contains a magnetic component that encodes the gauge orbit of parallel transports along the string, as well as an electric charge that decomposes the symmetries of the gauge actions on the corresponding strings \cite{Bullivant:2020xhy}. In the following, we shall utilise the following graphical notation to refer to such 1-morphisms:
\begin{equation}
	\stringHom{1} \, ,
\end{equation}
where the nomenclature is that of \eqref{eq:convPoints}. Given a triple $(\mc A_\phi, \mc B_\psi, \mc C_\varphi)$ of separable algebra objects the composition of two 1-morphisms $M_{\mc A_\phi \mc B_\psi}$ and $M_{\mc B_\psi \mc C_\varphi}$ specified by a choice of $(\mc A_\phi, \mc B_\psi)$- and $(\mc B_\psi,\mc C_\varphi)$-bimodules, respectively, results in a 1-morphism specified by a choice of $(\mc A_\phi, \mc C_\varphi)$-bimodule $M_{\mc A_\phi \mc B_\psi} \otimes_{\mc B_\psi} M_{\mc B_\psi \mc C_\varphi}$, whose precise definition can be found in \cite{Bullivant:2020xhy}. Graphically, we represent this composition via horizontal stacking as
\begin{equation}
	\stringHom{2} \, = \, \stringHom{3} \, .
\end{equation}
Similarly, morphisms of $\VectGr$-module functors
 are depicted as per the following example:
\begin{equation}
	\stringHom{3} \, \Rightarrow \, \stringHom{4} \, .
\end{equation}
As mentioned earlier, the associativity of the composition would need to be accounted for in order to make this graphical calculus more precise.

We showed in the previous section the equivalence $\msf{Dim} \, \MOD(\VectGr) \cong \ms Z(\VectGr)$, which relies on the fact that objects $(V,R_{V,-}) \in \Ob(\ms Z(\VectGr))$ are identified with (possibly zero) $\VectGr$-module endofunctors of all $\VectGr$-module categories that satisfy the pseudo-naturality conditions given in eq.~\eqref{eq:dimBraiding}. Graphically, given a choice of $\VectGr$-module category we depict the special kind of endofunctor identified with an object $(V,R_{V,-}) \in \Ob(\ms Z(\VectGr))$ as
\begin{equation}
	\label{eq:spe1mor}
	\stringHom{5} \, \equiv \, \loopHom{1}{1} \, .
\end{equation}
Let us clarify why we choose to represent such 1-morphisms in $\MOD(\VectGr)$ as loops. We reviewed above that objects in $\ms Z(\VectGr)$, or equivalently $\Mod(\GrAlg)$, encode the loop-like bulk excitations of the Hamiltonian realisation. More specifically, such an object is identified with the dyonic quantum number of a loop-like excitation that is constrained by the presence of a threading flux. The magnetic component of this loop-like excitation is measured by a closed holonomy that originates from the parallel transports along two string-like excitations whose endpoints were matched so as to obtain a loop-like object. On the other hand, the module category $\Mod_{\VectGr}(\mc A_\phi)$ encodes among other things a set of possible threading fluxes. In other words, the 1-morphisms under consideration correspond to loop-like excitations whose threading flux is constrained by the source object. Given two such 1-morphisms, the fusion of the corresponding loop-like excitations can then be depicted as the following composition
\begin{equation*}
	\loopHom{1}{4} \, = \, \stringHom{6} \, = \, \loopHom{1}{6} \, ,
\end{equation*}
such that only loop-like excitations with compatible threading fluxes can be fused.

Given two 1-morphisms of the form \eqref{eq:spe1mor}, the isomorphism \eqref{eq:dimBraidingSpe} induces the braiding isomorphism for the corresponding objects in $\ms Z(\VectGr)$, which in turns yields the following 2-isomorphism in $\MOD(\VectGr)$:
\begin{equation}
	\loopHom{1}{4}\,  \xRightarrow{\sim \,} \,  \loopHom{1}{5} \, .
\end{equation}
Given that the braiding isomorphism $R_{(V,R_{V,-}),(W,R_{W,-})}$ in $\ms Z(\VectGr)$ defines a representation of the braid group $\ms B_n$ and that $\ms B_n \simeq \ms L \ms N_n$, the 2-isomorphism presented above yields a representation of the linear necklace group $\ms L \ms N_n$.
More specifically, given any $n$-tuple of $\VectGr$-module endofunctors of $\Mod_{\VectGr}(\mc A_\phi)$ induced by objects $\{(V_{i},R_{V_i,-})\}_{i=1,\ldots,n}\subset {\rm Ob}(\ms{Z}(\VectGr))$, we can define an action of $\ms{B}_{n}$ on the object 
\begin{equation}
	-\otimes V_{1}\otimes\cdots V_{i}\otimes V_{i+1}\otimes \cdots V_{n}\in {\rm Ob}(\mathsf{End}_{\MOD(\VectGr)}(\Mod_{\VectGr}(\mc A_{\phi})))
\end{equation} 
by identifying the generator $\sigma_{i}\in \ms{B}_{n}$ with the natural isomorphism
\begin{align}
	\nn
	&{\rm id}_{-}\otimes {\rm id}_{V_{1}} \otimes \cdots\otimes R_{V_{i},V_{i+1}}\otimes \cdots \otimes {\rm id}_{V_{n}}:
	\\
	& \q -\otimes V_{1}\otimes\cdots \otimes V_{i}\otimes V_{i+1}\otimes \cdots \otimes V_{n}
	\xRightarrow{\sim \,}
	-\otimes V_{1}
	\otimes
	\cdots 
	\otimes
	V_{i}
	\otimes
	V_{i+1}
	\otimes
	\cdots \otimes V_{n}
\end{align}
in $\Hom(\mathsf{End}_{\MOD(\VectGr)}(\Mod_{\VectGr}(A_{\phi})))$, which permutes the $i$-th and ($i$+1)-th objects in the $n$-fold tensor product. It follows from the coherence relations satisfied by the braidings $R_{-,-}$ in the centre $\ms{Z}(\VectGr)$ that such $\VectGr$-natural isomorphisms satisfy the braid group relations {\small B1} and {\small B2}.
This process has an interpretation in terms of the \emph{Aharonov-Bohm} effect whereby a charge localised on a loop-like excitation is acted upon via the flux component of the other loop-like excitation \cite{preskill1993chromatic,deWildPropitius:1995hk,Bullivantthesis}.

\subsection{Loop braid group}\label{loopbraidrelations}

We conclude this section by showing that our construction yields representations of another motion group. Let $\Sigma= \mathbb D^3$ and $\Xi \subset \Sigma$ the disjoint union of $n$ pairwise unlinked `loops' obtained by removing the `necklace' $K_0$ from the $n$-component linear necklace. The motion group $\mathtt{Mot}(\mathbb D^3, \Xi)$ is known as the loop braid group $\ms L \ms B_n$. It admits a presentation in terms of generators $\{s_i,\sigma_i\}_{i=1,\ldots,n-1}$ subject to the relations
\begin{align*}
	\text{Braid group relations:}
	&\begin{cases}
		&\text{\small B1.} \;\; \sigma_{i}\sigma_{j}=\sigma_{j}\sigma_{i} \, ,\q {\rm for} \; |i-j|>1,
		\\
		&\text{\small B2.} \;\; \sigma_{i}\sigma_{i+1}\sigma_{i}=\sigma_{i+1}\sigma_{i}\sigma_{i+1},
	\end{cases}
	\\
	\text{Symmetric group relations:}
	&\begin{cases}
		&\text{\small S1.} \;\; s_{i}s_{j}=s_{j}s_{i} \, \q  {\rm for} \; |i-j|>1,
		\\
		&\text{\small S2.} \;\; s_{i}s_{i+1}s_{i}=s_{i+1}s_{i}s_{i+1},
		\\
		&\text{\small S3.} \;\; s_{i}^2=1,
	\end{cases}
	\\
	\text{Mixed relations:}
	&\begin{cases}
		&\text{\small M1.} \;\; \sigma_{i}s_{j}=s_{j}\sigma_{i} \, , \q {\rm for} \; |i-j|>1,
		\\
		&\text{\small M2.} \;\; s_{i}s_{i+1}\sigma_{i}=\sigma_{i+1}s_{i}s_{i+1},
		\\
		&\text{\small M3.} \;\; s_{i}\sigma_{i+1}\sigma_{i}=\sigma_{i+1}\sigma_{i}s_{i+1}.
	\end{cases}
\end{align*}
The generators $\sigma_i$ and $s_i$ correspond in this context to elementary motions which can be  visualized in terms of worldsheets of the corresponding loops as
\begin{align}
	\sigma_i = \loopTube{0.27}{1}\;\;  \ldots \loopBraid{0.27}{1} \ldots \;\; \loopTube{0.27}{n}
	\q\q {\rm with} \q
	\loopBraid{0.27}{1} \leftrightarrow \movieLoop{0.27}{1} \, ,
\end{align}	
and
\begin{align}
	s_i =\loopTube{0.27}{1} \;\; \ldots \loopBraid{0.27}{2} \ldots \;\; \loopTube{0.27}{n}
	\q\q {\rm with} \q
	\loopBraid{0.27}{2} \leftrightarrow \movieLoop{0.27}{2}  \, ,
\end{align}
respectively. We represented on the r.h.s. the `movie' diagrams that display projections of time slices of the corresponding four-dimensional worldsheets following the conventions of \cite{Baez:2006un}. Both moves correspond to the exchange of two loops. However, $\sigma_{i}$ corresponds to the loop $i$+1 braiding with loop $i$ by passing through its centre, whereas $s_{i}$ corresponds to the exchange of loops $i$ and $i$+1 without threading through each other. The graphical presentation provided above can be conveniently used to graphically represent the eight defining relations of the group \cite{Baez:2006un}. Furthermore, we can easily show that the braid group $\ms B_{n}$ is the subgroup of $\ms L \ms B_n$ generated by $\{\sigma_{i}\}_{i=1,...,n-1}$, while the symmetric group $\ms S_n$ is the subgroup generated by $\{s_{i}\}_{i=1,...,n-1}$. 

By construction, the loop braid group differs as a motion group from the linear necklace group by the fact that the pairwise unlinked loops are not linked to the necklace which passes through the $x$-axis, allowing for the symmetric exchanges governed by the generators $\{s_i\}_{i=1,\ldots,n-1}$. Physically, the role of the necklace is played by the threading flux. This suggests that loop-like excitations whose threading fluxes are trivial should yield representations of the loop braid group. Given the decomposition \eqref{eq:decompMod}, considering loop-like excitations with trivial threading fluxes amounts to restricting the direct sum to connected components of objects in $\Lambda^2G$ of the form $\mathbbm 1 \! \xrightarrow{g}$ with $g \in G$ so that $\Mod(\GrAlg)$ reduces to
\begin{equation}
	\label{eq:double}
	\bigoplus_{[\mathbbm 1 \xrightarrow{g}]\in \uppi_0(\Lambda^2 G)} \!\!\!\! \msf{Rep}(\mathtt{Aut}(\mathbbm 1 \! \xrightarrow{g})) \cong
	\bigoplus_{[g] \in \uppi_0 (\Lambda G)} \!\! \msf{Rep}(\mathtt{Aut}(g)) \cong \Mod(\mathbb C[\Lambda G]) \cong \ms Z(\msf{Vec}_G) \, ,
\end{equation}
where we used the fact that the loop-groupoid cocycle $\msf t(\pi)$ is normalised so that it equals the unit in $\rU(1)$ whenever any of the arguments is an identity morphism. In the equation above, $[g] \in \uppi_0(\Lambda G)$ simply corresponds to a \emph{conjugacy class} of $G$ and $\mathtt{Aut}(g)$ to the corresponding \emph{stabilizer} subgroup. As explained earlier, when thinking of $\Mod(\GrAlg)$ as the dimension of $\MOD(\VectGr)$, a choice of threading flux is constrained by a choice of object. We recover the loop-like excitations with trivial threading flux by choosing the $\VectGr$-module category to be the category of module objects over the algebra object $\mc A^0 \equiv (\bigoplus_{g \in G} \mathbb C_{\mathbbm 1 \xrightarrow{g}},m,u)$ with
\begin{equation}
	\begin{array}{ccccl}
		m & : & \mc{A}^0 \otimes \mc{A}^0 & \to & \mc{A}^0
		\\
		& : & \; \mathbb C_{\mathbbm 1\xrightarrow{g}} \otimes \mathbb C_{\mathbbm 1 \xrightarrow{g'}} &\mapsto & \mathbb C_{\mathbbm 1 \xrightarrow{gg'}}
	\end{array} 
	\q {\rm and} \q 
	u(\mathbbm{1}_{\VectGr}):=\mathbb C_{\mathbbm 1 \xrightarrow{\mathbbm 1}}\, .
\end{equation}
Notice that $\VectGr$-module endofunctors of the category of module objects over $\mc A^0$ induced from $n$-tuple objects $\{(V_i,R_{V_i,-})\}_{i=1,\ldots,n}$ in the centre $\ms{Z}(\VectGr))$ are only non-zero if for all $i=1,\ldots,n$ we have the following isomorphism of objects in $\VectGr$:
\begin{align}
	\label{eq:fluxtrivialcond}
	V_{i}\otimes \mathbb{C}_{\mathbbm{1}\xrightarrow{\mathbbm{1}}}
	\simeq
	V_{i} \, .
\end{align}
The full braided fusion subcategory of $\ms{Z}(\VectGr)$ given by objects satisfying the condition \eqref{eq:fluxtrivialcond} can then be checked to be equivalent, as a braided fusion category, to $\ms{Z}(\Vect_{G})$.

As previously, we define an action of $\ms{B}_{n}$ on the $n$-fold monoidal product $- \otimes V_1 \otimes \cdots \otimes V_n$ of objects in $\mathsf{End}_{\MOD(\VectGr)}(\Mod_{\VectGr}(\mc{A}^0))$ induced from $n$-tuple objects $\{(V_i,R_{V_i,-})\}_{i=1,\ldots,n} \subset \Ob(\ms{Z}(\VectGr))$. Let us now demonstrate that the module endofunctor $- \otimes V_1 \otimes \cdots \otimes V_n$ admits an action of the symmetric group $\ms{S}_{n}$ in addition to that of $\ms{B}_{n}$. Given the generator $s_{i}\in\ms{S}_{n}$, we identify the corresponding natural isomorphism of $\VectGr$-module functors defined on the single simple object of $\Mod_{\VectGr}(\mc A^0)$ provided by the regular $\mc A^0$-module object $(\mc{A}^0,m)$, and a fortiori all objects, via
\begin{align}
	\nn
	&\mc{A}^0\otimes V_{1}\otimes\cdots \otimes V_{i}\otimes V_{i+1}\otimes \cdots \otimes V_{n}
	\\
	\nn
	& \q = 
	\bigoplus_{\substack{
			g,\{g_{i}\}\in G
		}}
	\mathbb{C}_{\mathbbm{1}\xrightarrow{g}}\otimes (V_1)_{\mathbbm{1}\xrightarrow{g_{1}}}\otimes\cdots \otimes (V_i)_{\mathbbm{1}\xrightarrow{g_{i}}}\otimes (V_{i+1})_{\mathbbm{1}\xrightarrow{g_{i+1}}}\otimes \cdots \otimes (V_n)_{\mathbbm{1}\xrightarrow{g_{n}}}
	\\
	\nn
	& \q \xRightarrow{\sim \, } \hspace{-4pt}
	\bigoplus_{\substack{
			\tilde{g},\{g_{i}\}\in G
	}}
	\mathbb{C}_{\mathbbm{1}\xrightarrow{\tilde g}}\otimes (V_1)_{\mathbbm{1}\xrightarrow{g_{1}}}\otimes\cdots \otimes (V_{i+1})_{\mathbbm{1}\xrightarrow{g_{i+1}}}\otimes (V_{i})_{\mathbbm{1}\xrightarrow{g_{i}}}\otimes \cdots \otimes (V_n)_{\mathbbm{1}\xrightarrow{g_{n}}}
	\\
	& \q =
	\mc{A}^0\otimes V_{1}\otimes\cdots \otimes  V_{i+1}\otimes V_{i}\otimes \cdots \otimes V_{n} \, ,
\end{align} 
where $\tilde{g}=gg_{1}\cdots g_{i+1}g_{i}g^{-1}_{i+1}g^{-1}_{i}\ldots g^{-1}_{1}$ and $\xRightarrow{\sim}$ denotes a grading preserving isomorphism. It is straightforward to verify that such an action satisfies the symmetric group relations {\small S1}, {\small S2} and {\small S3}. Furthermore, a direct computation demonstrates that the actions of $\ms{B}_{n}$ and $\ms{S}_{n}$ satisfy the mixed relations {\small M1}, {\small M2} and {\small M3}, thus establishing a representation of $\ms{LB}_{n}$ inside $\mathsf{End}_{\MOD(\VectGr)}(\Mod_{\VectGr}(\mc{A}^0))$. Let us emphasize that in general, for an arbitrary choice of group $G$ and  $[\pi]\in H^{4}(G,\rU(1))$, the symmetric group action will fail to define a morphism in $\mathsf{End}_{\MOD(\VectGr)}(\Mod_{\VectGr}(\mc{A}_{\phi}))$ for a choice of algebra object $A_{\phi}$ not Morita equivalent to $\mc{A}^0$. Physically, the action of the symmetric group defined above is interpreted as the exchange of two loop-like excitations whose internal Hilbert spaces are transposed while acting as the identity on the corresponding states.

Note that it had been established before that braid group representations arising from the data of $\ms{Z}(\Vect_{G})$ could be lifted to representations of the loop braid group \cite{kadar2017local,ainsworth2014optimising,Bullivant:2018pju,Bullivant:2019fmk}. However, it was not known in which categorical framework such extended representations existed. Here, we presented one answer by demonstrating that such representations of $\ms{LB}_{n}$ are defined in terms of $\VectGr$-module endofunctors of $\mathsf{Mod}_{\VectGr}(\mc{A}^0)$ by analogy with bulk loop-like excitations in the lattice Hamiltonian realisation of (3+1)d Dijkgraaf-Witten theory.

\bigskip\bigskip
\begin{center}
	\textbf{Acknowledgements}
\end{center}

\noindent
CD is funded by the Deutsche Forschungsgemeinschaft (DFG, German Research Foundation) under Germany’s Excellence Strategy – EXC-2111 – 390814868. AB acknowledges financial support from Science Foundation Ireland through Principal Investigator Award 16/IA/4524.

\newpage
\renewcommand*\refname{\hfill References \hfill}
\bibliographystyle{JHEP}
\bibliography{ref_cat}

\end{document}